\documentclass[reprint,superscriptaddress]{revtex4-2}
\usepackage[utf8]{inputenc}
\usepackage{color}
\usepackage{array}
\usepackage{booktabs}
\usepackage{multirow}
\usepackage{amsmath}
\usepackage{amssymb}
\usepackage{graphicx}
\usepackage[colorlinks=true,linkcolor=blue,anchorcolor=red,citecolor=blue,urlcolor=blue]{hyperref}

\makeatletter

\providecommand{\tabularnewline}{\\}
\let\Re\relax
\let\Im\relax
\DeclareMathOperator{\Re}{Re}
\DeclareMathOperator{\Im}{Im}
\DeclareMathOperator{\Tr}{Tr}

\makeatother

\begin{document}
\title{Impurity and dispersion effects on the linear magnetoresistance in the quantum limit}
\author{Shuai Li}
\affiliation{Department of Physics, Harbin Institute of Technology, Harbin 150001, China}
\affiliation{Shenzhen Institute for Quantum Science and Engineering and Department of Physics, Southern University of Science and Technology (SUSTech), Shenzhen 518055, China}

\affiliation{Quantum Science Center of Guangdong-Hong Kong-Macao Greater Bay Area (Guangdong), Shenzhen 518045, China}

\affiliation{Shenzhen Key Laboratory of Quantum Science and Engineering, Shenzhen 518055, China}

\affiliation{International Quantum Academy, Shenzhen 518048, China}
\author{Hai-Zhou Lu}
\email{luhz@sustech.edu.cn}

\affiliation{Shenzhen Institute for Quantum Science and Engineering and Department of Physics, Southern University of Science and Technology (SUSTech), Shenzhen 518055, China}

\affiliation{Quantum Science Center of Guangdong-Hong Kong-Macao Greater Bay Area (Guangdong), Shenzhen 518045, China}

\affiliation{Shenzhen Key Laboratory of Quantum Science and Engineering, Shenzhen 518055, China}

\affiliation{International Quantum Academy, Shenzhen 518048, China}

\author{X. C. Xie}
\affiliation{International Center for Quantum Materials, School of Physics, Peking University, Beijing 100871, China}
\affiliation{Institute for Nanoelectronic Devices and Quantum Computing, Fudan University, Shanghai 200433, China}
\affiliation{Hefei National Laboratory, Hefei 230088, China}

\begin{abstract}
Magnetoresistance, that is, the change of the resistance with the magnetic field, is usually a quadratic function of the field strength. A linear magnetoresistance usually reveals extraordinary properties of a system. In the quantum limit where only the lowest Landau band is occupied, a quantum linear magnetoresistance was believed to be the signature of the Weyl fermions with 3D linear dispersion. Here, we comparatively investigate the quantum-limit magnetoresistance of systems with different band dispersions as well as different types of impurities. We find that the magnetoresistance can also be linear for the quadratic energy dispersion. We show that both longitudinal and transverse magnetoresistance can be linear if long-range-Gaussian-type impurities dominate, but Coulomb-type impurities can only induce linear transverse magnetoresistance. 
Moreover, we find a negative longitudinal magnetoresistance in massless Dirac fermions, regardless of the impurity type, as a result of the combined effect of the linear dispersion and the scattering mechanism. 
Our findings well explain some of the linear magnetoresistance observed in the experiments and provide insights to the understanding of quantum-limit magnetoresistance.
\end{abstract}
\maketitle

\section{INTRODUCTION}

Magnetoresistance effects can often reveal nontrivial characteristics of emergent materials. Among them, the linear magnetoresistance is of particular interest. Specifically, while the magnetoresistance of ordinary materials generally increases quadratically with the magnetic field, it increases linearly for some topological materials \cite{AbrikosovA_PRB_1998}. The linear magnetoresistance has been attracting considerable attention since it was found in silver chalcogenides \cite{XuR_N_1997} and later in various systems, including Dirac/Weyl semimetals \cite{HeL_PRL_2014,NarayananA_PRL_2015,XiangZ_PRL_2015,FengJ_PRB_2015,ZhaoY_PRX_2015,LiangT_NM_2015,LiH_NC_2016,TakiguchiK_NC_2020,ZhuW_PRB_2022}, nodal-line semimetals \cite{LahaA_PRB_2020,YangJ_NL_2021}, graphene \cite{WuJ_C_2019}, superconductors \cite{NiuQ_NC_2017,GiraldoGalloP_S_2018,SarkarT_SA_2019,ZhangW_PRB_2020,MaksimovicN_PRX_2020}, density-wave materials \cite{KolincioK_PRL_2020}, and magnetic topological materials \cite{LeiX_PRB_2020,CampbellD_nQM_2021}. 

\begin{table*}
\caption{\label{tab:Theories-of-linear}Comparison of the theories on the linear magnetoresistance in the literature and this paper.}

\begin{ruledtabular}

\begin{tabular}{ccccc}
Ref. & Field direction & Origin of the linear magnetoresistivity & Field strength & Mechanism\tabularnewline
\midrule
\cite{AbrikosovA_PRB_1998,KlierJ_PRB_2015,XiaoX_PRB_2017,KoenyeV_PRB_2018,RodionovY_PRB_2020} & Transverse & Linear dispersion with Coulomb-type impurities  & Quantum limit & Quantum\tabularnewline
\cite{ParishM_N_2003,ParishM_PRB_2005,HuJ_PRB_2007,XuJ_JoAP_2008,RamakrishnanN_PRB_2017,KisslingerF_PRB_2017,ChenS_CPB_2022} & Transverse & Carrier density fluctuations in inhomogeneous systems  & Strong  & Classical\tabularnewline
\cite{AlekseevP_PRL_2015,AlekseevP_PRB_2017} & Transverse & Near charge neutrality of finite-size samples  & Strong  & Classical\tabularnewline
\cite{SongJ_PRB_2015} & Transverse & Guiding center diffusion in a smooth random potential  & Strong  & Semiclassical\tabularnewline
\cite{XiaoC_PRB_2020} & Transverse & Intra-scattering semiclassics of Bloch electrons  & Weak & Semiclassical\tabularnewline
\midrule 
\multirow{4}{*}{This paper} & Transverse & Quadratic dispersion with long-Gaussian-type impurities & \multirow{4}{*}{Quantum limit} & \multirow{4}{*}{Quantum}\tabularnewline
 & Longitudinal & Quadratic dispersion with long-Gaussian-type impurities &  & \tabularnewline
 & Transverse & Quadratic dispersion with Coulomb-type impurities  &  & \tabularnewline
 & Transverse & Linear dispersion with long-Gaussian-type impurities  &  & \tabularnewline
\end{tabular}

\end{ruledtabular}
\end{table*}

%In recent years, the magnetoresistance of topological materials has been intensively studied. 
Many explanations have been proposed for the origin of the linear magnetoresistance (see Table~\ref{tab:Theories-of-linear}). Abrikosov's quantum magnetoresistance theory \cite{AbrikosovA_PRB_1998} is believed to be a signature for the 3D massless Weyl fermions \cite{HeL_PRL_2014,XiangZ_PRL_2015,FengJ_PRB_2015,ZhaoY_PRX_2015,LiangT_NM_2015,LiH_NC_2016,TakiguchiK_NC_2020}; the classical theory of Parish and Littlewood \cite{ParishM_N_2003} is often used to explain the linear magnetoresistance in highly inhomogeneous systems \cite{NarayananA_PRL_2015,KhouriT_PRL_2016,WuJ_C_2019,ZhuW_PRB_2022,MallikS_NL_2022}; the classical theory of Alekseev \emph{et al}. \cite{AlekseevP_PRL_2015} can explain the linear magnetoresistance in compensated systems \cite{VasilevaG_PRB_2016,Chandan_JoPDAP_2020}; the semiclassical theories by Song \emph{et al}. \cite{SongJ_PRB_2015} and Xiao \emph{et al}. \cite{XiaoC_PRB_2020} explain the linear magnetoresistance under semiclassical strong and weak magnetic fields, respectively. However, the previous theories focus on the transverse linear magnetoresistance, leaving the field dependence of the corresponding longitudinal magnetoresistance undiscussed. The observations of the longitudinal linear  magnetoresistance \cite{WangY_NC_2016,LahaA_PRB_2020,Chandan_JoPDAP_2020,ZhuW_PRB_2022} remain unexplained. In addition, the impurity effects on the magnetoresistance of different electronic structures remains unexplored in a comprehensive way. A comparative study on the linear magnetoresistance along both the longitudinal and transverse magnetic-field directions for different types of impurity and electronic structure is highly needed. 

In this paper, we systematically investigate the longitudinal and transverse magnetoresistance of typical three-dimensional (3D) energy dispersions (including massless Dirac fermions, Dirac/Weyl semimetals, and conventional electron gas) in the strong magnetic field quantum limit. We consider weak but different kinds of impurity potentials. 
Some distinct field dependencies of magnetoresistance are found. For massless Dirac fermions, there is always a negative longitudinal magnetoresistance regardless of the impurity type; however, the corresponding transverse magnetoresistance varies with the impurity type. This is a characteristic of the lowest Landau band with linear dispersion. 
For Dirac/Weyl semimetals and electron gas, both longitudinal and transverse magnetoresistance can be linear in magnetic field if long-range-Gaussian-type impurities dominate; but Coulomb-type impurities can only lead to linear transverse magnetoresistance. These field dependencies of magnetoresistance well explain some experimental observations \cite{WangY_NC_2016,LahaA_PRB_2020,Chandan_JoPDAP_2020,ZhuW_PRB_2022}. Furthermore, we present a clear and standard procedure to find the quantum-limit magnetoresistance, and we give many general formulas that can be easily applied to other systems.

The paper is organized as follows. In Sec.~\ref{sec:Field-dependence}, we summarize the field dependencies of the quantum-limit magnetoresistance for different systems with different impurity types. In Sec.~\ref{sec:MC-in-QL}, we present models for massless Dirac fermions, Dirac/Weyl semimetals, and electron gas, along with their Landau bands and eigenvectors. The impurity models are also introduced. Then, we derive the corresponding magnetoconductivities and analyze them in detail. In Sec.~\ref{sec:conclusion-and-discussion}, we conclude with a discussion of the results we find and some possible future research directions. Calculation details are provided in Appendices~\ref{sec:The-critical-magnetic}--\ref{sec:Hall-conductivity-in}. 

\section{Summary of the results on the quantum-limit magnetoresistance \label{sec:Field-dependence}}

Magnetoresistance can be significantly different for systems with different energy band dispersions. Here, three representative 3D systems are investigated, massless Dirac fermions with linear dispersion, two-node Dirac/Weyl semimetals where the two nodes merge together at higher energies, and electron gas with quadratic dispersion. Under magnetic fields, 3D energy dispersion develops into one-dimensional Landau bands, which disperse along the magnetic field direction. When the magnetic field is strong enough, only the lowest Landau band is occupied, i.e., the system enters the quantum-limit regime. For massless Dirac fermions, the lowest Landau band is linear, while both two-node Dirac/Weyl semimetals and electron gas have the quadratically-dispersed lowest Landau band. 

The longitudinal magnetoresistivity is inversely proportional to the longitudinal magnetoconductivity, i.e., $\rho_{zz}=1/\sigma_{zz}$ (the magnetic field is along $z$-direction). The transverse magnetoresistivity is related to magnetoconductivities through $\rho_{xx}=\sigma_{xx}/(\sigma_{xy}^{2}+\sigma_{xx}^{2})$. While the leading order of Hall conductivity is intrinsic, the longitudinal and transverse magnetoconductivities are strongly dependent on the transport/scattering time. Therefore, impurity effect plays a pivotal role in determining the field dependence of the magnetoresistance. Three types of impurity potentials are studied here: delta potential, Gaussian potential, and screened Coulomb potential. These potentials can model point defects in crystals, such as vacancies and interstitials \cite{Ashcroft_1976,XuY_NC_2017}. Specifically, the screened Coulomb potential can model the charged defects.

\begin{table}[b]
\caption{\label{tab:LMR}Magnetic field dependence of the longitudinal magnetoresistance in the quantum limit. Three different types of impurity potentials are listed here: delta potential, long-range Gaussian potential, and screened Coulomb potential. $B$ is the magnetic field strength. For two-node Dirac/Weyl semimetals or electron gas with screened-Coulomb-type impurities, the longitudinal magnetoresistance is proportional to $B^{a}$, where $a$ can be positive or negative, depending on the screening length.}

\begin{ruledtabular}

\begin{tabular}{>{\raggedright}p{0.5\columnwidth}>{\raggedright}m{0.15\columnwidth}>{\raggedright}p{0.15\columnwidth}>{\raggedright}p{0.15\columnwidth}}
 & \multirow{2}{0.15\columnwidth}{Delta} & Long & Screened\tabularnewline
 &  & Gaussian & Coulomb\tabularnewline
 \hline 
Massless Dirac fermions & $B^{-1}$ & $B^{-1}$ & $B^{-1}$\tabularnewline
Two-node Dirac/Weyl & \multirow{2}{0.15\columnwidth}{$B^{2}$} & \multirow{2}{0.15\columnwidth}{$B$} & \multirow{2}{0.15\columnwidth}{$B^{a}$}\tabularnewline
semimetals and electron gas &  &  & \tabularnewline
\end{tabular}

\end{ruledtabular}
\end{table}

With the magnetoconductivities derived from the Kubo formula, we present in Table~\ref{tab:LMR} the magnetic field dependence of the quantum-limit longitudinal magnetoresistance for different systems when different types of impurity scattering dominate. The corresponding analytical expressions and detailed analysis are shown in Sec.~\ref{sec:MC-in-QL}. 

\begin{figure}
\includegraphics{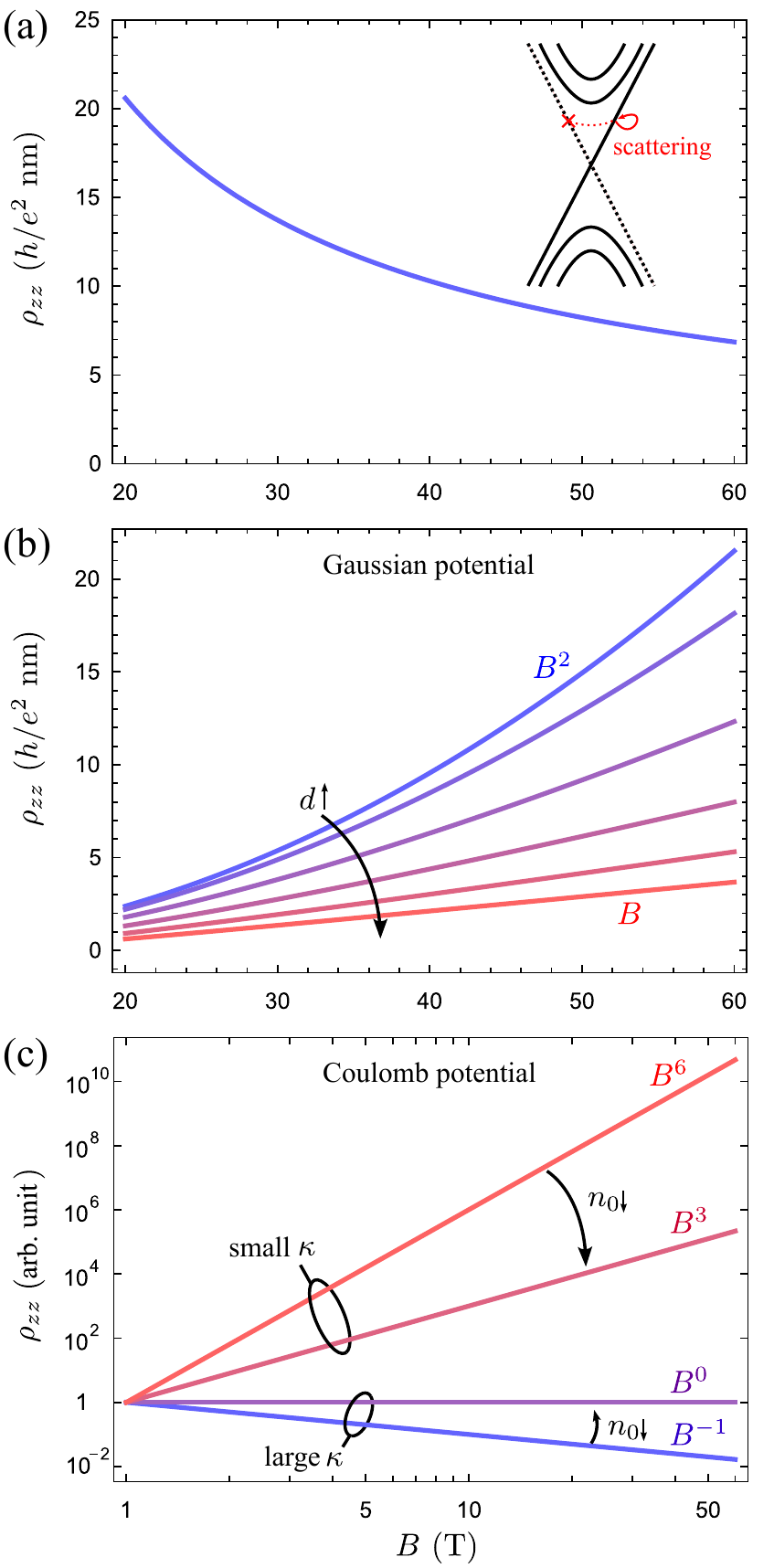}

\caption{\label{fig:Magnetic-field-dependence-LMR}Magnetic field dependence of the quantum-limit longitudinal magnetoresistivity of (a) massless Dirac fermions, (b) two-node Dirac/Weyl semimetals when the Gaussian-type impurities dominate, and (c) two-node Dirac/Weyl semimetals when the screened-Coulomb-type impurities dominate. (a) The magnetoresistivity is independent on the impurity type; the mean free path is taken as 10~nm. The inset shows that the scattering between the lowest Landau bands with different chirality is not allowed. (b) The decay length $d$ of the Gaussian potential is taken as $0,1,2,3,4$ and 5~nm. The carrier concentration $n_{0}$ is taken as $10^{-4}$~$\mathrm{nm^{-3}}$; the impurity parameter $n_{i}u_{0}^{2}$ is taken as 1 $\mathrm{eV^{2}\,nm^{3}}$, and the model parameter $M=5$~$\mathrm{eV\,nm^{2}}$. (c) Sketch for the field dependence of magnetoresistivity under different limits. $n_{0}$ is the carrier concentration, and  $\kappa$ is the reciprocal Debye screening radius.}
\end{figure}

For massless Dirac fermions, the longitudinal magnetoresistivity always decreases with increasing magnetic field, regardless of the impurity type. The reason is that the backward scattering (scattering between $k_{F}$ and $-k_{F}$) is prohibited in the linearly-dispersed lowest Landau band, as shown in the inset of Fig.~\ref{fig:Magnetic-field-dependence-LMR}(a). Therefore, the magnetoresistivity is only affected by the field-independent band broadening effect. This is guaranteed by the vertex correction in the calculation. If there is a finite mass term, the lowest Landau bands of opposite chirality will couple together \cite{WangH_PRB_2018,KoenyeV_PRB_2018}, and the backward scattering is allowed. The impurity configuration will then play a role in the field dependence of the longitudinal magnetoresistance.

For two-node Dirac/Weyl semimetals and electron gas, their longitudinal magnetoresistance have the same the field dependence in the quantum limit, due to the similar quadratic dispersion of their lowest Landau bands. As shown in Fig.~\ref{fig:Magnetic-field-dependence-LMR}(b), the magnetoresistivity is quadratic in $B$ for delta-type impurities (when the decay length of Gaussian potential is set to zero, i.e., $d=0$), and it evolves from the quadratic form to a linear form when the decay length of Gaussian potential is increased. When the screened-Coulomb-type impurities dominate, the magnetoresistivity is generally non-monotonic. However, in the long screening length limit, i.e., small reciprocal Debye screening radius $\kappa$, the magnetoresistivity increases with increasing magnetic field; in the short screening length limit, i.e., large $\kappa$, negative magnetoresistance exists, but becomes weakly $B$ dependent at extreme low carrier concentration. The magnetic field dependence of the longitudinal magnetoresistance in the presence of the screened-Coulomb-type impurities is illustrated in Fig.~\ref{fig:Magnetic-field-dependence-LMR}(c).

\begin{figure}
\includegraphics{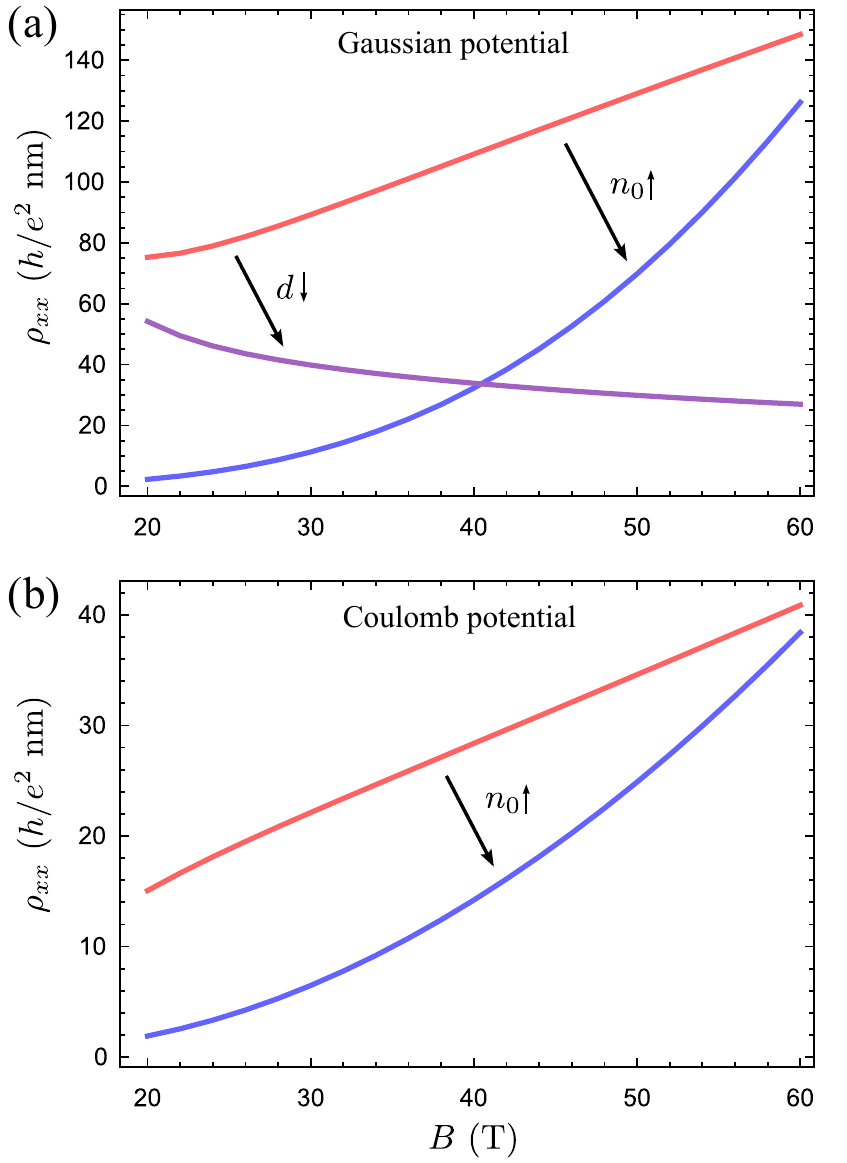}

\caption{\label{fig:Magnetic-field-dependence-TMR}Transverse magnetoresistivity of two-node Dirac/Weyl semimetals in the quantum limit with (a) Gaussian-type impurities dominance, and (b) screened-Coulomb-type impurities dominance. The carrier concentration $n_{0}$ is taken as $10^{-5}$~$\mathrm{nm^{-3}}$ (red line) and $10^{-4}$~$\mathrm{nm^{-3}}$ (blue line). For Gaussian potential, the decay length $d$ is taken as 5 nm (red line) and 1 nm (purple line), and the impurity parameter $n_{i}u_{0}^{2}$ is taken as 1 $\mathrm{eV^{2}\,nm^{3}}$. For screened Coulomb potential, the relative permittivity $\varepsilon_{r}$ is taken as 10, and impurity concentration $n_{i}$ is taken as $0.1n_{0}$. Model parameters $M=5$~$\mathrm{eV\,nm^{2}}$, $A=0.5$~$\mathrm{eV\,nm}$, and $k_{w}=0.1$~$\mathrm{nm^{-1}}$.}
\end{figure}

For all three systems studied, both long-Gaussian-type and screened-Coulomb-type impurities can bring linear transverse magnetoresistance at low carrier concentration. Figure~\ref{fig:Magnetic-field-dependence-TMR} shows the transverse magnetoresistivity of two-node Dirac/Weyl semimetals. When increasing the carrier concentration, the magnetoresistivity deviates from the linear $B$ dependence for both Gaussian and screened Coulomb potentials. Delta-type impurities, on the contrary, result in a linear transverse magnetoconductivity at low carrier concentration, and the corresponding transverse magnetoresistance decreases with increasing magnetic field. 

\section{Calculation of the magnetoconductivity in the quantum limit \label{sec:MC-in-QL}}

\subsection{Models and Landau bands}

With a zero-mass term, the Dirac Hamiltonian is decoupled into two parts with opposite chirality \cite{Shen_2017,WangH_PRB_2018,KoenyeV_PRB_2018}. Each part describes a massless Weyl cone
\begin{equation}
H^{L}=\hbar v_{F}\mathbf{k}\cdot\boldsymbol{\sigma},\label{eq:H-L}
\end{equation}
where $\hbar$ is the reduced Planck constant, $v_{F}$ is the Fermi velocity, $\mathbf{k}=\left(k_{x},k_{y},k_{z}\right)$ is the wave vector, and $\boldsymbol{\sigma}=\left(\sigma_{x},\sigma_{y},\sigma_{z}\right)$ is the vector form of the Pauli matrix. This low-energy effective Hamiltonian is favored among many studies because its simplicity allows a lot analytical deductions. Furthermore, it directly models materials with a single node around the Fermi energy, such as $\mathrm{Ag_{2}Se}$ \cite{XuR_N_1997,AbrikosovA_PRB_1998} and $\mathrm{ZrTe_{5}}$ \cite{ChenR_PRL_2015,ManzoniG_PRL_2016}. To directly capture two Weyl nodes, a two-node model is often used,
\begin{equation}
H^{Q}=A\left(k_{x}\sigma_{x}+k_{y}\sigma_{y}\right)+M\left(k_{w}^{2}-k_{x}^{2}-k_{y}^{2}-k_{z}^{2}\right)\sigma_{z},\label{eq:H-Q}
\end{equation}
where $A,M,$ and $k_{w}$ are model parameters. This Hamiltonian describes two Weyl nodes separated along $k_{z}$ with a distance of $k_{w}$. In the momentum space, these two Weyl nodes act as a source and a sink of Berry curvature, respectively \cite{LuH_PRB_2015}. In addition, this two-node model supports special surface states, i.e., the Fermi arc \cite{ZhangS_NJoP_2016}. Combined with its time-reversal partner, Eq.~(\ref{eq:H-Q}) can model two-node Dirac semimetals like $\mathrm{Cd_{3}As_{2}}$ and $\mathrm{Na_{3}Bi}$ \cite{ArmitageN_RMP_2018}. When the magnetic field is extremely strong, the lowest Landau band of the time-reversal partner of Eq.~(\ref{eq:H-Q}) is buried in the Fermi sea; therefore, it is not considered in the following calculations of magnetoconductivities.

Under the $z$-directional magnetic field, $\mathbf{B}=\left(0,0,B\right)$, the vector potential in the Landau gauge is $\mathbf{A}=\left(-yB,0,0\right)$. For both Eq.~(\ref{eq:H-L}) and Eq.~(\ref{eq:H-Q}), the eigenvectors have the below form: 
\begin{align}
\left|k_{x},k_{z},n+\right\rangle  & =\begin{pmatrix}\cos\frac{\theta_{k_{z},n}}{2}\left|n-1\right\rangle \\
\sin\frac{\theta_{k_{z},n}}{2}\left|n\right\rangle 
\end{pmatrix}\left|k_{x},k_{z}\right\rangle ,\nonumber \\
\left|k_{x},k_{z},n-\right\rangle  & =\begin{pmatrix}-\sin\frac{\theta_{k_{z},n}}{2}\left|n-1\right\rangle \\
\cos\frac{\theta_{k_{z},n}}{2}\left|n\right\rangle 
\end{pmatrix}\left|k_{x},k_{z}\right\rangle ,\nonumber \\
\left|k_{x},k_{z},0\right\rangle  & =\begin{pmatrix}0\\
\left|0\right\rangle 
\end{pmatrix}\left|k_{x},k_{z}\right\rangle ,\label{eq:eigenvectors}
\end{align}
where integer $n\ge1$. The matrix representation and Dirac notation are used here. The band index is denoted by $n\pm$ or $0$. For the one-node model, $\cos\theta_{k_{z},n}=k_{z}/\sqrt{k_{z}^{2}+2n/l_{B}^{2}}$ with the magnetic length $l_{B}=\sqrt{\hbar/(eB)}$ ($e$ is the electron charge); the Landau bands can be found as
\begin{align}
E_{k_{z},n\pm}^{L} & =\pm\hbar v_{F}\sqrt{k_{z}^{2}+\frac{2n}{l_{B}^{2}}},\nonumber \\
E_{k_{z},0}^{L} & =-\hbar v_{F}k_{z}.\label{eq:dispersion-L}
\end{align}
For the two-node model, $\cos\theta_{k_{z},n}=M_{n}/\sqrt{M_{n}^{2}+2nA^{2}/l_{B}^{2}}$ with $M_{n}=M\left(k_{w}^{2}-k_{z}^{2}\right)-n\omega$ and $\omega=2M/l_{B}^{2}$; the Landau bands can be found as
\begin{align}
E_{k_{z},n\pm}^{Q} & =\frac{\omega}{2}\pm\sqrt{M_{n}^{2}+2n\frac{A^{2}}{l_{B}^{2}}},\nonumber \\
E_{k_{z},0}^{Q} & =\frac{\omega}{2}+M\left(k_{z}^{2}-k_{w}^{2}\right).\label{eq:dispersion-Q}
\end{align}
The expressions of Landau bands are independence of the quantum number $k_{x}$; each bands has a degeneracy of $N_{L}=eB/h$, where $h$ is the Planck constant.

Compared to the above two models, the Hamiltonian of 3D electron gas is much simpler 
\begin{equation}
H^{EG}=\frac{\hbar^{2}k^{2}}{2m},
\end{equation}
where $k=\sqrt{k_{x}^{2}+k_{y}^{2}+k_{z}^{2}}$, and $m$ is the effective mass. It only has a simple quadratically-dispersed energy band. Under the magnetic field, the electron gas has Landau bands 
\begin{equation}
E_{k_{z},n}^{EG}=\frac{\hbar^{2}k_{z}^{2}}{2m}+\frac{\hbar^{2}}{ml_{B}^{2}}\left(n+\frac{1}{2}\right)
\end{equation}
with the eigenvectors $\left|n\right\rangle \left|k_{x},k_{z}\right\rangle $ (here $n=0,1,2,\ldots$). The lowest Landau band has the same $k_{z}$ dependence and magnetic field dependence as the two-node model.

In a small magnetic field, all the Landau bands are sticking together; since both the spacing between Landau bands and the degeneracy of Landau bands increase with the field strength, the Fermi energy crosses fewer bands as the magnetic field increases. When the Fermi energy is at the band bottom of $n=1$ conduction band, the critical magnetic field $B_{c}$ is reached. For magnetic fields higher than this critical value, only the lowest Landau band is occupied by the carriers. In the quantum limit, if the system has a fixed carrier concentration $n_{0}$, the Fermi wave vector is $\left|k_{F}\right|=\pi n_{0}/N_{L}$ for the quadratic lowest Landau band cases and $\left|k_{F}\right|=2\pi n_{0}/N_{L}$ for the one-node model. Then, from $\min\left(E_{k_{z},1+}\right)=E_{k_{F},0}$, the electron gas model and one-node model have $B_{c}=\frac{\hbar}{e}\left(\sqrt{2}\pi^{2}n_{0}\right)^{2/3}$ and $B_{c}=\frac{\hbar}{e}\left(2\sqrt{2}\pi^{2}n_{0}\right)^{2/3}$, respectively. This expression does not depend on any model parameters; it can be used to determine the value of the carrier concentration or critical field in the experiment. For the two-node model, there is no simple analytic expression for the critical field, and the model parameters can affect the results (see Appendix~\ref{sec:The-critical-magnetic}). 

\subsection{Different types of impurity potential}

Generally, the impurity potential can be written as
\begin{equation}
V\left(\mathbf{r}\right)=\sum_{i}U\left(\mathbf{r}-\mathbf{R}_{i}\right),
\end{equation}
where $U\left(\mathbf{r}-\mathbf{R}_{i}\right)$ denotes the potential of the $i$-th impurity that centered at $\mathbf{R}_{i}$. If the potential is point-like (i.e., the potential energy is finite at $\mathbf{r}=\mathbf{R}_{i}$ and zero elsewhere), it can be modeled by 
\begin{eqnarray}
U(\mathbf{r}-\mathbf{R}_i)=    u_{0}\delta\left(\mathbf{r}-\mathbf{R}_{i}\right),
\end{eqnarray}
where $u_{0}$ is an energy constant describing the strength of the potential. The Fourier transform of this delta potential is simply $u_{0}$. Although analytical derivations are easier when the delta function is used, the impurity potential cannot be perfectly point-like in realistic materials. To model impurity potentials with a finite range, the Gaussian-type impurities are used in the studies on the electron density of states \cite{HalperinB_PR_1966,SamathiyakanitV_JoPCSSP_1974,SaitohM_JoPCSSP_1974,SayakanitV_PRB_1979}, yielding a qualitative agreement with the experimental results. The Gaussian-type impurities have also been used in the transport studies \cite{VargiamidisV_PRB_2005,YuanS_PRB_2010,SuzuuraH_PRB_2016}, but not for calculating the linear magnetoresistance. The Gaussian potential reads
\begin{eqnarray}
U(\mathbf{r}-\mathbf{R}_i)=  u_{0}\left(\frac{1}{d\sqrt{2\pi}}\right)^{3}e^{-\frac{\left|\mathbf{r}-\mathbf{R}_{i}\right|^{2}}{2d^{2}}},  
\end{eqnarray}
where $d$ is the decay length. It describes potentials that have the maximum energy at $\mathbf{r}=\mathbf{R}_{i}$, and the energy decays when the position is away from $\mathbf{R}_{i}$. Its Fourier transform is $u_{0}e^{-\frac{q^{2}d^{2}}{2}}$, which reduces to the delta potential when taking $d=0$. Another often used finite range potential is the screened Coulomb potential
\begin{eqnarray}
U(\mathbf{r}-\mathbf{R}_i)=\frac{e^{2}}{4\pi\varepsilon\left|\mathbf{r}-\mathbf{R}_{i}\right|}e^{-\kappa\left|\mathbf{r}-\mathbf{R}_{i}\right|} ,   
\end{eqnarray}
where $\varepsilon=\varepsilon_{0}\varepsilon_{r}$ is the absolute permittivity, $\varepsilon_{0}$ is the vacuum permittivity, $\varepsilon_{r}$ is the relative permittivity, and $\kappa$ is the reciprocal Debye screening radius. Its Fourier transform is $\frac{e^{2}}{\varepsilon\left(q^{2}+\kappa^{2}\right)}$. 

\subsection{Longitudinal magnetoconductivity in the quantum limit\label{subsec:Longitudinal-MC}}

In the limit of zero temperature, with one-loop diagram approximation, the longitudinal and transverse magnetoconductivity can be found from 
\begin{align}
\sigma_{\alpha\alpha} & =\frac{\pi\hbar e^{2}}{V}\sum_{u,u'}A_{u}\left(E_{F}\right)A_{u'}\left(E_{F}\right)\left|\left\langle u'\right|v_{\alpha}\left|u\right\rangle \right|^{2},\label{eq:s_general}
\end{align}
where $V$ is the volume of the system, $v_{\alpha}$ is the velocity operator, $\alpha=x,y$ or $z$, and $u$ denotes the quantum number, including $k_{x},k_{z}$, and the band index. The spectral function $A_{u}\left(E_{F}\right)=\frac{i}{2\pi}\left[G_{u}^{R}\left(E_{F}\right)-G_{u}^{A}\left(E_{F}\right)\right]$, the retarded/advanced Green function $G_{u}^{R/A}\left(E_{F}\right)=1/\left[E_{F}-E_{u}\pm i\hbar/\tau_{u}\left(E_{F}\right)\right]$, and $\tau_{u}\left(E_{F}\right)$ is the scattering time. The detailed deductions can be found in Appendix~\ref{sec:Kubo-formula-for}.

For the longitudinal magnetoconductivity, $v_{z}=\frac{1}{\hbar}\frac{\partial H}{\partial k_{z}}$, and $\left\langle u'\right|v_{z}\left|u\right\rangle $ is non zero only when $u'=u$. The leading term of the square of the spectral function $A_{u}\left(E_{F}\right)$ is $\frac{1}{2\pi^{2}}G_{u}^{R}\left(E_{F}\right)G_{u}^{A}\left(E_{F}\right)$, which is approximately $\frac{1}{\pi\hbar}\tau_{u}\left(E_{F}\right)\delta\left(E_{F}-E_{u}\right)$ \cite{Datta_2009,LuH_PRB_2015}. The delta function is centered at $E_{u}=E_{F}$, which means that only the bands crossing $E_{F}$ have non zero contribution. In the quantum limit, the longitudinal magnetoconductivity is found as
\begin{equation}
\sigma_{zz}=\frac{e^{2}}{h}N_{L}\sum_{i}\left|v_{k_{F}^{i},0}\right|\tau_{k_{F}^{i},0},\label{eq:sigma_zz}
\end{equation}
where $k_{F}^{i}$ denotes the $i$-th Fermi wave vector, the Fermi velocity $v_{k_{F}^{i},0}=\frac{1}{\hbar}\frac{\partial E_{k_{z},0}}{\partial k_{z}}|_{k_{z}=k_{F}^{i}}$, and $\tau_{k_{F}^{i},0}\equiv\tau_{k_{F}^{i},0}\left(E_{F}\right)$ is the scattering time of the state with $k_{F}^{i}$ in the lowest Landau band. For the lowest Landau band of the one-node model, there is only one $k_{F}$ (negative value), and $\sigma_{zz}^{L}=\frac{e^{2}}{h}N_{L}\left|v_{F}\right|\tau_{k_{F},0}$. If $\tau_{k_{F},0}$ has no magnetic field dependence, $\sigma_{zz}^{L}$ is simply proportional to $B$. For the lowest Landau band with quadratic dispersion, there are $\pm k_{F}$ (the notation $k_{F}$ denotes the positive value for the case of quadratic lowest Landau band in this paper), and $\sigma_{zz}^{Q}=2\frac{e^{2}}{h}N_{L}v_{k_{F},0}\tau_{k_{F},0}$. In contrast to the linear lowest Landau band, the Fermi velocity here is proportional to $k_{F}$. This results in that the Fermi velocity is inversely proportional to $B$ if the carrier concentration is fixed. Therefore, when $\tau_{k_{F},0}$ is not magnetic-field dependent, $\sigma_{zz}^{Q}$ is $B$ independent with a fixed carrier concentration. When the Fermi energy is fixed instead, $\sigma_{zz}^{Q}$ is proportional to $B$.

The scattering time is related to the imaginary part of self-energy through $\frac{\hbar}{2\tau_{u}\left(E_{F}\right)}=-\Im\left[\Sigma_{u}^{R}\left(E_{F}\right)\right]$. Generally, the self-energy is contributed by the scattering from impurities, electrons, and phonons; here, we consider the case in which impurity scattering dominates. In the Born approximation, the self-energy of the state $u$ can be found by
\begin{equation}
\Sigma_{u}^{R}\left(E_{F}\right)=\sum_{u'}\frac{\left|\left\langle u'\right|\hat{V}\left|u\right\rangle \right|^{2}}{E_{F}-E_{u'}+i\eta},\label{eq:self-energy}
\end{equation}
where $\eta$ is a positive infinitesimal quantity, and the impurity potential $V\left(\mathbf{r}\right)=\left\langle \mathbf{r}\right|\hat{V}\left|\mathbf{r}\right\rangle $. The impurity average, $\left\langle V\left(\mathbf{r}_{1}\right)V\left(\mathbf{r}_{2}\right)\right\rangle _{imp}$ , is needed in the calculation, which gives \cite{Mahan_2000} $n_{i}\frac{\int d\boldsymbol{q}}{\left(2\pi\right)^{3}}u\left(\boldsymbol{q}\right)u\left(-\boldsymbol{q}\right)e^{i\boldsymbol{q}\left(\mathbf{r}_{1}-\mathbf{r}_{2}\right)}$, where $n_{i}$ denotes the concentration of the impurity, and $u\left(\boldsymbol{q}\right)$ is the Fourier transform of $U\left(\mathbf{r}-\mathbf{R}_{i}\right)$. From the Cauchy relation, one has $\Im\left[1/(E_{F}-E_{u'}+i\eta)\right]=-\pi\delta\left(E_{F}-E_{u'}\right)$; therefore, in the quantum limit, only the lowest Landau band contributes to the self-energy. After some cumbersome but straightforward calculations (details in Appendix~\ref{subsec:Self-energy-for-LLB}), the scattering time can be found as
\begin{align}
\frac{\hbar}{2\tau_{k_{F},0}}= & \pi n_{i}\frac{\int d\boldsymbol{q}}{\left(2\pi\right)^{3}}\left[\delta\left(E_{F}-E_{k_{F}-q_{z},0}\right)\right.\nonumber \\
 & \left.\times u\left(\boldsymbol{q}\right)u\left(-\boldsymbol{q}\right)e^{-\frac{1}{2}\left(q_{x}^{2}+q_{y}^{2}\right)l_{B}^{2}}\right].\label{eq:tau0_general}
\end{align}
This is the general form of the scattering time of the lowest Landau band. After substituting the specific expressions of impurity potentials, it reduces to 
\begin{align}
\frac{\hbar}{2\tau_{k_{F},0}^{G}}= & \frac{n_{i}u_{0}^{2}}{4\pi l_{B}^{2}\left(1+2d^{2}/l_{B}^{2}\right)}\nonumber \\
 & \times\int_{-\infty}^{\infty}dq_{z}\delta\left(E_{F}-E_{k_{F}-q_{z},0}\right)e^{-q_{z}^{2}d^{2}},\label{eq:tau0_G}\\
\frac{\hbar}{2\tau_{k_{F},0}^{C}}= & \frac{n_{i}e^{4}l_{B}^{2}}{16\pi\varepsilon^{2}}\int_{-\infty}^{\infty}dq_{z}\left\{ \delta\left(E_{F}-E_{k_{F}-q_{z},0}\right)\right.\nonumber \\
 & \left.\times\mathcal{F}_{1}\left[\frac{l_{B}^{2}}{2}\left(q_{z}^{2}+\kappa^{2}\right)\right]\right\} ,\label{eq:tau0_C}
\end{align}
for the Gaussian and screened Coulomb potentials, respectively. Here, $\mathcal{F}_{1}\left[x\right]=\frac{1}{x}+e^{x}E_{i}\left[-x\right]$, and $E_{i}\left[-x\right]=-\int_{x}^{\infty}\frac{1}{t}e^{-t}dt$ is the exponential integral function. The behavior of $\mathcal{F}_{1}\left[x\right]$ is studied in Appendix~\ref{subsec:Self-energy-for-LLB}.

\begin{figure}
\includegraphics{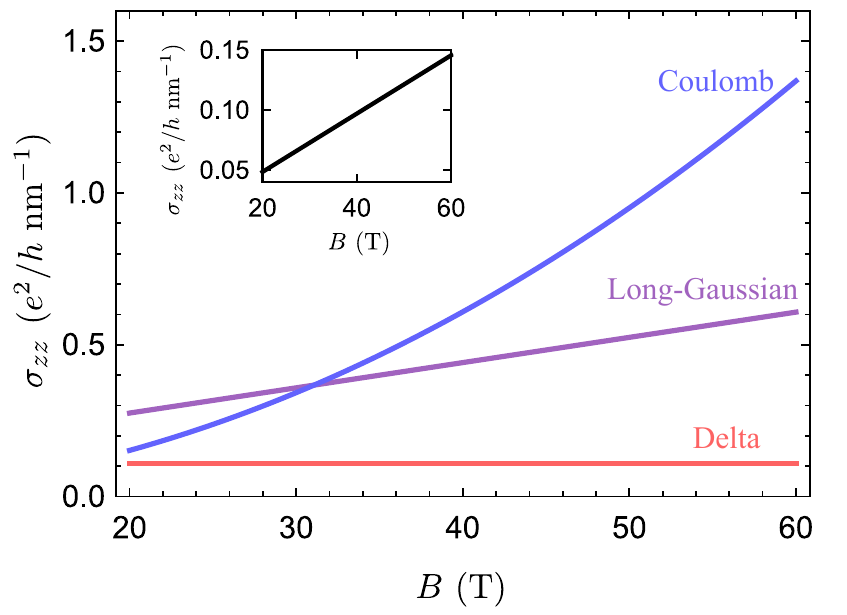}
\caption{\label{fig:LMC-1}Magnetic field dependence of the longitudinal magnetoconductivity for the one-node model. The red, purple, and blue lines represent cases where delta-type, long-Gaussian-type, and screened-Coulomb-type impurities dominate, respectively. The inset shows the result after the vertex correction, which is independent of the impurity type. The decay length, $d$, is taken as 5 nm for the long Gaussian potential. For the screened Coulomb potential, the impurity concentration, $n_i$, is taken as $3\times 10^{-4}$~$\mathrm{nm^{-3}}$; the relative permittivity, $\varepsilon_{r}$, is taken as 50. Other parameters are the same as those in Fig.~\ref{fig:Magnetic-field-dependence-LMR}.}
\end{figure}

Combining the expressions of scattering time and band dispersion with Eq.~(\ref{eq:sigma_zz}), the longitudinal magnetoconductivity for the one-node model can be found,
\begin{align}
\sigma_{zz}^{L,G} & =\frac{e^{2}}{h}\frac{\left(\hbar v_{F}\right)^{2}}{n_{i}u_{0}^{2}}\left[1+2\left(\frac{d}{l_{B}}\right)^{2}\right],\label{eq:s_zz_L,G}\\
\sigma_{zz}^{L,C} & =\frac{e^{2}}{h}\frac{\varepsilon^{2}}{n_{i}e^{4}l_{B}^{4}}\frac{4\left(\hbar v_{F}\right)^{2}}{\mathcal{F}_{1}\left[\frac{l_{B}^{2}}{2}\kappa^{2}\right]}.\label{eq:s_zz_L,C}
\end{align}
The longitudinal magnetoconductivity is inversely proportional to the impurity concentration and strength. For the field dependence, only $l_{B}$ and $\kappa$ are $B$ dependent in above expressions. Therefore, one can easily found that: for delta-type impurities, i.e., $d=0$ in Eq.~(\ref{eq:s_zz_L,G}), the longitudinal magnetoconductivity is field independent, as shown by the red line in Fig.~\ref{fig:LMC-1}; once $d\ne0$, there will be a linear-$B$ term in $\sigma_{zz}^{L,G}$, and this linear-$B$ term dominates when $2d^{2}\gg l_{B}^{2}$, as shown by the purple line in Fig.~\ref{fig:LMC-1}; for the screened-Coulomb-type impurities ($\kappa\propto\sqrt{B}$ for the one-node model, see Appendix~\ref{sec:Calculation-of-kappa}), $\mathcal{F}_{1}\left[l_{B}^{2}\kappa^{2}/2\right]$ is independent on magnetic field, and the longitudinal magnetoconductivity is proportional to $B^{2}$, as shown by the blue line in Fig.~\ref{fig:LMC-1}.

\begin{figure*}
\includegraphics{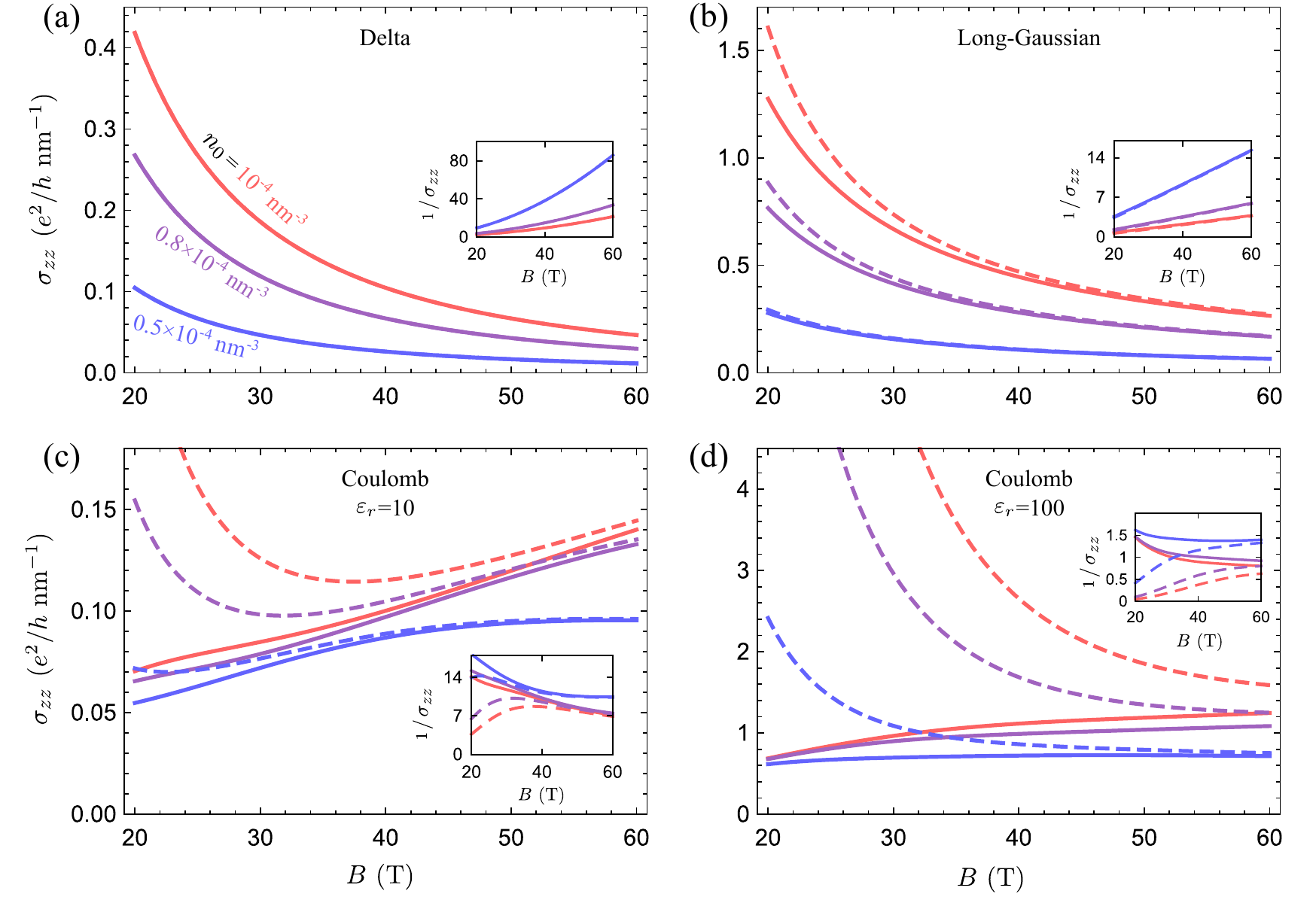}
\caption{\label{fig:LMC-2}Magnetic field dependence of the longitudinal magnetoconductivity for the two-node model. (a) and (b) are cases for delta-type, and long-Gaussian-type impurities. (c) and (d) are cases for screened-Coulomb-type impurities with $\varepsilon_{r}=10$ and $\varepsilon_{r}=100$, respectively. The carrier concentration, $n_0$, is taken as $10^{-4}$~$\mathrm{nm^{-3}}$ (red line), $8\times 10^{-5}$~$\mathrm{nm^{-3}}$ (purple line), and $5\times 10^{-5}$~$\mathrm{nm^{-3}}$ (blue line); dashed lines represent results after the vertex correction, which completely overlay with the solid lines in (a). Insets in (a-d) show the corresponding magnetic field dependence of the inverse of the longitudinal magnetoconductivity. For the screened Coulomb potential, the impurity concentration, $n_i$, is taken as $0.1n_0$. Other parameters are the same as those in Fig.~\ref{fig:LMC-1}.}
\end{figure*}

In the same way, but using the dispersion expressions of the two-node model, the longitudinal magnetoconductivity of the two-node model can be found,
\begin{align}
\sigma_{zz}^{Q,G} & =\frac{e^{2}}{h}\frac{\left(\hbar v_{k_{F},0}\right)^{2}}{n_{i}u_{0}^{2}}\frac{2\left(1+2d^{2}/l_{B}^{2}\right)}{1+e^{-4k_{F}^{2}d^{2}}},\label{eq:s_zz_Q,G}\\
\sigma_{zz}^{Q,C} & =\frac{e^{2}}{h}\frac{\varepsilon^{2}}{n_{i}e^{4}l_{B}^{4}}\frac{8\left(\hbar v_{k_{F},0}\right)^{2}}{\mathcal{F}_{1}\left[\frac{l_{B}^{2}}{2}\left(4k_{F}^{2}+\kappa^{2}\right)\right]+\mathcal{F}_{1}\left[\frac{l_{B}^{2}}{2}\kappa^{2}\right]}.\label{eq:s_zz_Q,C}
\end{align}
Unlike the case of one-node model, the Fermi velocity $v_{k_{F},0}$ is $B$ dependent here. Equation~(\ref{eq:s_zz_Q,G}) is a general expression with a Gaussian decay length $d$; when the decay length is taken as zero, it reduces to Eq.~(24) in Ref.~\cite{LuH_PRB_2015}. From Eq.~(\ref{eq:s_zz_Q,G}), one can find that: for delta-type impurities ($d=0$), the longitudinal magnetoconductivity has a simple $B^{-2}$ dependence, as shown in Fig.~\ref{fig:LMC-2}(a); for Gaussian-type impurities, there is one more $B^{-1}$ dependent term in the longitudinal magnetoconductivity ($e^{-4k_{F}^{2}d^{2}}\approx1$ for small $k_{F}$), and it dominates when $2d^{2}\gg l_{B}^{2}$, as shown in Fig.~\ref{fig:LMC-2}(b). The field dependence of the longitudinal magnetoconductivity in the case of screened-Coulomb-type impurity is complicated, as depicted in Figs.~\ref{fig:LMC-2}(c) and \ref{fig:LMC-2}(d). Generally, Eq.~(\ref{eq:s_zz_Q,C}) shows a non-monotonic $B$ dependence, which is contained in $\mathcal{F}_{1}\left[l_{B}^{2}\left(4k_{F}^{2}+\kappa^{2}\right)/2\right]+\mathcal{F}_{1}\left[l_{B}^{2}\kappa^{2}/2\right]$ [note that $\left(v_{k_{F},0}/l_{B}^{2}\right)^{2}$ is $B$-independent]. The reciprocal Debye screening length $\kappa\propto B$ for the quadratically-dispersed lowest Landau band (see Appendix~\ref{sec:Calculation-of-kappa}). When $4l_{B}^{2}k_{F}^{2}\ll l_{B}^{2}\kappa^{2}$, the longitudinal magnetoconductivity is in proportion to $\mathcal{F}_{1}^{-1}\left[l_{B}^{2}\kappa^{2}/2\right]$; if $l_{B}^{2}\kappa^{2}/2\ll1$, it is proportional to $B$; if $l_{B}^{2}\kappa^{2}/2\gg1$, it is proportional to $B^{2}$. When $4l_{B}^{2}k_{F}^{2}\gg l_{B}^{2}\kappa^{2}$, the above results still hold; this is because that $\mathcal{F}_{1}\left[l_{B}^{2}\left(4k_{F}^{2}+\kappa^{2}\right)/2\right]\propto\left(2l_{B}^{2}k_{F}^{2}\right)^{-1}$ and $\mathcal{F}_{1}\left[l_{B}^{2}\kappa^{2}/2\right]\propto\left(l_{B}^{2}\kappa^{2}/2\right)^{-1}$ lead to $\mathcal{F}_{1}\left[l_{B}^{2}\kappa^{2}/2\right]\gg\mathcal{F}_{1}\left[l_{B}^{2}\left(4k_{F}^{2}+\kappa^{2}\right)/2\right]$. 

For 3D electron gas, its longitudinal magnetoconductivity share the same expression with the two-node model. This is due to the fact that: after the replacement $\hbar^{2}/\left(2m\right)\rightarrow M$, the expression of the lowest Landau band of electron gas only has a constant energy difference, $Mk_{w}^{2}$, compared to that of the two-node model. From Eqs.~(\ref{eq:sigma_zz}) and (\ref{eq:tau0_general}), one can find that a constant energy difference in the lowest Landau band does not affect the result.

When the carrier concentration is low and the magnetic field is extremely strong, the carriers will squeeze at the band bottom of the quadratically-dispersed lowest Landau band. In this case, the field dependence of Eq.~(\ref{eq:s_zz_Q,C}) need to be corrected. This is due to the delta function approximation in the deduction. The delta functions in $G_{u}^{R}\left(E_{F}\right)G_{u}^{A}\left(E_{F}\right)$, $\hbar/\left(2\tau_{k_{F},0}\right)$ and $\kappa^{2}$ are obtained by making the finite Lorentz-type broadening approximately equal to zero, which is appropriate in most cases. However, when the carriers squeeze at the band bottom ($k_{F}\rightarrow0$), an incorrect factor, which is $\propto B$, is introduced in the integrals, see details in Appendix~\ref{sec:Correction-when-the}. As $G_{u}^{R}\left(E_{F}\right)G_{u}^{A}\left(E_{F}\right)\approx\frac{2\pi}{\hbar}\tau_{u}\left(E_{F}\right)\delta\left(E_{F}-E_{u}\right)$, this incorrect factor cancels out in the expression of longitudinal magnetoconductivity. Therefore, only the $B$ dependence of $\kappa^{2}$ in Eq.~(\ref{eq:s_zz_Q,C}) need to be corrected, which leads to $\mathcal{F}_{1}\left[l_{B}^{2}\kappa^{2}/2\right]$ proportional to $B^{0}$. 

\begin{figure}
\includegraphics{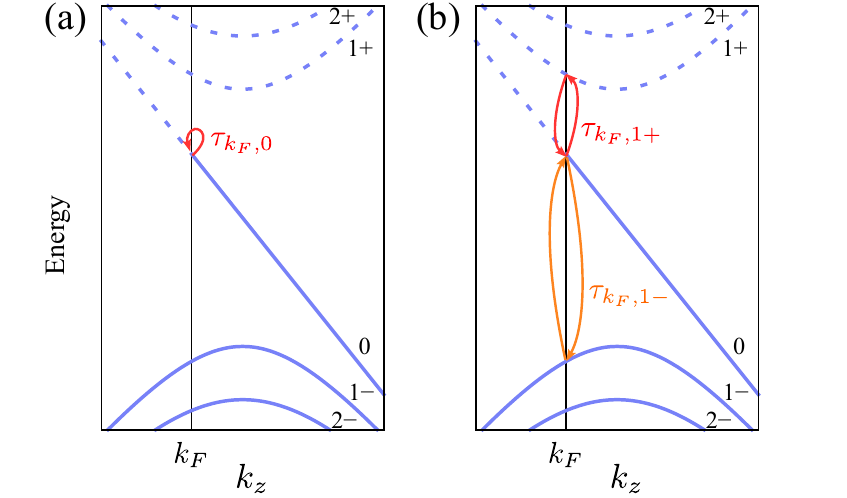}

\caption{\label{fig:scattering}Illustrations of the scattering process for (a) $\tau_{k_{F},0}$ and (b) $\tau_{k_{F},1\pm}$. The scattering in (a) is between the degenerate states with the same $k_{z}$, which does not affect the longitudinal transport.}
\end{figure}

The conductivity formula, Eq.~(\ref{eq:s_general}), corresponds to a bubble diagram in the Feynman diagram technique. By making the vertex correction, ladder diagrams, in which the impurity scattering links the Green functions on both sides of the bubble, can be concluded \cite{Mahan_2000}. Instead of directly calculating the conductivity from the bubble and ladder diagrams, the vertex correction for the longitudinal magnetoconductivity can be made \cite{Mahan_2000,LuH_PRB_2015} by adding $\left(1-v_{k_{F}-q_{z},0}/v_{k_{F},0}\right)$ in the integral of Eq.~(\ref{eq:tau0_general}), i.e.,
\begin{align}
\frac{\hbar}{2\tau'_{k_{F},0}}= & \pi n_{i}\frac{\int d\boldsymbol{q}}{\left(2\pi\right)^{3}}\left[\left(1-\frac{v_{k_{F}-q_{z},0}}{v_{k_{F},0}}\right)\delta\left(E_{F}-E_{k_{F}-q_{z},0}\right)\right.\nonumber \\
 & \left.\times u\left(\boldsymbol{q}\right)u\left(-\boldsymbol{q}\right)e^{-\frac{1}{2}\left(q_{x}^{2}+q_{y}^{2}\right)l_{B}^{2}}\right]\label{eq:transport-time}
\end{align}
This additional factor weights the scattering. In the above expression, $q_{z}$ indicates the momentum change after the scattering, and the delta function guarantees the energy has not changed during the scattering. At the Fermi surface, only $q_{z}=0$ is possible for the one-node model, resulting in Eq.~(\ref{eq:transport-time}) being zero, see Fig.~\ref{fig:scattering}(a). This means that the longitudinal magnetoconductivity, Eq.~(\ref{eq:sigma_zz}), would not be impaired by the impurity scattering. When the full massless Dirac Hamiltonian is considered, there is another lowest Landau band with opposite chirality. However, these two lowest Landau bands are decoupled, and scattering between them is not allowed, as shown in the inset of Fig.~\ref{fig:Magnetic-field-dependence-LMR}(a). Therefore, with a constant $\tau_{k_{F},0}$ (the self-energy of Green function will not be zero, there are still other contributions to band broadening, for example, the temperature effect), the longitudinal magnetoconductivity $\sigma_{zz}^{L}$ is proportional to $B$, regardless of the impurity type, as shown in the inset of Fig.~\ref{fig:LMC-1}. For the two-node model and electron gas, in additional to $0$, $q_{z}$ can also take $2k_{F}$, i.e., states scattering between $k_{F}$ and $-k_{F}$, which gives $\left(1-v_{k_{F}-q_{z},0}/v_{k_{F},0}\right)=2$. In the final expression of the longitudinal magnetoconductivity, $(1+e^{-4k_{F}^{2}d^{2}})$ changes to $2e^{-4k_{F}^{2}d^{2}}$ in Eq.~(\ref{eq:s_zz_Q,G}), and $\mathcal{F}_{1}\left[l_{B}^{2}\left(4k_{F}^{2}+\kappa^{2}\right)/2\right]+\mathcal{F}_{1}\left[l_{B}^{2}\kappa^{2}/2\right]$ changes to $2\mathcal{F}_{1}\left[l_{B}^{2}\left(4k_{F}^{2}+\kappa^{2}\right)/2\right]$ in Eq.~(\ref{eq:s_zz_Q,C}). This does not affect the $B$ dependence of Eq.~(\ref{eq:s_zz_Q,G}). However, for screened-Coulomb-type potential, the vertex correction eliminates $\mathcal{F}_{1}\left[l_{B}^{2}\kappa^{2}/2\right]$ in Eq. (\ref{eq:s_zz_Q,C}), leading to $\sigma_{zz}^{Q,C}\propto\mathcal{F}_{1}^{-1}\left[l_{B}^{2}\left(4k_{F}^{2}+\kappa^{2}\right)/2\right]$, which can be proportional to $B^{-3}$ or $B^{-6}$ when $4l_{B}^{2}k_{F}^{2}\gg l_{B}^{2}\kappa^{2}$. The dashed lines in Fig.~\ref{fig:LMC-2} exhibit the results after vertex correction.

The above results and discussions show that bringing the two nodes far apart in the two-node model can approach the one-node model if we linearize the dispersion in the two-node model and ignore the scattering between $k_{F}$ and $-k_{F}$.

\subsection{Transverse magnetoconductivity in the quantum limit}

Different from the longitudinal magnetoconductivity, in Eq.~(\ref{eq:s_general}), $\left\langle u'\right|v_{x}\left|u\right\rangle $ for the transverse magnetoconductivity is only non zero when the Landau indexes (i.e., 0 or $n$ ) in $u'$ and $u$ differ $1$. In the quantum limit, one has 
\begin{align}
\sigma_{xx}\approx & \sigma_{xx,1+}+\sigma_{xx,1-},\nonumber \\
\sigma_{xx,1\pm}= & \hbar e^{2}N_{L}\int_{-\infty}^{\infty}\left[A_{k_{z},0}\left(E_{F}\right)A_{k_{z},1\pm}\left(E_{F}\right)\right.\nonumber \\
 & \left.\times\left|\left\langle k_{x},k_{z},0\right|v_{x}\left|k_{x},k_{z},1\pm\right\rangle \right|^{2}\right]dk_{z},\label{eq:sigma_xx}
\end{align}
for two-band models. The spectral functions in the above expression are approximately delta functions. In the quantum limit, the Fermi energy does not cross $E_{k_{z},1\pm}$, resulting in that $A_{k_{z},1\pm}\left(E_{F}\right)$ can substantially reduce the value of transverse magnetoconductivity compared to $A_{k_{z},0}\left(E_{F}\right)$. Therefore, we take $A_{k_{z},0}\left(E_{F}\right)\approx\delta\left(E_{F}-E_{k_{z},0}\right)$ and $A_{k_{z},1\pm}\left(E_{F}\right)\approx\frac{1}{\pi}\frac{1}{\left(E_{F}-E_{k_{z},1\pm}\right)^{2}}\frac{\hbar}{2\tau_{k_{z},1\pm}\left(E_{F}\right)}$ in the following calculation \cite{AbrikosovA_PRB_1998}. 

The impurity effect is included by $\tau_{k_{F},1\pm}\equiv\tau_{k_{F},1\pm}\left(E_{F}\right)$. Unlike longitudinal magnetoconductivity, the transverse magnetoconductivity in the quantum limit is more related to the scattering time of the Landau band of index $\text{1}$. It can be found from Eq.~(\ref{eq:self-energy}) that in the quantum limit, the term, of which $u'$ has the lowest Landau index $0$, dominates $\Im\left[\Sigma_{k_{F},1\pm}^{R}\left(E_{F}\right)\right]$.  After some cumbersome but straightforward calculations (see Appendix~\ref{subsec:Self-energy-for-1LB}), one has

\begin{widetext}
\begin{equation}
\frac{\hbar}{2\tau_{k_{F},1+}}=\pi n_{i}\left(\sin\frac{\theta_{k_{F},1}}{2}\right)^{2}\frac{\int d\boldsymbol{q}}{\left(2\pi\right)^{3}}\delta\left(E_{F}-E_{k_{F}-q_{z},0}\right)u\left(\boldsymbol{q}\right)u\left(-\boldsymbol{q}\right)e^{-\frac{1}{2}\left(q_{x}^{2}+q_{y}^{2}\right)l_{B}^{2}}\frac{\left(q_{x}^{2}+q_{y}^{2}\right)l_{B}^{2}}{2}.\label{eq:tau1_general}
\end{equation}
This is the general form of the scattering time of $1+$ band. After substituting the specific expressions of impurity potentials, the above expression reduces to
\begin{align}
\frac{\hbar}{2\tau_{k_{F},1+}^{G}}= & \frac{n_{i}u_{0}^{2}}{4\pi l_{B}^{2}\left(1+2d^{2}/l_{B}^{2}\right)^{2}}\left(\sin\frac{\theta_{k_{F},1}}{2}\right)^{2}\int_{-\infty}^{\infty}dq_{z}\delta\left(E_{F}-E_{k_{F}-q_{z},0}\right)e^{-q_{z}^{2}d^{2}},\label{eq:tau_1p_G}\\
\frac{\hbar}{2\tau_{k_{F},1+}^{C}}= & \frac{n_{i}e^{4}l_{B}^{2}}{16\pi\varepsilon^{2}}\left(\sin\frac{\theta_{k_{F},1}}{2}\right)^{2}\int_{-\infty}^{\infty}dq_{z}\delta\left(E_{F}-E_{k_{F}-q_{z},0}\right)\mathcal{F}_{2}\left[\frac{l_{B}^{2}}{2}\left(q_{z}^{2}+\kappa^{2}\right)\right],\label{eq:tau_1p_C}
\end{align}
for Gaussian and screened Coulomb potentials, respectively. Here, $\mathcal{F}_{2}\left[x\right]=-1-\left(1+x\right)e^{x}E_{i}\left[-x\right]$. The behavior of $\mathcal{F}_{2}\left[x\right]$ is studied in Appendix~\ref{subsec:Self-energy-for-1LB}.

\end{widetext}

For linearly-dispersed lowest Landau band  with a small carrier concentration, $1+$ and $1-$ bands contribute to transverse magnetoconductivity approximating equally, i.e., $\sigma_{xx,1+}^{L}\approx\sigma_{xx,1-}^{L}$ (see Appendix~\ref{subsec:Transverse-magnetoconductivity}), one has 
\begin{equation}
\sigma_{xx}^{L,G}\approx\frac{e^{2}}{h}\frac{n_{i}u_{0}^{2}}{\left(4\pi\hbar v_{F}\right)^{2}l_{B}^{2}}\frac{1}{\left(1+2d^{2}/l_{B}^{2}\right)^{2}},\label{eq:s_xx_L,G}
\end{equation}
and 
\begin{equation}
\sigma_{xx}^{L,C}\approx\frac{e^{2}}{h}\frac{n_{i}e^{4}l_{B}^{2}}{\left(8\pi\hbar v_{F}\right)^{2}\varepsilon^{2}}\mathcal{F}_{2}\left[\frac{l_{B}^{2}}{2}\kappa^{2}\right].\label{eq:s_xx_L,C}
\end{equation}

\begin{figure}
\includegraphics{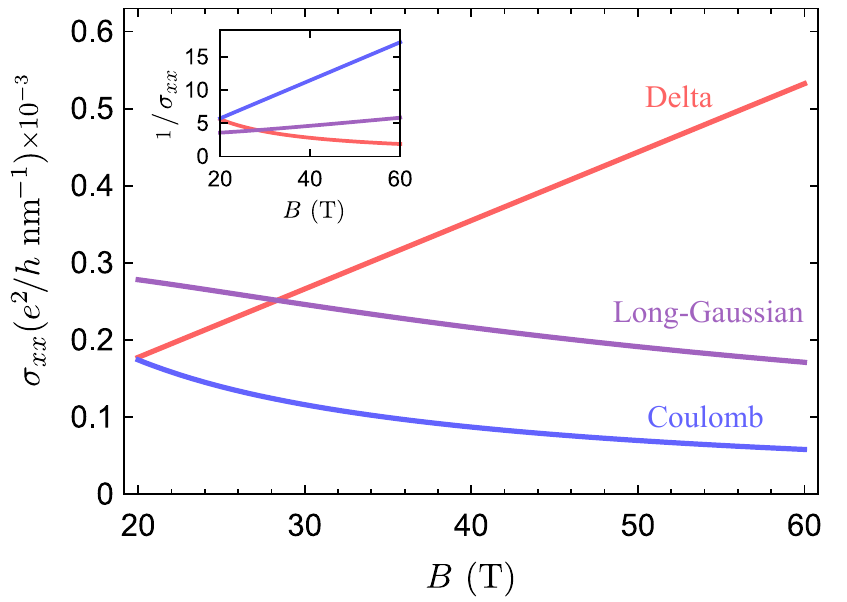}
\caption{\label{fig:TMC-1}Magnetic field dependence of the transverse magnetoconductivity for the one-node model. The red, purple, and blue lines represent cases where delta-type, long-Gaussian-type, and screened-Coulomb-type impurities dominate, respectively. The decay length, $d$, is taken as 5 nm for the long Gaussian potential. For the screened Coulomb potential, the impurity concentration, $n_i$, is taken as $10^{-3}$~$\mathrm{nm^{-3}}$; the relative permittivity, $\varepsilon_{r}$, is taken as 50. Other parameters are the same as those in Fig.~\ref{fig:Magnetic-field-dependence-TMR}.}
\end{figure}

From Eq.~(\ref{eq:s_xx_L,G}), one can find that: contrary to longitudinal magnetoconductivity, the transverse magnetoconductivity is proportional to the impurity concentration and strength; $\sigma_{xx}^{L,G}$ is proportional to $B$ when delta-type impurities dominate, and the field dependence changes to $B^{-1}$ when the long-range-Gaussian-type impurities ($2d^{2}\gg l_{B}^{2}$) dominate; for the case of screened-Coulomb-type impurities, $\sigma_{xx}^{L,C}$ is proportional to $B^{-1}$. These behaviors are illustrated in Fig.~\ref{fig:TMC-1}. The $B^{-1}$ dependence of the transverse magnetoconductivity in the case of screened Coulomb potential was first found by Abrikosov \cite{AbrikosovA_PRB_1998}. Here, the general result for the transverse magnetoconductivity of the one-node model with screened Coulomb potential is given in Appendix~\ref{subsec:Transverse-magnetoconductivity}. Equation (\ref{eq:s_xx_L,C}) is for the case of small carrier concentrations, and it reduces to Eq.~(36) in Ref.~[1] when a large relative permittivity is taken.

\begin{figure*}
\includegraphics{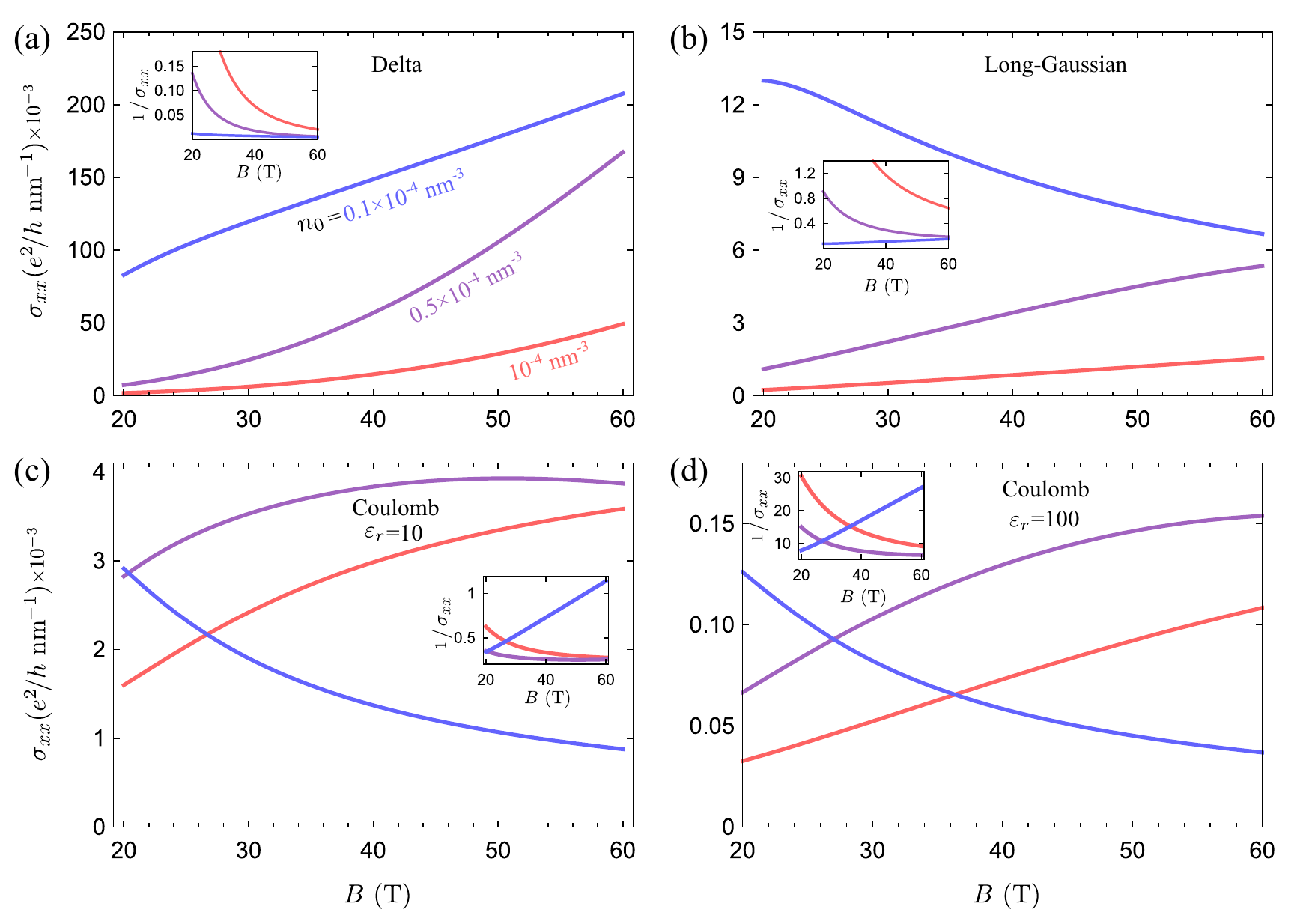}
\caption{\label{fig:TMC-2}Magnetic field dependence of the transverse magnetoconductivity for the two-node model. (a) and (b) are cases for delta-type, and long-Gaussian-type impurities. (c) and (d) are cases for screened-Coulomb-type impurities with $\varepsilon_{r}=10$ and $\varepsilon_{r}=100$, respectively. The carrier concentration, $n_0$, is taken as $10^{-4}$~$\mathrm{nm^{-3}}$ (red line), $5\times 10^{-5}$~$\mathrm{nm^{-3}}$ (purple line), and $10^{-5}$~$\mathrm{nm^{-3}}$ (blue line). Insets in (a-d) show the corresponding magnetic field dependence of the inverse of the transverse magnetoconductivity. For the screened Coulomb potential, the impurity concentration, $n_i$, is taken as $0.1n_0$. Other parameters are the same as those in Fig.~\ref{fig:TMC-1}.}
\end{figure*}

The transverse magnetoconductivity of the two-node model can be found in the same procedure. It is more parameter-dependent than its longitudinal magnetoconductivity (see Appendix~\ref{subsec:Transverse-magnetoconductivity}). Approximately, one has 

\begin{align}
\sigma_{xx}^{Q,G}\approx & \frac{e^{2}}{h}\frac{n_{i}u_{0}^{2}}{8\pi^{2}\left(\hbar v_{k_{F},0}\right)^{2}l_{B}^{2}}\frac{\left(1+e^{-4k_{F}^{2}d^{2}}\right)}{\left(1+2d^{2}/l_{B}^{2}\right)^{2}},\label{eq:s_xx_Q,G}\\
\sigma_{xx}^{Q,C}\approx & \frac{e^{2}}{h}\frac{n_{i}e^{4}l_{B}^{2}}{32\pi^{2}\varepsilon^{2}\left(\hbar v_{k_{F},0}\right)^{2}}\nonumber \\
 & \times\left\{ \mathcal{F}_{2}\left[\frac{l_{B}^{2}}{2}\left(4k_{F}^{2}+\kappa^{2}\right)\right]+\mathcal{F}_{2}\left[\frac{l_{B}^{2}}{2}\kappa^{2}\right]\right\} .\label{eq:s_xx_Q,C}
\end{align}
When the decay length in Eq.~(\ref{eq:s_xx_Q,G}) is taken as zero (the delta potential case), it reduces to Eq.~(41) in Ref.~\cite{LuH_PRB_2015}, and it is proportional to $B^{3}$. In the case of long-range Gaussian potential ($2d^{2}\gg l_{B}^{2}$), Eq.~(\ref{eq:s_xx_Q,G}) is proportional to $B$. When the carrier concentration is low, the correction is required, resulting in $\sigma_{xx}^{Q,G}\propto B$ for the case of delta-type impurities and $\sigma_{xx}^{Q,G}\propto B^{-1}$ for the case of long-range-Gaussian-type impurities. These magnetic field dependence of the transverse magnetoresistance with varying carrier concentrations are illustrated in Fig.~\ref{fig:TMC-2}(a) and (b). For the case of Coulomb potential, like the longitudinal magnetoconductivity, $\sigma_{xx}^{Q,C}$ is also non-monotonically $B$ dependent, as depicted in Fig.~\ref{fig:TMC-2}(c) and (d). The $B$ dependence of Eq.~(\ref{eq:s_xx_Q,C}) is contained in $l_{B}^{2}\left\{ \mathcal{F}_{2}\left[l_{B}^{2}\left(4k_{F}^{2}+\kappa^{2}\right)/2\right]+\mathcal{F}_{2}\left[l_{B}^{2}\kappa^{2}/2\right]\right\} /\left(v_{k_{F},0}\right)^{2}$. When $4l_{B}^{2}k_{F}^{2}\ll l_{B}^{2}\kappa^{2}$, Eq.~(\ref{eq:s_xx_Q,C}) is in proportion to $l_{B}^{2}\mathcal{F}_{2}\left[l_{B}^{2}\kappa^{2}/2\right]/\left(v_{k_{F},0}\right)^{2}$. If $l_{B}^{2}\kappa^{2}/2\ll1$, it is approximately proportional to $B$; if $l_{B}^{2}\kappa^{2}/2\gg1$, it is proportional to $B^{-1}$. When $4l_{B}^{2}k_{F}^{2}\gg l_{B}^{2}\kappa^{2}$, one has $\mathcal{F}_{2}\left[2l_{B}^{2}k_{F}^{2}\right]\ll\mathcal{F}_{2}\left[l_{B}^{2}\kappa^{2}/2\right]$, and the result is the same as when $4l_{B}^{2}k_{F}^{2}\ll l_{B}^{2}\kappa^{2}$. However, when the carrier concentration is low, the correction makes the field dependence in $\mathcal{F}_{2}\left[l_{B}^{2}\kappa^{2}/2\right]$ disappear, leading to $\sigma_{xx}^{Q,C}\propto B^{-1}$.

The analytical expression of the transverse magnetoconductivity of the electron gas is similar to that of the two-node model (see Appendix~\ref{subsec:Transverse-magnetoconductivity} for details), but it is less dependent on parameters.

Unlike the longitudinal magnetoconductivity, which relies on the scattering between the states of the lowest Landau band, the transverse magnetoconductivity is decided by the scattering between the lowest Landau band and the bands with Landau index $1$. Figure~\ref{fig:scattering}(b) shows the scattering process of $\tau_{k_{F},1\pm}$ for the one-node model. Different from $\tau_{k_{F},0}$ and $\tau_{k_{F},1+}$, which have the scattering processes between states with same sign of $v_{z}$, the scattering process of $\tau_{k_{F},1-}$ is between states with opposite signs of $v_{z}$. The transverse magnetoconductivity is not decided by $v_{z}$, and the its vertex correction is different from that of the longitudinal magnetoconductivity. It has been verified in Ref.~\cite{KlierJ_PRB_2015} that the vertex correction cannot make dramatic changes to the transverse magnetoconductivity.

\section{conclusions and discussion \label{sec:conclusion-and-discussion}}

We have investigated the longitudinal and transverse magnetoresistance of three different systems (one-node model, two-node model, and electron gas) with three types of impurities (delta type, Gaussian type, and screened-Coulomb type), i.e., totally of $2\times3\times3=18$ situations. Among these, three situations have been previously explored in Refs.~\cite{AbrikosovA_PRB_1998,LuH_PRB_2015}, specifically, the transverse linear magnetoresistance of the one-node model with the screened Coulomb impurity potential \cite{AbrikosovA_PRB_1998} and the longitudinal and transverse magnetoresistance of the two-node model with the delta-type impurity potential \cite{LuH_PRB_2015}. More importantly, here our focus is the four new cases of linear magnetoresistance compared to that in Ref.~\cite{AbrikosovA_PRB_1998} (as shown in Table~\ref{tab:Theories-of-linear}), while no linear magnetoresistance is addressed in Ref.~\cite{LuH_PRB_2015}. Furthermore, we have presented a standard procedure for finding the magnetoresistance in the quantum limit, and given many general formulas. Equation~(\ref{eq:sigma_zz}) is the general formula for the longitudinal magnetoconductivity in the quantum limit. With the Fermi velocity of 3D electron gas, it reduces to the classical Drude formula \cite{MurzinS_P_2000}. The transport time caused by impurities with arbitrary potential can be derived from the general expression Eq.~(\ref{eq:transport-time}). For transverse magnetoconductivity, general formulas of two-band model, Eqs.~(\ref{eq:sigma_xx}) and (\ref{eq:tau1_general}), have been found. After small modifications, they can be directly applied to single-band models. Utilizing these general formulas, the quantum-limit magnetoresistance of a new system can be easily investigated. 

We have already discussed in detail in Sec.~\ref{sec:Field-dependence} that different impurity potentials and band structures can lead to negative or positive magnetoresistance, and here we further extend on three points. First, the linear magnetoresistance we found can occur in both longitudinal and transverse directions or only in one direction, depending on the system studied. Different from those classical theories of linear magnetoresistance \cite{ParishM_N_2003,ParishM_PRB_2005,HuJ_PRB_2007,XuJ_JoAP_2008,AlekseevP_PRL_2015,SongJ_PRB_2015,RamakrishnanN_PRB_2017,KisslingerF_PRB_2017,AlekseevP_PRB_2017,XiaoC_PRB_2020,ChenS_CPB_2022}, our results are closely related to the band structure and impurity type. Second, the quantum-limit longitudinal magnetoresistance of massless Dirac fermions is special because of the combined effect of the linear dispersion and scattering mechanism. In real materials, there is very likely to be a small mass term in the single-Dirac-cone semimetals. Even a very small mass term can couple together the lowest Landau bands of opposite chirality. In this case, the magnetoconductivity will be described by Eqs.~(\ref{eq:s_zz_Q,G},\ref{eq:s_zz_Q,C}) and (\ref{eq:s_xx_Q,G},\ref{eq:s_xx_Q,C}) but with a constant $v_{0,k_{F}}$ (thus still distinguishable from the quadratic lowest Landau band). Nevertheless, the massless Dirac fermions are still possible to be explored in the acoustic \cite{MaG_NRP_2019} and photonic crystals \cite{OzawaT_RMP_2019}. Third, in some cases we studied, the longitudinal magnetoresistance can be negative. In the earlier theories \cite{NielsenH_PLB_1983,SonD_PRB_2013,BurkovA_PRL_2014}, the negative longitudinal magnetoresistance was attributed to the chiral anomaly. Later, it was found that there is no necessity to use the chiral anomaly to explain the negative longitudinal magnetoresistance. For example, the negative longitudinal magnetoresistance has been observed in topological insulators \cite{DaiX_PRL_2017}, but there is no well-defined chirality, not to mention the chiral anomaly \cite{AndreevA_PRL_2018}. The same treatment used in this paper can also be used to address the negative longitudinal magnetoresistance in the quantum limit. For the one-node model, the negative longitudinal magnetoresistance is found in the quantum limit \cite{NielsenH_PLB_1983,SonD_PRB_2013}, but where the scattering time ($\tau$) is assumed to be a constant. Later, it is found that the dependence of the scattering time on the magnetic field can be considered, by including the disorder scattering \cite{LuH_PRB_2015}. In this paper, we show that for the one-node model, the scattering time has no magnetic field dependence after including the vertex correction to the velocity from the disorder scattering, so we have a negative magnetoresistance in Fig.~\ref{fig:Magnetic-field-dependence-LMR}(a), but the microscopic mechanism is subtly different from the earlier prediction \cite{NielsenH_PLB_1983,SonD_PRB_2013}. For the two-node model, we find that the longitudinal magnetoresistance is generally positive [Figs.~\ref{fig:Magnetic-field-dependence-LMR}(b) and \ref{fig:Magnetic-field-dependence-LMR}(c)], but it can be negative at a large carrier concentration in the presence of the short-screening-length Coulomb potential [Fig.~\ref{fig:Magnetic-field-dependence-LMR}(c) large $\kappa$].

Finally, our results apply to the case where the impurity strength is weak. Further studies can be carried out on the quantum-limit transport under strong disorder. It has been reported that strong disorder can induce resonance states to the spectrum \cite{BlackSchafferA_PRB_2015,XuY_NC_2017}, which further modifies the signature of the optical conductivity \cite{LiS_FoP_2017,PiresJ__2022}. The effect of strong disorder to Landau bands is unknown, and the corresponding magnetoresistance remains to be explored.
\begin{acknowledgments}
This work was supported by the National Key R\&D Program of China (Grant No. 2022YFA1403700), the Innovation Program for Quantum Science and Technology (Grant No. 2021ZD0302400), the National Natural Science Foundation of China (Grant No. 11925402), Guangdong province (Grants No. 2020KCXTD001 and No. 2016ZT06D348), the Science, Technology and Innovation Commission of Shenzhen Municipality (Grants No. ZDSYS20170303165926217, No. JAY20170412152620376, and No. KYTDPT20181011104202253). The numerical calculations were supported by Center for Computational Science and Engineering of SUSTech.
\end{acknowledgments}

\begin{widetext}

\appendix

\section{Critical magnetic field for the two-node model\label{sec:The-critical-magnetic}}

$k_{z}$ of the minimum $E_{k_{z},1+}^{Q}$ varies with the magnetic field. When $k_{w}^{2}>2/l_{B}^{2}$, the minimum $E_{k_{z},1+}^{Q}$ has $k_{z}=\pm\sqrt{k_{w}^{2}-2/l_{B}^{2}}$. Then, the relation between the critical magnetic field $B_{c}$ and the carrier concentration $n_{0}$ is 
\begin{equation}
\sqrt{2\frac{A^{2}}{l_{B_c}^{2}M^{2}}}=\left(2\pi^{2}l_{B_c}^{2}n_{0}\right)^{2}-k_{w}^{2},
\end{equation}
where $l_{B_c}=\sqrt{\hbar/(eB_c)}$. When $k_{w}^{2}\le2/l_{B}^{2}$, the minimum $E_{k_{z},1+}^{Q}$ has $k_{z}=0$. Then, one has 
\begin{equation}
\sqrt{\left(k_{w}^{2}-\frac{2}{l_{B_c}^{2}}\right)^{2}+2\frac{A^{2}}{l_{B_c}^{2}M^{2}}}=\left(2\pi^{2}l_{B_c}^{2}n_{0}\right)^{2}-k_{w}^{2}.
\end{equation}

\section{Relation of $n_{0}$ and $E_{F}$ in the quantum limit}

In the case of the one-node model, the lowest Landau band has an equal number of states above and below zero energy. This allows for the definition of carrier concentration as the difference between the number of occupied states above zero energy and the number of holes below zero energy, see Fig.~\ref{fig:Illustrations-of-Ef}(a). On the other hand, for the two-node model, the lowest Landau band is quadratic, and thus the carrier concentration is defined using the traditional method employed in the case of electron gases, i.e., the number of all occupied states, see Fig.~\ref{fig:Illustrations-of-Ef}(b).

For the one-node model, the relationship between the Fermi energy $E_{F}$ and carrier concentration $n_{0}$ can be found through their relations with the Fermi wave vector $k_{F}$ as
\begin{equation}
n_{0}=N_{L}\frac{\left|k_{F}\right|}{2\pi},
\end{equation}
\begin{align}
E_{F} & =\hbar v_{F}\left|k_{F}\right|\nonumber \\
 & =4\pi^{2}\hbar v_{F}l_{B}^{2}n_{0}.
\end{align}

Similarly, for the two-node model,

\begin{equation}
n_{0}=N_{L}\frac{2\left|k_{F}\right|}{2\pi},
\end{equation}
\begin{align}
E_{F} & =\frac{\omega}{2}+M\left(k_{F}^{2}-k_{w}^{2}\right)\nonumber \\
 & =M\left[\frac{1}{l_{B}^{2}}+\left(2\pi^{2}n_{0}l_{B}^{2}\right)^{2}-k_{w}^{2}\right].
\end{align}
If the carrier concentration of the two-node model is defined as the occupied states above zero energy, one has 
\begin{equation}
n_{0}=N_{L}\frac{2\left(\left|k_{F}\right|-\left|k_{z}^{0}\right|\right)}{2\pi},
\end{equation}
\begin{equation}
k_{z}^{0}=\pm\sqrt{k_{w}^{2}-1/l_{B}^{2}},
\end{equation}
where $k_{z}^{0}$ is the value of $k_{z}$ when $E_{k_{z},0}^{Q}$ equals zero ($B\le k_{w}^{2}\hbar/e$). Although this modification will alter the field dependence of $k_{F}$, it will not eliminate it. Consequently, the magnetoresistance expression of the two-node model will still have a $B$-dependent $v_{k_{F},0}$. When $B>k_{w}^{2}\hbar/e$, the entire lowest Landau band will shift to energies above zero. Defining the carrier concentration of the two-node model in this way requires the total number of occupied states in the lowest Landau band to vary with the magnetic field strength to keep the number of occupied states above zero energy fixed. In summary, defining the carrier concentration in this way is not suitable for the two-node model.

\begin{figure}
\includegraphics{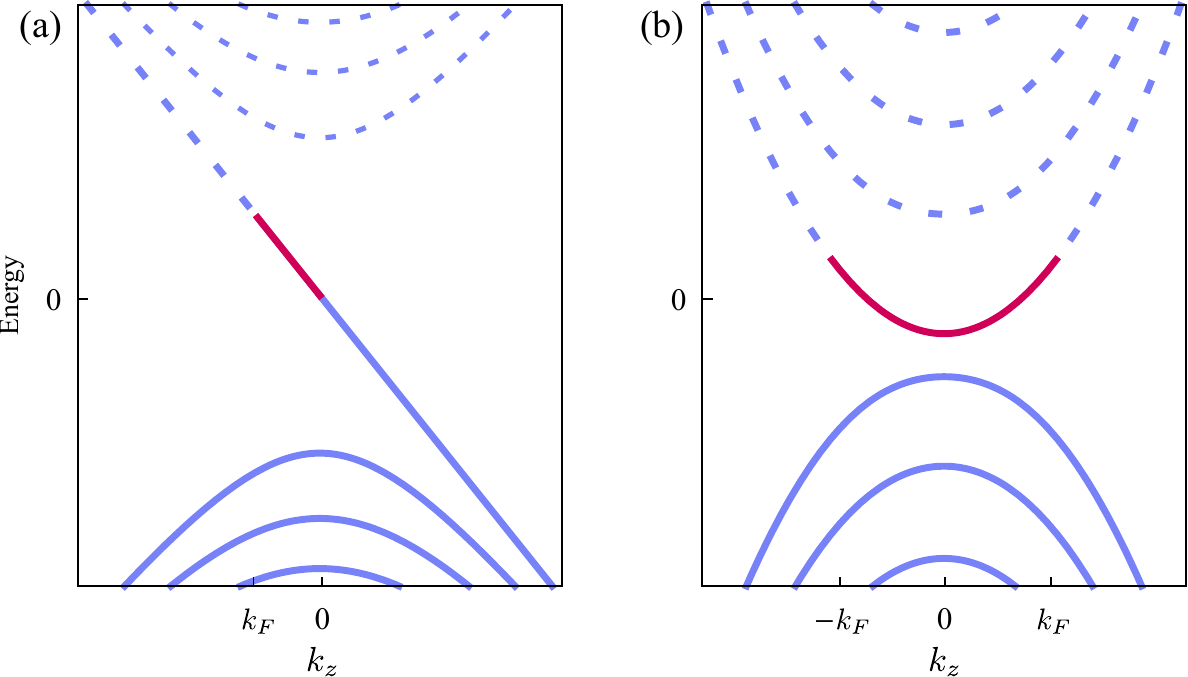}
\caption{\label{fig:Illustrations-of-Ef}Illustrations of the relation of $n_{0}$ and $E_{F}$ in the quantum limit for (a) one-node model and (b) two-node model. Solid lines and dashed lines denote the occupied and unoccupied states, respectively. The red solid lines represent the carriers in the system.}
\end{figure}

\section{Green function and self-energy for Landau bands\label{sec:Green-function-and}}

A general Hamiltonian reads
\begin{align}
\hat{H} & =\hat{H}_{0}+\hat{V},\\
\hat{H}_{0}\left|u\right\rangle  & =E_{u}\left|u\right\rangle ,
\end{align}
where $\hat{V}$ is the operator of the impurity potential. By its definition, Matsubara Green function is 
\begin{align}
\hat{G}\left(i\omega_{m}\right) & =\left(i\omega_{m}-\hat{H}\right)^{-1}\nonumber \\
 & =\left\{ \left[\hat{G_{0}}\left(i\omega_{m}\right)\right]^{-1}-\hat{V}\right\} ^{-1}\nonumber \\
 & =\left\{ 1-\hat{G_{0}}\left(i\omega_{m}\right)\hat{V}\right\} ^{-1}\hat{G_{0}}\left(i\omega_{m}\right),\label{eq:GF-definition}
\end{align}
where $i\omega_{m}/\hbar$ are the imaginary frequency. Equivalently, the above expression can be written as 
\begin{equation}
\hat{G}\left(i\omega_{m}\right)=\hat{G_{0}}\left(i\omega_{m}\right)+\hat{G_{0}}\left(i\omega_{m}\right)\hat{V}\hat{G_{0}}\left(i\omega_{m}\right)+\hat{G_{0}}\left(i\omega_{m}\right)\hat{V}\hat{G_{0}}\left(i\omega_{m}\right)\hat{V}\hat{G_{0}}\left(i\omega_{m}\right)+\cdots.\label{eq:GF-expand}
\end{equation}

In the non perturbation energy basis $\left|u\right\rangle $, the non diagonal terms of $\left\langle u\right|\hat{G}\left|u'\right\rangle $ is generally non zero. The quantum number $u$ contains band index $b$ and momentum $\mathbf{k}$. After the impurity average \cite{Mahan_2000}, $\left\langle b,\mathbf{k}\right|\hat{G}\left|b',\mathbf{k}'\right\rangle $ reduces to $\left\langle b,\mathbf{k}\right|\hat{G}\left|b',\mathbf{k}\right\rangle $, but the interaction brought by the impurity potential between states with different band indexes still exists. Assuming that states will not be scattered from one band to another, only the diagonal terms remain, i.e., only the effect of band broadening from impurity is considered. One has

\begin{align}
G_{u}\left(i\omega_{m}\right) & =\left\langle u\right|\hat{G}\left(i\omega_{m}\right)\left|u\right\rangle \nonumber \\
 & =\left\langle u\right|\hat{G_{0}}\left(i\omega_{m}\right)\left|u\right\rangle +\left\langle u\right|\hat{G_{0}}\left(i\omega_{m}\right)\hat{V}\hat{G_{0}}\left(i\omega_{m}\right)\left|u\right\rangle +\left\langle u\right|\hat{G_{0}}\left(i\omega_{m}\right)\hat{V}\hat{G_{0}}\left(i\omega_{m}\right)\hat{V}\hat{G_{0}}\left(i\omega_{m}\right)\left|u\right\rangle +\cdots,\label{eq:GF-expand2}
\end{align}
where the first term is the non perturbation part, and the second term gives a constant energy shift that can be included in the Fermi energy. Note that the notion of impurity average on Green functions and the following correlation functions is omitted. In the first Born approximation, the self-energy can be extracted from the third term,
\begin{align}
\left\langle u\right|\hat{G_{0}}\left(i\omega_{m}\right)\hat{V}\hat{G_{0}}\left(i\omega_{m}\right)\hat{V}\hat{G_{0}}\left(i\omega_{m}\right)\left|u\right\rangle  & =\frac{1}{i\omega_{m}-E_{u}}\left\langle u\right|\hat{V}G_{0}\left(i\omega_{m}\right)\hat{V}\left|u\right\rangle \frac{1}{i\omega_{m}-E_{u}}\nonumber \\
 & =\frac{1}{i\omega_{m}-E_{u}}\sum_{u'}\left\langle u\right|\hat{V}\left|u'\right\rangle \frac{1}{i\omega_{m}-E_{u'}}\left\langle u'\right|\hat{V}\left|u\right\rangle \frac{1}{i\omega_{m}-E_{u}}.
\end{align}
That is 
\begin{equation}
\Sigma_{u}\left(i\omega_{m}\right)=\sum_{u'}\frac{\left|\left\langle u'\right|\hat{V}\left|u\right\rangle \right|^{2}}{i\omega_{m}-E_{u'}}.
\end{equation}
Utilizing the Dyson equation \cite{Mahan_2000}, one has the approximating Green function 
\begin{equation}
G_{u}\left(i\omega_{m}\right)\approx\frac{1}{i\omega_{m}-E_{u}-\Sigma_{u}\left(i\omega_{m}\right)},
\end{equation}
and its operator form
\begin{equation}
\hat{G}\left(i\omega_{m}\right)=\sum_{u}\frac{\left|u\right\rangle \left\langle u\right|}{i\omega_{m}-E_{u}-\Sigma_{u}\left(i\omega_{m}\right)}.\label{eq:Matsubara-Green-operator}
\end{equation}

\subsection{Self-energy for the lowest Landau band\label{subsec:Self-energy-for-LLB}}

For the lowest Landau band, the retarded self-energy is 
\begin{equation}
\Sigma_{k_{z},0}^{R}\left(E_{F}\right)=\sum_{k_{x}',k_{z}',b'}\frac{\left|\left\langle k_{x}',k_{z}',b'\right|\hat{V}\left|k_{x},k_{z},0\right\rangle \right|^{2}}{E_{F}-E_{k_{z}',b'}+i\eta},
\end{equation}
where $b$ indicates the band index. In the quantum limit, only $b'=0$ contributes to the imaginary part of the self-energy. One has
\begin{equation}
\left\langle k_{x}',k_{z}',0\right|\hat{V}\left|k_{x},k_{z},0\right\rangle =\int\int d\mathbf{r}d\mathbf{r}'\left\langle k_{x}',k_{z}',0\right|\left.\mathbf{r}\right\rangle \left\langle \mathbf{r}\right|\hat{V}\left|\mathbf{r}'\right\rangle \left\langle \mathbf{r}'\right|\left.k_{x},k_{z},0\right\rangle ,\label{eq:scattering-matrix00}
\end{equation}
with $\left\langle \mathbf{r}\right|\hat{V}\left|\mathbf{r}'\right\rangle =V\left(\mathbf{r}\right)\delta\left(\mathbf{r}-\mathbf{r}'\right)$, and 
\begin{align}
\left\langle \mathbf{r}\right|\left.k_{x},k_{z},0\right\rangle  & =\left\langle x,z\right|\left.k_{x},k_{z}\right\rangle \begin{pmatrix}0\\
\left\langle y\right|\left.0\right\rangle 
\end{pmatrix}\nonumber \\
 & =\frac{1}{\sqrt{L_{x}L_{z}}}e^{i\left(xk_{x}+zk_{z}\right)}\begin{pmatrix}0\\
\phi_{0}\left(y,k_{x}\right)
\end{pmatrix},
\end{align}
where
\begin{equation}
\phi_{n}\left(y,k_{x}\right)=\frac{1}{\sqrt{n!2^{n}\ell_{B}\sqrt{\pi}}}e^{-\frac{1}{2}\left(\frac{y}{\ell_{B}}-k_{x}\ell_{B}\right)^{2}}H_{n}\left(\frac{y}{\ell_{B}}-k_{x}\ell_{B}\right),
\end{equation}
with Hermite polynomials $H_{n}\left(x\right)$. Together with the complex conjugate of Eq.~(\ref{eq:scattering-matrix00}), one has 
\begin{align}
\left|\left\langle k_{x}',k_{z}',0\right|\hat{V}\left|k_{x},k_{z},0\right\rangle ^{2}\right|= & n_{i}\int\frac{d\boldsymbol{q}}{\left(2\pi\right)^{3}}u\left(\boldsymbol{q}\right)u\left(-\boldsymbol{q}\right)\delta_{k_{z}',k_{z}-q_{z}}\delta_{k_{x}',k_{x}-q_{x}}\nonumber \\
 & \times\int dydy'e^{iq_{y}\left(y-y'\right)}\phi_{0}\left(y,k_{x}\right)\phi_{0}\left(y,k_{x}'\right)\phi_{0}\left(y',k_{x}\right)\phi_{0}\left(y',k_{x}'\right).
\end{align}
Here the impurity average is taken, $\left<V\left(\mathbf{r}\right)V\left(\mathbf{r}'\right)\right>_{imp}=n_{i}\frac{\int d\boldsymbol{q}}{\left(2\pi\right)^{3}}u\left(\boldsymbol{q}\right)u\left(-\boldsymbol{q}\right)e^{i\boldsymbol{q}\left(\mathbf{r}-\mathbf{r}'\right)}$. After integrating $y$ and $y'$, 
\begin{equation}
\int dye^{\pm iq_{y}y}\phi_{0}\left(y,k_{x}\right)\phi_{0}\left(y,k_{x}'\right)=e^{-\frac{1}{4}\left[\left(k_{x}-k_{x}'\right)^{2}+q_{y}^{2}\right]l_{B}^{2}\pm\frac{1}{2}iq_{y}\left(k_{x}+k_{x}'\right)l_{B}^{2}},
\end{equation}
one has 
\begin{equation}
\Im\left[\Sigma_{k_{F},0}^{R}\left(E_{F}\right)\right]=-\pi n_{i}\frac{\int d\boldsymbol{q}}{\left(2\pi\right)^{3}}\delta\left(E_{F}-E_{k_{F}-q_{z},0}\right)u\left(\boldsymbol{q}\right)u\left(-\boldsymbol{q}\right)e^{-\frac{1}{2}\left(q_{x}^{2}+q_{y}^{2}\right)l_{B}^{2}}.
\end{equation}

For Gaussian potential, $u\left(\boldsymbol{q}\right)=u_{0}e^{-\frac{q^{2}d^{2}}{2}}$, the integral over $q_{x},q_{y}$ in above expression can be performed, 
\begin{align}
\frac{\int\int dq_{x}dq_{y}}{\left(2\pi\right)^{2}}e^{-\left(q_{x}^{2}+q_{y}^{2}\right)d^{2}}e^{-\frac{1}{2}\left(q_{x}^{2}+q_{y}^{2}\right)l_{B}^{2}} & =\frac{1}{2\pi}\int_{0}^{\infty}q_{\parallel}e^{-q_{\parallel}^{2}\left(d^{2}+\frac{1}{2}l_{B}^{2}\right)}dq_{\parallel}\nonumber \\
 & =\frac{1}{2\pi\left(2d^{2}+l_{B}^{2}\right)},
\end{align}
where $q_{\parallel}=\sqrt{q_{x}^{2}+q_{y}^{2}}$ (the integral is calculated in the polar coordination). Then, one has Eq.~(\ref{eq:tau0_G}).

To find Eq.~(\ref{eq:tau0_C}), one need to use Coulomb potential $u\left(\boldsymbol{q}\right)=\frac{e^{2}}{\varepsilon\left(q^{2}+\kappa^{2}\right)}$, 
\begin{align}
\frac{\int\int dq_{x}dq_{y}}{\left(2\pi\right)^{2}}\frac{e^{4}}{\varepsilon^{2}\left(q_{x}^{2}+q_{y}^{2}+q_{z}^{2}+\kappa^{2}\right)^{2}}e^{-\frac{1}{2}\left(q_{x}^{2}+q_{y}^{2}\right)l_{B}^{2}} & =\frac{e^{4}}{2\pi\varepsilon^{2}}\int_{0}^{\infty}q_{\parallel}\frac{1}{\left(q_{\parallel}^{2}+q_{z}^{2}+\kappa^{2}\right)}e^{-\frac{1}{2}q_{\parallel}^{2}l_{B}^{2}}dq_{\parallel}\nonumber \\
 & =\frac{e^{4}l_{B}^{2}}{8\pi\varepsilon^{2}}\mathcal{F}_{1}\left[\frac{l_{B}^{2}}{2}\left(q_{z}^{2}+\kappa^{2}\right)\right],
\end{align}
where $\mathcal{F}_{1}\left[x\right]=\frac{1}{x}+e^{x}E_{i}\left[-x\right]$ and $E_{i}\left[-x\right]=-\int_{x}^{\infty}\frac{1}{t}e^{-t}dt$. The power function fitting of $\mathcal{F}_{1}\left[x\right]$ is shown in Fig.~\ref{fig:Fitting-of}(a).

\begin{figure*}
\includegraphics{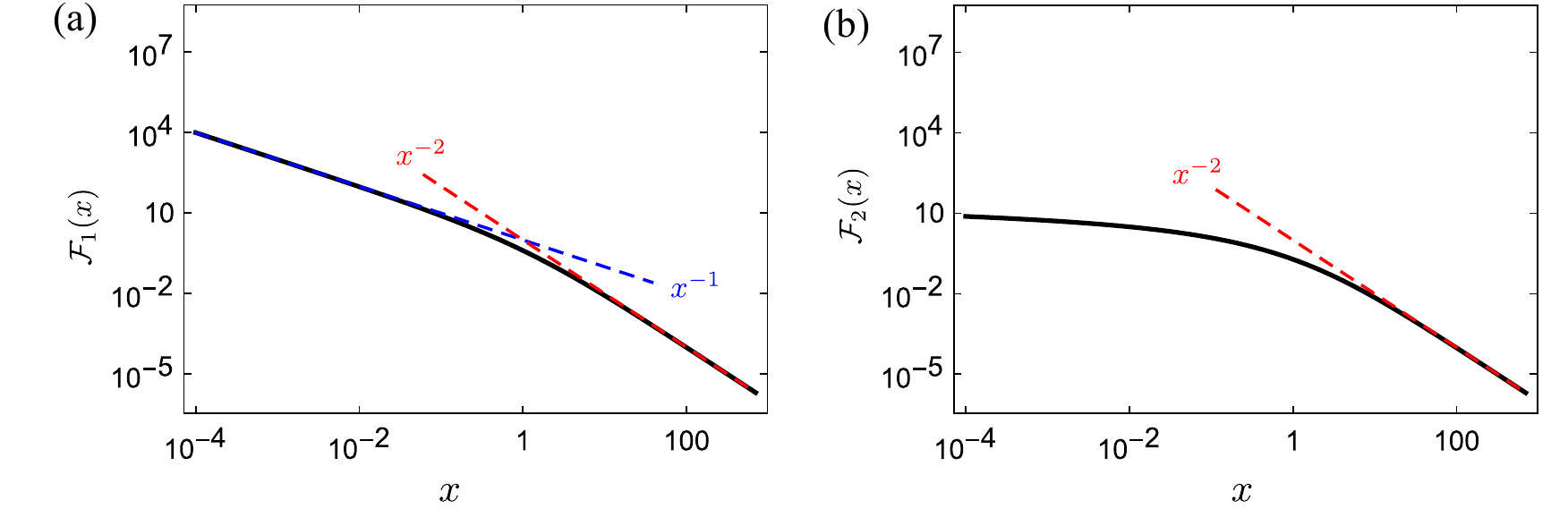}

\caption{\label{fig:Fitting-of}Power function fitting of (a) $\mathcal{F}_{1}\left[x\right]$ and (b) $\mathcal{F}_{2}\left[x\right]$.}

\end{figure*}

\subsection{Self-energy for the n=1 Landau band\label{subsec:Self-energy-for-1LB}}

The self-energy for the $b=1+$ and $1-$ bands in the quantum limit can be found in the same procedure.

For $\Sigma_{k_{z},1+}^{R}\left(E_{F}\right)$, 
\begin{align}
\left\langle \mathbf{r}\right|\left.k_{x},k_{z},1+\right\rangle  & =\left\langle x,z\right|\left.k_{x},k_{z}\right\rangle \begin{pmatrix}\cos\frac{\theta_{k_{z},1}}{2}\left\langle y\right|\left.0\right\rangle \\
\sin\frac{\theta_{k_{z},1}}{2}\left\langle y\right|\left.1\right\rangle 
\end{pmatrix}\nonumber \\
 & =\frac{1}{\sqrt{L_{x}L_{z}}}e^{i\left(xk_{x}+zk_{z}\right)}\begin{pmatrix}\cos\frac{\theta_{k_{z},1}}{2}\phi_{0}\left(y,k_{x}\right)\\
\sin\frac{\theta_{k_{z},1}}{2}\phi_{1}\left(y,k_{x}\right)
\end{pmatrix},
\end{align}
and
\begin{equation}
\int dye^{\pm iq_{y}y}\phi_{1}\left(y,k_{x}\right)\phi_{0}\left(y,k_{x}'\right)=\frac{1}{\sqrt{2}}l_{B}\left(k_{x}-k_{x}'\pm iq_{y}\right)e^{-\frac{1}{4}\left[\left(k_{x}-k_{x}'\right)^{2}+q_{y}^{2}\right]l_{B}^{2}\pm\frac{1}{2}iq_{y}\left(k_{x}+k_{x}'\right)l_{B}^{2}}.
\end{equation}
Therefore, 
\begin{equation}
\Im\left[\Sigma_{k_{F},1+}^{R}\left(E_{F}\right)\right]=-\pi n_{i}\frac{\int d\boldsymbol{q}}{\left(2\pi\right)^{3}}\delta\left(E_{F}-E_{k_{F}-q_{z},0}\right)\left(\sin\frac{\theta_{k_{F},1}}{2}\right)^{2}u\left(\boldsymbol{q}\right)u\left(-\boldsymbol{q}\right)e^{-\frac{1}{2}\left(q_{x}^{2}+q_{y}^{2}\right)l_{B}^{2}}\frac{\left(q_{x}^{2}+q_{y}^{2}\right)l_{B}^{2}}{2}.
\end{equation}
The integrals of $q_{x}$ and $q_{y}$ for Gaussian and Coulomb potentials are more complicated compared to that in $\Im\left[\Sigma_{k_{F},0}^{R}\left(E_{F}\right)\right]$, 
\begin{align}
\frac{\int\int dq_{x}dq_{y}}{\left(2\pi\right)^{2}}e^{-\left(q_{x}^{2}+q_{y}^{2}\right)d^{2}}e^{-\frac{1}{2}\left(q_{x}^{2}+q_{y}^{2}\right)l_{B}^{2}}\frac{\left(q_{x}^{2}+q_{y}^{2}\right)l_{B}^{2}}{2} & =\frac{l_{B}^{2}}{4\pi}\int_{0}^{\infty}q_{\parallel}^{3}e^{-q_{\parallel}^{2}\left(d^{2}+\frac{1}{2}l_{B}^{2}\right)}dq_{\parallel}\nonumber \\
 & =\frac{l_{B}^{2}}{2\pi\left(2d^{2}+l_{B}^{2}\right)^{2}},
\end{align}
and
\begin{align}
\frac{\int\int dq_{x}dq_{y}}{\left(2\pi\right)^{2}}\frac{e^{4}}{\varepsilon^{2}\left(q_{x}^{2}+q_{y}^{2}+q_{z}^{2}+\kappa^{2}\right)^{2}}e^{-\frac{1}{2}\left(q_{x}^{2}+q_{y}^{2}\right)l_{B}^{2}}\frac{\left(q_{x}^{2}+q_{y}^{2}\right)l_{B}^{2}}{2} & =\frac{e^{4}l_{B}^{2}}{4\pi\varepsilon^{2}}\int_{0}^{\infty}q_{\parallel}^{3}\frac{1}{\left(q_{\parallel}^{2}+q_{z}^{2}+\kappa^{2}\right)}e^{-\frac{1}{2}q_{\parallel}^{2}l_{B}^{2}}dq_{\parallel}\nonumber \\
 & =\frac{e^{4}l_{B}^{2}}{8\pi\varepsilon^{2}}\mathcal{F}_{2}\left[\frac{l_{B}^{2}}{2}\left(q_{z}^{2}+\kappa^{2}\right)\right],
\end{align}
where $\mathcal{F}_{2}\left[x\right]=-1-\left(1+x\right)e^{x}E_{i}\left[-x\right]$. The power function fitting of $\mathcal{F}_{2}\left[x\right]$ is shown in Fig.~\ref{fig:Fitting-of}(b).

\section{Kubo formula for magnetoconductivity\label{sec:Kubo-formula-for}}

The conductivity is related to the retarded current-current correlation function $\Pi_{\alpha\alpha}^{R}$, 
\begin{equation}
\Re\left[\sigma_{\alpha\beta}\right]=-\lim_{\Omega\rightarrow0}\frac{\hbar}{\Omega}\Im\left[\Pi_{\alpha\beta}^{R}\left(\Omega\right)\right].\label{eq:Kubo-correlation}
\end{equation}
From the bubble diagram, one has 
\begin{equation}
\Pi_{\alpha\beta}\left(i\Omega\right)=\frac{e^{2}k_{B}T}{V}\sum_{m}\Tr\left[v_{\alpha}\hat{G}\left(i\omega_{m}\right)v_{\beta}\hat{G}\left(i\omega_{m}+i\Omega\right)\right],
\end{equation}
where $i\Omega/\hbar$ is the imaginary frequencies, $k_{B}$ is the Boltzmann constant, $T$ is temperature. After substituting Eq.~(\ref{eq:Matsubara-Green-operator}) into above expression, one has
\begin{align}
\Pi_{\alpha\beta}\left(i\Omega\right) & =\frac{e^{2}k_{B}T}{V}\sum_{m}\sum_{u,u'}\frac{\Tr\left[v_{\alpha}\left|u\right\rangle \left\langle u\right|v_{\beta}\left|u'\right\rangle \left\langle u'\right|\right]}{\left[i\omega_{m}-E_{u}-\Sigma_{u}\left(i\omega_{m}\right)\right]\left[i\omega_{m}+i\Omega-E_{u'}-\Sigma_{u'}\left(i\omega_{m}\right)\right]}\nonumber \\
 & =\frac{e^{2}k_{B}T}{V}\sum_{m}\sum_{u,u'}\frac{\left\langle u'\right|v_{\alpha}\left|u\right\rangle \left\langle u\right|v_{\beta}\left|u'\right\rangle }{\left[i\omega_{m}-E_{u}-\Sigma_{u}\left(i\omega_{m}\right)\right]\left[i\omega_{m}+i\Omega-E_{u'}-\Sigma_{u'}\left(i\omega_{m}\right)\right]},
\end{align}
where the trace operation is performed using the fact that the trace is invariant under cyclic permutations, and the trace of a inner product gives the inner product itself. 

To sum $m$, make use of the identity $\frac{1}{i\omega_{m}-E_{u}-\Sigma_{u}\left(i\omega_{m}\right)}=\int d\omega_{1}\frac{A_{u}\left(\omega_{1}\right)}{i\omega_{m}-\omega_{1}}$ and $k_{B}T\sum_{m}\frac{1}{\left(i\omega_{m}-\omega_{1}\right)\left(i\omega_{m}+i\Omega-\omega_{2}\right)}=\frac{n_{F}\left(\omega_{1}\right)-n_{F}\left(\omega_{2}\right)}{\omega_{1}-\omega_{2}+i\Omega}$, where $n_{F}\left(\omega_{1}\right)$ is the Fermi-Dirac distribution function \cite{Mahan_2000}. Then, the retarded correlation function can be found by making the analytical continuation $i\Omega\rightarrow\Omega+i0^{+}$. By taking its imaginary part (only the part $\frac{1}{\omega_{1}-\omega_{2}+\Omega+i0^{+}}$ is complex in the retarded correlation function when $\alpha=\beta$), one has
\begin{align}
\Im\left[\Pi_{\alpha\alpha}^{R}\left(\Omega\right)\right] & =\frac{e^{2}}{V}\sum_{u,u'}\int\int d\omega_{1}d\omega_{2}\left[-\pi\delta\left(\omega_{1}-\omega_{2}+\Omega\right)\right]\left[n_{F}\left(\omega_{1}\right)-n_{F}\left(\omega_{2}\right)\right]A_{u}\left(\omega_{1}\right)A_{u'}\left(\omega_{2}\right)\left|\left\langle u'\right|v_{\alpha}\left|u\right\rangle \right|^{2}\nonumber \\
 & =-\frac{\pi e^{2}}{V}\sum_{u,u'}\int d\omega_{1}\left[n_{F}\left(\omega_{1}\right)-n_{F}\left(\omega_{1}+\Omega\right)\right]A_{u}\left(\omega_{1}\right)A_{u'}\left(\omega_{1}+\Omega\right)\left|\left\langle u'\right|v_{\alpha}\left|u\right\rangle \right|^{2}.
\end{align}
Note that $\left\langle u'\right|v_{\alpha}\left|u\right\rangle \left\langle u\right|v_{\alpha}\left|u'\right\rangle =\left|\left\langle u'\right|v_{\alpha}\left|u\right\rangle \right|^{2}$. After taking the above expression into Eq.~(\ref{eq:Kubo-correlation}) and making $\lim_{\Omega\rightarrow0}\frac{\left[n_{F}\left(\omega_{1}\right)-n_{F}\left(\omega_{1}+\Omega\right)\right]}{\Omega}=-\frac{dn_{F}\left(\omega_{1}\right)}{d\omega_{1}}$, in the zero temperature limit [$-\frac{dn_{F}\left(\omega_{1}\right)}{d\omega_{1}}\rightarrow\delta\left(\omega_{1}-E_{F}\right)$], one has
\begin{equation}
\Re\left[\sigma_{\alpha\alpha}\right]=\frac{\pi\hbar e^{2}}{V}\sum_{u,u'}A_{u}\left(E_{F}\right)A_{u'}\left(E_{F}\right)\left|\left\langle u'\right|v_{\alpha}\left|u\right\rangle \right|^{2}.
\end{equation}

\subsection{Longitudinal magnetoconductivity}

For $\sigma_{zz}$, the magnetic field is in $z$ direction. Therefore, the wave vector $k_{z}$ is not Landau quantized. One has 
\begin{align}
\left\langle u\right|v_{z}\left|u'\right\rangle  & =\left\langle u\right|\frac{1}{\hbar}\frac{\partial H}{\partial k_{z}}\left|u'\right\rangle \nonumber \\
 & =\frac{1}{\hbar}\frac{\partial E_{u}}{\partial k_{z}}\delta_{u,u'}.
\end{align}
Because $\frac{1}{V}\sum_{k_{x},k_{z}}=\frac{1}{L_{y}}\int\int\frac{dk_{x}dk_{z}}{\left(2\pi\right)^{2}}$ and $\frac{1}{L_{y}}\int_{0}^{^{eBL_{y}/\hbar}}\frac{dk_{x}}{2\pi}=\frac{eB}{h}$ (the guiding center lies between $0$ and $L_{y}$, $0\le\frac{\hbar}{eB}k_{x}\le L_{y}$, giving the limits of $k_{x}$), the summation $\frac{1}{V}\sum_{k_{x},k_{z}}=N_{L}\int\frac{dk_{z}}{2\pi}$.

With the approximation $\left[A_{u}\left(E_{F}\right)\right]^{2}\approx\frac{1}{\pi\hbar}\tau_{u}\left(E_{F}\right)\delta\left(E_{F}-E_{u}\right)$, the quantum-limit longitudinal magnetoconductivity is
\begin{align}
\sigma_{zz} & \approx\pi\hbar e^{2}N_{L}\int_{-\infty}^{\infty}\frac{dk_{z}}{2\pi}\left[\frac{1}{\pi\hbar}\tau_{k_{z},0}\left(E_{F}\right)\delta\left(E_{F}-E_{k_{z},0}\right)\right]\left|\frac{1}{\hbar}\frac{\partial E_{k_{z},0}}{\partial k_{z}}\right|^{2}\nonumber \\
 & =e^{2}N_{L}\int_{-\infty}^{\infty}\frac{dk_{z}}{2\pi}\tau_{k_{z},0}\left(E_{F}\right)\left[\sum_{i}\frac{\delta\left(k_{z}-k_{F}^{i}\right)}{\left|\frac{\partial E_{k_{z},0}}{\partial k_{z}}\right|_{k_{z}=k_{F}^{i}}}\right]\left|\frac{1}{\hbar}\frac{\partial E_{k_{z},0}}{\partial k_{z}}\right|^{2}\nonumber \\
 & =\frac{e^{2}}{h}N_{L}\sum_{i}\tau_{k_{F}^{i},0}\left(E_{F}\right)\left|v_{k_{F}^{i},0}\right|,
\end{align}
where $v_{k_{F}^{i},0}=\frac{1}{\hbar}\frac{\partial E_{k_{z},0}}{\partial k_{z}}|_{k_{z}=k_{F}^{i}}$. For parabolic bands, one has $v_{-k_{F},0}=-v_{k_{F},0}$ and $\tau_{k_{F},0}\left(E_{F}\right)=\tau_{-k_{F},0}\left(E_{F}\right)$. 

\subsection{Transverse magnetoconductivity\label{subsec:Transverse-magnetoconductivity}}

For the transverse magnetoconductivity, 
\begin{align}
\sigma_{xx} & =\frac{\pi\hbar e^{2}}{V}\sum_{u,u'}A_{u}\left(E_{F}\right)A_{u'}\left(E_{F}\right)\left|\left\langle u\right|v_{x}\left|u'\right\rangle \right|^{2},
\end{align}
one need to calculate $\left\langle u\right|v_{x}\left|u'\right\rangle $. For the one-node model, one has $v_{x}=v_{F}\sigma_{x}$, which makes $\left\langle u\right|v_{x}\left|u'\right\rangle $ non zero only if the Landau indexes (i.e., 0 or $n$ ) of $u'$ and $u$ differ $1$. Therefore, the dominant terms of the quantum-limit magnetoconductivity are $\sigma_{xx,1+}^{L}$ (when the band index $b=0,b'=1+$ and $b=1+,b'=0$) and $\sigma_{xx,1-}^{L}$ (when $b=0,b'=1-$ and $b=1-,b'=0$). One has 

\begin{equation}
\left|\left\langle k_{x},k_{z},0\right|v_{x}\left|k_{x}',k_{z}',1+\right\rangle \right|^{2}=v_{F}^{2}\left(\cos\frac{\theta_{k_{z},1}}{2}\right)^{2}\delta_{k_{z},k_{z}'}\delta_{k_{x},k_{x}'}.
\end{equation}
With the approximation $A_{k_{z},0}\left(E_{F}\right)\approx\delta\left(E_{F}-E_{k_{z},0}\right)$ and $A_{k_{z},1\pm}\left(E_{F}\right)\approx\frac{1}{\pi}\frac{1}{\left(E_{F}-E_{k_{z},1\pm}\right)^{2}}\frac{\hbar}{2\tau_{k_{z},1\pm}\left(E_{F}\right)}$ ,
\begin{align}
\sigma_{xx,1+}^{L} & =\frac{\hbar e^{2}v_{F}^{2}}{\pi}N_{L}\int_{-\infty}^{\infty}\delta\left(E_{F}-E_{k_{z},0}\right)\left[\left(\frac{\cos\frac{\theta_{k_{z},1}}{2}}{E_{F}-E_{k_{z},1+}}\right)^{2}\frac{\hbar}{2\tau_{k_{z},1+}}\right]dk_{z}\nonumber \\
 & =\frac{e^{2}v_{F}}{\pi}N_{L}\left[\left(\frac{\cos\frac{\theta_{k_{F},1}}{2}}{E_{F}-E_{k_{F},1+}}\right)^{2}\frac{\hbar}{2\tau_{k_{F},1+}}\right].
\end{align}
Combining the above expression with Eqs.~(\ref{eq:tau_1p_G}) and (\ref{eq:tau_1p_C}), one has 
\begin{equation}
\sigma_{xx,1+}^{L,G}=\frac{e^{2}}{h}\frac{n_{i}u_{0}^{2}}{\left(2\pi l_{B}^{2}\right)^{2}}\frac{1}{\left(1+2d^{2}/l_{B}^{2}\right)^{2}}f_{1+}^{L},
\end{equation}
\begin{equation}
\sigma_{xx,1+}^{L,C}=\frac{e^{2}}{h}\frac{n_{i}e^{4}}{\left(4\pi\right)^{2}\varepsilon^{2}}\mathcal{F}_{2}\left[\frac{l_{B}^{2}\kappa^{2}}{2}\right]f_{1+}^{L},
\end{equation}
with 
\begin{equation}
f_{1+}^{L}=\left(\frac{\cos\frac{\theta_{k_{F},1}}{2}\sin\frac{\theta_{k_{F},1}}{2}}{E_{F}-E_{k_{F},1+}}\right)^{2}.
\end{equation}
With an observation on Eqs.~(\ref{eq:sigma_xx},\ref{eq:tau1_general}) and (\ref{eq:eigenvectors}), one can find $\sigma_{xx,1-}$ by replacing $E_{k_{F},1+}$ with $E_{k_{F},1-}$ in above expressions. When the carrier concentration is low, one has $\frac{2}{l_{B}}\gg k_{F}$ under the strong magnetic field, resulting in $f_{1\pm}^{L}\approx\frac{l_{B}^{2}}{8\left(\hbar v_{F}\right)^{2}}$. 

For the two-node model, $v_{x}=\frac{A}{\hbar}\sigma_{x}-\frac{2M}{\hbar}\left[\frac{1}{\sqrt{2}l_{B}}\left(a+a^{\dagger}\right)\right]\sigma_{z}$, in the same procedure, one has
\begin{align}
\left|\left\langle k_{x},k_{z},0\right|v_{x}\left|k_{x}',k_{z}',1+\right\rangle \right|^{2} & =\frac{1}{\hbar^{2}}\left(A\cos\frac{\theta_{k_{z},1}}{2}+\frac{\sqrt{2}M}{l_{B}}\sin\frac{\theta_{k_{z},1}}{2}\right)^{2}\delta_{k_{z},k_{z}'}\delta_{k_{x},k_{x}'},
\end{align}
and
\begin{equation}
\sigma_{xx,1+}^{Q}=\frac{e^{2}}{\pi}N_{L}\frac{2}{\left|v_{k_{F},0}\right|}\left[\left(\frac{\frac{A}{\hbar}\cos\frac{\theta_{k_{F},1}}{2}+\frac{\sqrt{2}M}{l_{B}\hbar}\sin\frac{\theta_{k_{F},1}}{2}}{E_{F}-E_{k_{F},1+}}\right)^{2}\frac{\hbar}{2\tau_{k_{F},1+}}\right],
\end{equation}
where the factor $2$ comes from $\pm k_{F}$. Note that $\tau_{k_{F},1+}=\tau_{-k_{F},1+}$ for parabolic bands. Combining the above expression with Eqs.~(\ref{eq:tau_1p_G}) and (\ref{eq:tau_1p_C}), one has

\begin{align}
\sigma_{xx,1+}^{Q,G} & =\frac{e^{2}}{h}\frac{2n_{i}u_{0}^{2}}{\left(2\pi l_{B}^{2}v_{k_{F},0}\right)^{2}}\frac{\left(1+e^{-4k_{F}^{2}d^{2}}\right)}{\left(1+2d^{2}/l_{B}^{2}\right)^{2}}f_{1+}^{Q},\\
\sigma_{xx,1+}^{Q,C} & =\frac{e^{2}}{h}\frac{n_{i}e^{4}}{2\varepsilon^{2}\left(2\pi v_{k_{F},0}\right)^{2}}\left\{ \mathcal{F}_{2}\left[\frac{l_{B}^{2}\left(4k_{F}^{2}+\kappa^{2}\right)}{2}\right]+\mathcal{F}_{2}\left[\frac{l_{B}^{2}\kappa^{2}}{2}\right]\right\} f_{1+}^{Q},
\end{align}
with
\begin{equation}
f_{1+}^{Q}=\left[\frac{\frac{A}{\hbar}\cos\frac{\theta_{k_{F},1}}{2}\sin\frac{\theta_{k_{F},1}}{2}+\frac{\sqrt{2}M}{l_{B}\hbar}\left(\sin\frac{\theta_{k_{F},1}}{2}\right)^{2}}{E_{F}-E_{k_{F},1+}}\right]^{2}.
\end{equation}
The exact results of the above expressions are parameter dependent. The parameter-dependent SdH oscillation of this model has been studied in Ref.~\cite{WangC_PRL_2016}. Here, the magnetic dependence is little affected by the choice of parameters. When $2M^{2}\gg A^{2}l_{B}^{2}$, one has $f_{1+}^{Q}=0$ and $f_{1-}^{Q}=\frac{1}{2}\left(\frac{l_{B}}{\hbar}\right)^{2}$; actually, this is the exact result for the 3D electron gas. When $2M^{2}\ll A^{2}l_{B}^{2}$, one has $f_{1\pm}^{Q}=\frac{1}{8}\left(\frac{l_{B}}{\hbar}\right)^{2}$; in this case, it is similar to that of the massless Dirac fermions. Numerical analysis shows that the dominant term of $f_{1\pm}^{Q}$ is always $\propto B^{-1}$. 

For electron gas, only the following substitutions need to be made,
\begin{align}
\begin{pmatrix}0\\
\left|0\right\rangle 
\end{pmatrix} & \rightarrow\left|0\right\rangle ,\\
\begin{pmatrix}\cos\frac{\theta_{k_{z},1}}{2}\left|0\right\rangle \\
\sin\frac{\theta_{k_{z},1}}{2}\left|1\right\rangle 
\end{pmatrix} & \rightarrow\left|1\right\rangle .
\end{align}
With $v_{x}=\frac{\hbar}{m}\frac{1}{\sqrt{2}l_{B}}\left(a+a^{\dagger}\right)$, one has $\left|\left\langle k_{x},k_{z},0\right|v_{x}\left|k_{x},k_{z},1\right\rangle \right|^{2}=\frac{1}{2}\left(\frac{\hbar}{ml_{B}}\right)^{2}$. Then, the transverse magnetoconductivity of the electron gas has the same expressions as the above two-node model, except that $f_{1+}^{Q}$ is replaced by $\frac{l_{B}^{2}}{2\hbar^{2}}$. .

\section{Calculation of the reciprocal Debye screening length\label{sec:Calculation-of-kappa} }

The reciprocal screening radius $\kappa$ for the lowest Landau band is given by \cite{AbrikosovA_PRB_1998}
\begin{align}
\kappa^{2} & =\frac{e^{2}k_{B}T}{\varepsilon V}\sum_{m,k_{x},k_{z}}\left(\frac{1}{i\omega_{m}-E_{k_{z},0}}\right)^{2}\nonumber \\
 & =\frac{e^{2}}{\varepsilon}N_{L}\int\frac{dk_{z}}{2\pi}\delta\left(E_{F}-E_{k_{z},0}\right),
\end{align}
where $k_{B}T\sum_{m}\left(\frac{1}{i\omega_{m}-E_{k_{z},0}}\right)^{2}=\delta\left(E_{F}-E_{k_{z},0}\right)$ in the zero-temperature limit. For the one-node model, one has $\kappa^{2}=\frac{e^{2}}{2\pi\varepsilon}N_{L}\frac{1}{\hbar v_{F}}$; for the parabolic lowest Landau band , $\kappa^{2}=\frac{e^{2}}{2\pi\varepsilon}N_{L}\frac{2}{\hbar v_{k_{F},0}}$. 

\section{Correction when the Fermi energy is near the band bottom\label{sec:Correction-when-the}}

The delta function comes from the approximation $G_{k_{z},0}^{R}\left(E_{F}\right)G_{k_{z},0}^{A}\left(E_{F}\right)$ or $A_{k_{z},0}^{R}\left(E_{F}\right)$ in the magnetoconductivity formula, and from taking the imaginary part of $G_{k_{z},0}^{R}$ during finding the scattering time. The approximation $\lim_{\eta\rightarrow0}\frac{1}{\pi}\frac{\eta}{x^{2}+\eta^{2}}\rightarrow\delta\left(x\right)$ is making the Lorentz type broadening approaches zero. Using the delta functions makes the analytic process possible. For the two-node model, 
\begin{align}
\int_{-\infty}^{\infty}\frac{1}{\pi}\frac{\eta}{\left[M\left(k_{z}^{2}-k_{F}^{2}\right)\right]^{2}+\eta^{2}}dk_{z}\approx & \int_{-\infty}^{\infty}\delta\left[M\left(k_{F}^{2}-k_{z}^{2}\right)\right]dk_{z}\nonumber \\
 & =\frac{1}{M\left|k_{F}\right|}.
\end{align}
It is natural that the above result diverges when $k_{F}\rightarrow0$. However, without making the approximation, one can directly get 
\begin{equation}
\int_{-\infty}^{\infty}\left[\frac{1}{\pi}\frac{\eta}{\left[M\left(k_{F}^{2}-k_{z}^{2}\right)\right]^{2}+\eta^{2}}\right]dk_{z}=\frac{1}{M}\sqrt{\frac{k_{F}^{2}+\sqrt{k_{F}^{4}+\left(\frac{\eta}{M}\right)^{2}}}{2\left(k_{F}^{4}+\left(\frac{\eta}{M}\right)^{2}\right)}}.\label{eq:integral}
\end{equation}
For $k_{F}^{2}\gg\frac{\eta}{M}$, it gives $\frac{1}{M\left|k_{F}\right|}$ (as the result from the delta function), but for $k_{F}^{2}\ll\frac{\eta}{M}$, it gives $\sqrt{\frac{1}{2\eta M}}$. The result of Eq.~(\ref{eq:integral}) will not diverge because the integrand is not a strict Dirac delta function but a delta-like function with a finite Lorentz type broadening. Therefore, for the case of very small $k_{F}$ ($k_{F}$ decrease with decreasing $n_{0}$ or increasing $B$), the results obtained using the delta-function approximation need to be corrected.

\section{Hall conductivity in the quantum limit\label{sec:Hall-conductivity-in}}

The Hall conductivity can be found from
\begin{align}
\sigma_{xy} & =\frac{\hbar e^{2}}{V}\sum_{k_{x},k_{z},b,b'\ne b}\frac{n_{F}\left(E_{k_{z},b}\right)-n_{F}\left(E_{k_{z},b'}\right)}{\left(E_{k_{z},b}-E_{k_{z},b'}\right)^{2}}\Im\left[\left\langle k_{x},k_{z},b\right|v_{x}\left|k_{x},k_{z},b'\right\rangle \left\langle k_{x},k_{z},b'\right|v_{y}\left|k_{x},k_{z},b\right\rangle \right],
\end{align}
where $v_{y}=\frac{A}{\hbar}\sigma_{y}-\frac{2M}{\hbar}\left[-i\frac{1}{\sqrt{2}l_{B}}\left(a^{\dagger}-a\right)\right]\sigma_{z}$. For the two-node model in the quantum limit, the Hall conductivity is 
\begin{align}
\sigma_{xy}= & \frac{e^{2}}{h}\frac{1}{\pi l_{B}^{2}}\int dk_{z}\left\{ \frac{n_{F}\left(E_{k_{z},1+}\right)-n_{F}\left(E_{k_{z},0}\right)}{\left(E_{k_{z},1+}-E_{k_{z},0}\right)^{2}}\left(A\cos\frac{\theta_{k_{z},1}}{2}+\frac{\sqrt{2}M}{l_{B}}\sin\frac{\theta_{k_{z},1}}{2}\right)^{2}\right.\nonumber \\
 & +\frac{n_{F}\left(E_{k_{z},1-}\right)-n_{F}\left(E_{k_{z},0}\right)}{\left(E_{k_{z},1-}-E_{k_{z},0}\right)^{2}}\left(-A\sin\frac{\theta_{k_{z},1}}{2}+\frac{\sqrt{2}M}{l_{B}}\cos\frac{\theta_{k_{z},1}}{2}\right)^{2}+\sum_{n\ge1}\left[\frac{n_{F}\left(E_{k_{z},\left(n+1\right)+}\right)-n_{F}\left(E_{k_{z},n-}\right)}{\left(E_{k_{z},\left(n+1\right)+}-E_{k_{z},n-}\right)^{2}}\right.\nonumber \\
 & \times\left(A\cos\frac{\theta_{k_{z},n+1}}{2}\cos\frac{\theta_{k_{z},n}}{2}+\frac{\sqrt{2n}M}{l_{B}}\cos\frac{\theta_{k_{z},n+1}}{2}\sin\frac{\theta_{k_{z},n}}{2}+\frac{\sqrt{2\left(n+1\right)}M}{l_{B}}\sin\frac{\theta_{k_{z},n+1}}{2}\cos\frac{\theta_{k_{z},n}}{2}\right)^{2}\nonumber \\
 & -\frac{n_{F}\left(E_{k_{z},n+}\right)-n_{F}\left(E_{k_{z},\left(n+1\right)-}\right)}{\left(E_{k_{z},n+}-E_{k_{z},\left(n+1\right)-}\right)^{2}}\left(-A\sin\frac{\theta_{k_{z},n}}{2}\sin\frac{\theta_{k_{z},n+1}}{2}+\frac{\sqrt{2n}M}{l_{B}}\cos\frac{\theta_{k_{z},n}}{2}\sin\frac{\theta_{k_{z},n+1}}{2}\right.\nonumber \\
 & \left.\left.\left.+\frac{\sqrt{2\left(n+1\right)}M}{l_{B}}\sin\frac{\theta_{k_{z},n}}{2}\cos\frac{\theta_{k_{z},n+1}}{2}\right)^{2}\right]\right\} .\label{eq:Hall}
\end{align}
The above expression includes bands with higher Landau indexes. Numerical results show that $\sigma_{xy}$ approximately equals to the value of the first term (contributed by band $1+$ and $0$), and a $B^{-1}$ dependence is found, see Fig.~\ref{fig:Hall-conductivity-of}.

\begin{figure}
\includegraphics{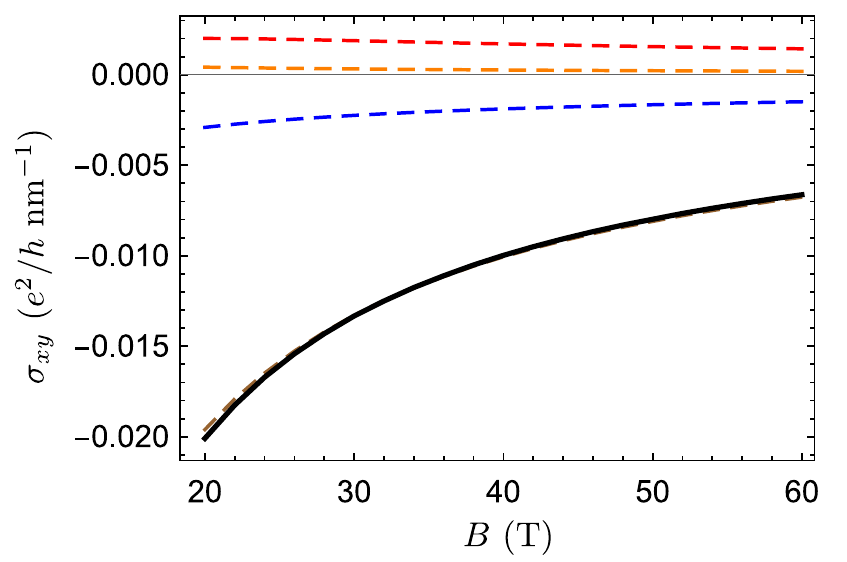}

\caption{\label{fig:Hall-conductivity-of}Hall conductivity of the two-node model. The solid black line is the result of Eq.~(\ref{eq:Hall}). Different terms of Eq.~(\ref{eq:Hall}) are plotted with colored dashed lines (the brown dashed line is for the first term). The carrier concentration $n_{0}$ is taken as $10^{-4}$~$\mathrm{nm^{-3}}$.}
\end{figure}

For the one-node model, one only need to replace $A$ and $M$ in above expression by $\hbar v_{F}$ and $0$. In addition, the expressions of Landau bands of the one-node model are simple. Therefore, one can analytically find that the Hall conductivity is approximately $-\frac{en_{0}}{B}$ when the Fermi energy is near the neutrality point. 

For the 3D electron gas, only bands of $E_{k_{z},0}$ and $E_{k_{z},1}$ involve the calculation. The Hall conductivity can be found as 
\begin{align}
\sigma_{xy} & =\frac{\hbar e^{2}}{V}\sum_{k_{x},k_{z}}\frac{2\left[n_{F}\left(E_{k_{z},1}\right)-n_{F}\left(E_{k_{z},0}\right)\right]}{\left(E_{k_{z},1}-E_{k_{z},0}\right)^{2}}\left(\frac{1}{\sqrt{2}l_{B}}\frac{\hbar}{m}\right)^{2}\nonumber \\
 & =\frac{e^{2}}{\hbar}l_{B}^{2}\frac{1}{V}\sum_{k_{x},k_{z}}\left[-n_{F}\left(E_{k_{z},0}\right)\right]\nonumber \\
 & =-\frac{en_{0}}{B}.
\end{align}

\end{widetext}

\bibliography{QLMR,QLMR-addition}

\begin{thebibliography}{72}%
\makeatletter
\providecommand \@ifxundefined [1]{%
 \@ifx{#1\undefined}
}%
\providecommand \@ifnum [1]{%
 \ifnum #1\expandafter \@firstoftwo
 \else \expandafter \@secondoftwo
 \fi
}%
\providecommand \@ifx [1]{%
 \ifx #1\expandafter \@firstoftwo
 \else \expandafter \@secondoftwo
 \fi
}%
\providecommand \natexlab [1]{#1}%
\providecommand \enquote  [1]{``#1''}%
\providecommand \bibnamefont  [1]{#1}%
\providecommand \bibfnamefont [1]{#1}%
\providecommand \citenamefont [1]{#1}%
\providecommand \href@noop [0]{\@secondoftwo}%
\providecommand \href [0]{\begingroup \@sanitize@url \@href}%
\providecommand \@href[1]{\@@startlink{#1}\@@href}%
\providecommand \@@href[1]{\endgroup#1\@@endlink}%
\providecommand \@sanitize@url [0]{\catcode `\\12\catcode `\$12\catcode
  `\&12\catcode `\#12\catcode `\^12\catcode `\_12\catcode `\%12\relax}%
\providecommand \@@startlink[1]{}%
\providecommand \@@endlink[0]{}%
\providecommand \url  [0]{\begingroup\@sanitize@url \@url }%
\providecommand \@url [1]{\endgroup\@href {#1}{\urlprefix }}%
\providecommand \urlprefix  [0]{URL }%
\providecommand \Eprint [0]{\href }%
\providecommand \doibase [0]{https://doi.org/}%
\providecommand \selectlanguage [0]{\@gobble}%
\providecommand \bibinfo  [0]{\@secondoftwo}%
\providecommand \bibfield  [0]{\@secondoftwo}%
\providecommand \translation [1]{[#1]}%
\providecommand \BibitemOpen [0]{}%
\providecommand \bibitemStop [0]{}%
\providecommand \bibitemNoStop [0]{.\EOS\space}%
\providecommand \EOS [0]{\spacefactor3000\relax}%
\providecommand \BibitemShut  [1]{\csname bibitem#1\endcsname}%
\let\auto@bib@innerbib\@empty
%</preamble>
\bibitem [{\citenamefont {Abrikosov}(1998)}]{AbrikosovA_PRB_1998}%
  \BibitemOpen
  \bibfield  {author} {\bibinfo {author} {\bibfnamefont {A.~A.}\ \bibnamefont
  {Abrikosov}},\ }\bibfield  {title} {\bibinfo {title} {Quantum
  magnetoresistance},\ }\href {https://doi.org/10.1103/PhysRevB.58.2788}
  {\bibfield  {journal} {\bibinfo  {journal} {Phys. Rev. B}\ }\textbf {\bibinfo
  {volume} {58}},\ \bibinfo {pages} {2788} (\bibinfo {year}
  {1998})}\BibitemShut {NoStop}%
\bibitem [{\citenamefont {Xu}\ \emph {et~al.}(1997)\citenamefont {Xu},
  \citenamefont {Husmann}, \citenamefont {Rosenbaum}, \citenamefont {Saboungi},
  \citenamefont {Enderby},\ and\ \citenamefont {Littlewood}}]{XuR_N_1997}%
  \BibitemOpen
  \bibfield  {author} {\bibinfo {author} {\bibfnamefont {R.}~\bibnamefont
  {Xu}}, \bibinfo {author} {\bibfnamefont {A.}~\bibnamefont {Husmann}},
  \bibinfo {author} {\bibfnamefont {T.~F.}\ \bibnamefont {Rosenbaum}}, \bibinfo
  {author} {\bibfnamefont {M.-L.}\ \bibnamefont {Saboungi}}, \bibinfo {author}
  {\bibfnamefont {J.~E.}\ \bibnamefont {Enderby}},\ and\ \bibinfo {author}
  {\bibfnamefont {P.~B.}\ \bibnamefont {Littlewood}},\ }\bibfield  {title}
  {\bibinfo {title} {Large magnetoresistance in non-magnetic silver
  chalcogenides},\ }\href {https://doi.org/10.1038/36306} {\bibfield  {journal}
  {\bibinfo  {journal} {Nature}\ }\textbf {\bibinfo {volume} {390}},\ \bibinfo
  {pages} {57} (\bibinfo {year} {1997})}\BibitemShut {NoStop}%
\bibitem [{\citenamefont {He}\ \emph {et~al.}(2014)\citenamefont {He},
  \citenamefont {Hong}, \citenamefont {Dong}, \citenamefont {Pan},
  \citenamefont {Zhang}, \citenamefont {Zhang},\ and\ \citenamefont
  {Li}}]{HeL_PRL_2014}%
  \BibitemOpen
  \bibfield  {author} {\bibinfo {author} {\bibfnamefont {L.~P.}\ \bibnamefont
  {He}}, \bibinfo {author} {\bibfnamefont {X.~C.}\ \bibnamefont {Hong}},
  \bibinfo {author} {\bibfnamefont {J.~K.}\ \bibnamefont {Dong}}, \bibinfo
  {author} {\bibfnamefont {J.}~\bibnamefont {Pan}}, \bibinfo {author}
  {\bibfnamefont {Z.}~\bibnamefont {Zhang}}, \bibinfo {author} {\bibfnamefont
  {J.}~\bibnamefont {Zhang}},\ and\ \bibinfo {author} {\bibfnamefont {S.~Y.}\
  \bibnamefont {Li}},\ }\bibfield  {title} {\bibinfo {title} {Quantum transport
  evidence for the three-dimensional {Dirac} semimetal phase in
  {${\mathrm{Cd}}_{3}{\mathrm{As}}_{2}$}},\ }\href
  {https://doi.org/10.1103/PhysRevLett.113.246402} {\bibfield  {journal}
  {\bibinfo  {journal} {Phys. Rev. Lett.}\ }\textbf {\bibinfo {volume} {113}},\
  \bibinfo {pages} {246402} (\bibinfo {year} {2014})}\BibitemShut {NoStop}%
\bibitem [{\citenamefont {Narayanan}\ \emph {et~al.}(2015)\citenamefont
  {Narayanan}, \citenamefont {Watson}, \citenamefont {Blake}, \citenamefont
  {Bruyant}, \citenamefont {Drigo}, \citenamefont {Chen}, \citenamefont
  {Prabhakaran}, \citenamefont {Yan}, \citenamefont {Felser}, \citenamefont
  {Kong}, \citenamefont {Canfield},\ and\ \citenamefont
  {Coldea}}]{NarayananA_PRL_2015}%
  \BibitemOpen
  \bibfield  {author} {\bibinfo {author} {\bibfnamefont {A.}~\bibnamefont
  {Narayanan}}, \bibinfo {author} {\bibfnamefont {M.~D.}\ \bibnamefont
  {Watson}}, \bibinfo {author} {\bibfnamefont {S.~F.}\ \bibnamefont {Blake}},
  \bibinfo {author} {\bibfnamefont {N.}~\bibnamefont {Bruyant}}, \bibinfo
  {author} {\bibfnamefont {L.}~\bibnamefont {Drigo}}, \bibinfo {author}
  {\bibfnamefont {Y.~L.}\ \bibnamefont {Chen}}, \bibinfo {author}
  {\bibfnamefont {D.}~\bibnamefont {Prabhakaran}}, \bibinfo {author}
  {\bibfnamefont {B.}~\bibnamefont {Yan}}, \bibinfo {author} {\bibfnamefont
  {C.}~\bibnamefont {Felser}}, \bibinfo {author} {\bibfnamefont
  {T.}~\bibnamefont {Kong}}, \bibinfo {author} {\bibfnamefont {P.~C.}\
  \bibnamefont {Canfield}},\ and\ \bibinfo {author} {\bibfnamefont {A.~I.}\
  \bibnamefont {Coldea}},\ }\bibfield  {title} {\bibinfo {title} {Linear
  magnetoresistance caused by mobility fluctuations in $n$-doped
  {${\mathrm{Cd}}_{3}{\mathrm{As}}_{2}$}},\ }\href
  {https://doi.org/10.1103/PhysRevLett.114.117201} {\bibfield  {journal}
  {\bibinfo  {journal} {Phys. Rev. Lett.}\ }\textbf {\bibinfo {volume} {114}},\
  \bibinfo {pages} {117201} (\bibinfo {year} {2015})}\BibitemShut {NoStop}%
\bibitem [{\citenamefont {Xiang}\ \emph {et~al.}(2015)\citenamefont {Xiang},
  \citenamefont {Zhao}, \citenamefont {Jin}, \citenamefont {Shang},
  \citenamefont {Ma}, \citenamefont {Ye}, \citenamefont {Lei}, \citenamefont
  {Wu}, \citenamefont {Xia},\ and\ \citenamefont {Chen}}]{XiangZ_PRL_2015}%
  \BibitemOpen
  \bibfield  {author} {\bibinfo {author} {\bibfnamefont {Z.~J.}\ \bibnamefont
  {Xiang}}, \bibinfo {author} {\bibfnamefont {D.}~\bibnamefont {Zhao}},
  \bibinfo {author} {\bibfnamefont {Z.}~\bibnamefont {Jin}}, \bibinfo {author}
  {\bibfnamefont {C.}~\bibnamefont {Shang}}, \bibinfo {author} {\bibfnamefont
  {L.~K.}\ \bibnamefont {Ma}}, \bibinfo {author} {\bibfnamefont {G.~J.}\
  \bibnamefont {Ye}}, \bibinfo {author} {\bibfnamefont {B.}~\bibnamefont
  {Lei}}, \bibinfo {author} {\bibfnamefont {T.}~\bibnamefont {Wu}}, \bibinfo
  {author} {\bibfnamefont {Z.~C.}\ \bibnamefont {Xia}},\ and\ \bibinfo {author}
  {\bibfnamefont {X.~H.}\ \bibnamefont {Chen}},\ }\bibfield  {title} {\bibinfo
  {title} {Angular-dependent phase factor of {Shubnikov--de} {Haas}
  oscillations in the {Dirac} semimetal
  {${\mathrm{Cd}}_{3}{\mathrm{As}}_{2}$}},\ }\href
  {https://doi.org/10.1103/PhysRevLett.115.226401} {\bibfield  {journal}
  {\bibinfo  {journal} {Phys. Rev. Lett.}\ }\textbf {\bibinfo {volume} {115}},\
  \bibinfo {pages} {226401} (\bibinfo {year} {2015})}\BibitemShut {NoStop}%
\bibitem [{\citenamefont {Feng}\ \emph {et~al.}(2015)\citenamefont {Feng},
  \citenamefont {Pang}, \citenamefont {Wu}, \citenamefont {Wang}, \citenamefont
  {Weng}, \citenamefont {Li}, \citenamefont {Dai}, \citenamefont {Fang},
  \citenamefont {Shi},\ and\ \citenamefont {Lu}}]{FengJ_PRB_2015}%
  \BibitemOpen
  \bibfield  {author} {\bibinfo {author} {\bibfnamefont {J.}~\bibnamefont
  {Feng}}, \bibinfo {author} {\bibfnamefont {Y.}~\bibnamefont {Pang}}, \bibinfo
  {author} {\bibfnamefont {D.}~\bibnamefont {Wu}}, \bibinfo {author}
  {\bibfnamefont {Z.}~\bibnamefont {Wang}}, \bibinfo {author} {\bibfnamefont
  {H.}~\bibnamefont {Weng}}, \bibinfo {author} {\bibfnamefont {J.}~\bibnamefont
  {Li}}, \bibinfo {author} {\bibfnamefont {X.}~\bibnamefont {Dai}}, \bibinfo
  {author} {\bibfnamefont {Z.}~\bibnamefont {Fang}}, \bibinfo {author}
  {\bibfnamefont {Y.}~\bibnamefont {Shi}},\ and\ \bibinfo {author}
  {\bibfnamefont {L.}~\bibnamefont {Lu}},\ }\bibfield  {title} {\bibinfo
  {title} {Large linear magnetoresistance in {Dirac} semimetal
  {${\mathrm{Cd}}_{3}{\mathrm{As}}_{2}$} with {F}ermi surfaces close to the
  {Dirac} points},\ }\href {https://doi.org/10.1103/PhysRevB.92.081306}
  {\bibfield  {journal} {\bibinfo  {journal} {Phys. Rev. B}\ }\textbf {\bibinfo
  {volume} {92}},\ \bibinfo {pages} {081306(R)} (\bibinfo {year}
  {2015})}\BibitemShut {NoStop}%
\bibitem [{\citenamefont {Zhao}\ \emph {et~al.}(2015)\citenamefont {Zhao},
  \citenamefont {Liu}, \citenamefont {Zhang}, \citenamefont {Wang},
  \citenamefont {Wang}, \citenamefont {Lin}, \citenamefont {Xing},
  \citenamefont {Lu}, \citenamefont {Liu}, \citenamefont {Wang}, \citenamefont
  {Brombosz}, \citenamefont {Xiao}, \citenamefont {Jia}, \citenamefont {Xie},\
  and\ \citenamefont {Wang}}]{ZhaoY_PRX_2015}%
  \BibitemOpen
  \bibfield  {author} {\bibinfo {author} {\bibfnamefont {Y.}~\bibnamefont
  {Zhao}}, \bibinfo {author} {\bibfnamefont {H.}~\bibnamefont {Liu}}, \bibinfo
  {author} {\bibfnamefont {C.}~\bibnamefont {Zhang}}, \bibinfo {author}
  {\bibfnamefont {H.}~\bibnamefont {Wang}}, \bibinfo {author} {\bibfnamefont
  {J.}~\bibnamefont {Wang}}, \bibinfo {author} {\bibfnamefont {Z.}~\bibnamefont
  {Lin}}, \bibinfo {author} {\bibfnamefont {Y.}~\bibnamefont {Xing}}, \bibinfo
  {author} {\bibfnamefont {H.}~\bibnamefont {Lu}}, \bibinfo {author}
  {\bibfnamefont {J.}~\bibnamefont {Liu}}, \bibinfo {author} {\bibfnamefont
  {Y.}~\bibnamefont {Wang}}, \bibinfo {author} {\bibfnamefont {S.~M.}\
  \bibnamefont {Brombosz}}, \bibinfo {author} {\bibfnamefont {Z.}~\bibnamefont
  {Xiao}}, \bibinfo {author} {\bibfnamefont {S.}~\bibnamefont {Jia}}, \bibinfo
  {author} {\bibfnamefont {X.~C.}\ \bibnamefont {Xie}},\ and\ \bibinfo {author}
  {\bibfnamefont {J.}~\bibnamefont {Wang}},\ }\bibfield  {title} {\bibinfo
  {title} {Anisotropic {F}ermi surface and quantum limit transport in high
  mobility three-dimensional {Dirac} semimetal
  {${\mathrm{Cd}}_{3}{\mathrm{As}}_{2}$}},\ }\href
  {https://doi.org/10.1103/PhysRevX.5.031037} {\bibfield  {journal} {\bibinfo
  {journal} {Phys. Rev. X}\ }\textbf {\bibinfo {volume} {5}},\ \bibinfo {pages}
  {031037} (\bibinfo {year} {2015})}\BibitemShut {NoStop}%
\bibitem [{\citenamefont {Liang}\ \emph {et~al.}(2015)\citenamefont {Liang},
  \citenamefont {Gibson}, \citenamefont {Ali}, \citenamefont {Liu},
  \citenamefont {Cava},\ and\ \citenamefont {Ong}}]{LiangT_NM_2015}%
  \BibitemOpen
  \bibfield  {author} {\bibinfo {author} {\bibfnamefont {T.}~\bibnamefont
  {Liang}}, \bibinfo {author} {\bibfnamefont {Q.}~\bibnamefont {Gibson}},
  \bibinfo {author} {\bibfnamefont {M.~N.}\ \bibnamefont {Ali}}, \bibinfo
  {author} {\bibfnamefont {M.}~\bibnamefont {Liu}}, \bibinfo {author}
  {\bibfnamefont {R.~J.}\ \bibnamefont {Cava}},\ and\ \bibinfo {author}
  {\bibfnamefont {N.~P.}\ \bibnamefont {Ong}},\ }\bibfield  {title} {\bibinfo
  {title} {Ultrahigh mobility and giant magnetoresistance in the {Dirac}
  semimetal{~}{${\mathrm{Cd}}_{3}{\mathrm{As}}_{2}$}},\ }\href
  {https://doi.org/10.1038/nmat4143} {\bibfield  {journal} {\bibinfo  {journal}
  {Nat. Mater.}\ }\textbf {\bibinfo {volume} {14}},\ \bibinfo {pages} {280}
  (\bibinfo {year} {2015})}\BibitemShut {NoStop}%
\bibitem [{\citenamefont {Li}\ \emph {et~al.}(2016)\citenamefont {Li},
  \citenamefont {He}, \citenamefont {Lu}, \citenamefont {Zhang}, \citenamefont
  {Liu}, \citenamefont {Ma}, \citenamefont {Fan}, \citenamefont {Shen},\ and\
  \citenamefont {Wang}}]{LiH_NC_2016}%
  \BibitemOpen
  \bibfield  {author} {\bibinfo {author} {\bibfnamefont {H.}~\bibnamefont
  {Li}}, \bibinfo {author} {\bibfnamefont {H.}~\bibnamefont {He}}, \bibinfo
  {author} {\bibfnamefont {H.-Z.}\ \bibnamefont {Lu}}, \bibinfo {author}
  {\bibfnamefont {H.}~\bibnamefont {Zhang}}, \bibinfo {author} {\bibfnamefont
  {H.}~\bibnamefont {Liu}}, \bibinfo {author} {\bibfnamefont {R.}~\bibnamefont
  {Ma}}, \bibinfo {author} {\bibfnamefont {Z.}~\bibnamefont {Fan}}, \bibinfo
  {author} {\bibfnamefont {S.-Q.}\ \bibnamefont {Shen}},\ and\ \bibinfo
  {author} {\bibfnamefont {J.}~\bibnamefont {Wang}},\ }\bibfield  {title}
  {\bibinfo {title} {Negative magnetoresistance in {Dirac} semimetal
  {${\mathrm{Cd}}_{3}{\mathrm{As}}_{2}$}},\ }\href
  {https://doi.org/10.1038/ncomms10301} {\bibfield  {journal} {\bibinfo
  {journal} {Nat. Commun.}\ }\textbf {\bibinfo {volume} {7}},\ \bibinfo {pages}
  {10301} (\bibinfo {year} {2016})}\BibitemShut {NoStop}%
\bibitem [{\citenamefont {Takiguchi}\ \emph {et~al.}(2020)\citenamefont
  {Takiguchi}, \citenamefont {Wakabayashi}, \citenamefont {Irie}, \citenamefont
  {Krockenberger}, \citenamefont {Otsuka}, \citenamefont {Sawada},
  \citenamefont {Nikolaev}, \citenamefont {Das}, \citenamefont {Tanaka},
  \citenamefont {Taniyasu},\ and\ \citenamefont
  {Yamamoto}}]{TakiguchiK_NC_2020}%
  \BibitemOpen
  \bibfield  {author} {\bibinfo {author} {\bibfnamefont {K.}~\bibnamefont
  {Takiguchi}}, \bibinfo {author} {\bibfnamefont {Y.~K.}\ \bibnamefont
  {Wakabayashi}}, \bibinfo {author} {\bibfnamefont {H.}~\bibnamefont {Irie}},
  \bibinfo {author} {\bibfnamefont {Y.}~\bibnamefont {Krockenberger}}, \bibinfo
  {author} {\bibfnamefont {T.}~\bibnamefont {Otsuka}}, \bibinfo {author}
  {\bibfnamefont {H.}~\bibnamefont {Sawada}}, \bibinfo {author} {\bibfnamefont
  {S.~A.}\ \bibnamefont {Nikolaev}}, \bibinfo {author} {\bibfnamefont
  {H.}~\bibnamefont {Das}}, \bibinfo {author} {\bibfnamefont {M.}~\bibnamefont
  {Tanaka}}, \bibinfo {author} {\bibfnamefont {Y.}~\bibnamefont {Taniyasu}},\
  and\ \bibinfo {author} {\bibfnamefont {H.}~\bibnamefont {Yamamoto}},\
  }\bibfield  {title} {\bibinfo {title} {Quantum transport evidence of {Weyl}
  fermions in an epitaxial ferromagnetic oxide},\ }\href
  {https://doi.org/10.1038/s41467-020-18646-8} {\bibfield  {journal} {\bibinfo
  {journal} {Nat. Commun.}\ }\textbf {\bibinfo {volume} {11}},\ \bibinfo
  {pages} {4969} (\bibinfo {year} {2020})}\BibitemShut {NoStop}%
\bibitem [{\citenamefont {Zhu}\ \emph {et~al.}(2022)\citenamefont {Zhu},
  \citenamefont {Cao}, \citenamefont {Guo}, \citenamefont {Li}, \citenamefont
  {Chen}, \citenamefont {Zhu}, \citenamefont {He}, \citenamefont {Huang},
  \citenamefont {Dong}, \citenamefont {Wang}, \citenamefont {Zhai},
  \citenamefont {Ou}, \citenamefont {Zhu}, \citenamefont {Lu}, \citenamefont
  {Li}, \citenamefont {Chen},\ and\ \citenamefont {Pan}}]{ZhuW_PRB_2022}%
  \BibitemOpen
  \bibfield  {author} {\bibinfo {author} {\bibfnamefont {W.~L.}\ \bibnamefont
  {Zhu}}, \bibinfo {author} {\bibfnamefont {Y.}~\bibnamefont {Cao}}, \bibinfo
  {author} {\bibfnamefont {P.~J.}\ \bibnamefont {Guo}}, \bibinfo {author}
  {\bibfnamefont {X.}~\bibnamefont {Li}}, \bibinfo {author} {\bibfnamefont
  {Y.~J.}\ \bibnamefont {Chen}}, \bibinfo {author} {\bibfnamefont {L.~J.}\
  \bibnamefont {Zhu}}, \bibinfo {author} {\bibfnamefont {J.~B.}\ \bibnamefont
  {He}}, \bibinfo {author} {\bibfnamefont {Y.~F.}\ \bibnamefont {Huang}},
  \bibinfo {author} {\bibfnamefont {Q.~X.}\ \bibnamefont {Dong}}, \bibinfo
  {author} {\bibfnamefont {Y.~Y.}\ \bibnamefont {Wang}}, \bibinfo {author}
  {\bibfnamefont {R.~Q.}\ \bibnamefont {Zhai}}, \bibinfo {author}
  {\bibfnamefont {Y.~B.}\ \bibnamefont {Ou}}, \bibinfo {author} {\bibfnamefont
  {G.~Q.}\ \bibnamefont {Zhu}}, \bibinfo {author} {\bibfnamefont {H.~Y.}\
  \bibnamefont {Lu}}, \bibinfo {author} {\bibfnamefont {G.}~\bibnamefont {Li}},
  \bibinfo {author} {\bibfnamefont {G.~F.}\ \bibnamefont {Chen}},\ and\
  \bibinfo {author} {\bibfnamefont {M.~H.}\ \bibnamefont {Pan}},\ }\bibfield
  {title} {\bibinfo {title} {Linear magnetoresistance induced by mobility
  fluctuations in iodine-intercalated tungsten ditelluride},\ }\href
  {https://doi.org/10.1103/PhysRevB.105.125116} {\bibfield  {journal} {\bibinfo
   {journal} {Phys. Rev. B}\ }\textbf {\bibinfo {volume} {105}},\ \bibinfo
  {pages} {125116} (\bibinfo {year} {2022})}\BibitemShut {NoStop}%
\bibitem [{\citenamefont {Laha}\ \emph {et~al.}(2020)\citenamefont {Laha},
  \citenamefont {Mardanya}, \citenamefont {Singh}, \citenamefont {Lin},
  \citenamefont {Bansil}, \citenamefont {Agarwal},\ and\ \citenamefont
  {Hossain}}]{LahaA_PRB_2020}%
  \BibitemOpen
  \bibfield  {author} {\bibinfo {author} {\bibfnamefont {A.}~\bibnamefont
  {Laha}}, \bibinfo {author} {\bibfnamefont {S.}~\bibnamefont {Mardanya}},
  \bibinfo {author} {\bibfnamefont {B.}~\bibnamefont {Singh}}, \bibinfo
  {author} {\bibfnamefont {H.}~\bibnamefont {Lin}}, \bibinfo {author}
  {\bibfnamefont {A.}~\bibnamefont {Bansil}}, \bibinfo {author} {\bibfnamefont
  {A.}~\bibnamefont {Agarwal}},\ and\ \bibinfo {author} {\bibfnamefont
  {Z.}~\bibnamefont {Hossain}},\ }\bibfield  {title} {\bibinfo {title}
  {Magnetotransport properties of the topological nodal-line semimetal
  {CaCdSn}},\ }\href {https://doi.org/10.1103/PhysRevB.102.035164} {\bibfield
  {journal} {\bibinfo  {journal} {Phys. Rev. B}\ }\textbf {\bibinfo {volume}
  {102}},\ \bibinfo {pages} {035164} (\bibinfo {year} {2020})}\BibitemShut
  {NoStop}%
\bibitem [{\citenamefont {Yang}\ \emph {et~al.}(2021)\citenamefont {Yang},
  \citenamefont {Song}, \citenamefont {Guo}, \citenamefont {Gao}, \citenamefont
  {Dong}, \citenamefont {Yu}, \citenamefont {Zheng}, \citenamefont {Kang},\
  and\ \citenamefont {Zhang}}]{YangJ_NL_2021}%
  \BibitemOpen
  \bibfield  {author} {\bibinfo {author} {\bibfnamefont {J.}~\bibnamefont
  {Yang}}, \bibinfo {author} {\bibfnamefont {Z.-Y.}\ \bibnamefont {Song}},
  \bibinfo {author} {\bibfnamefont {L.}~\bibnamefont {Guo}}, \bibinfo {author}
  {\bibfnamefont {H.}~\bibnamefont {Gao}}, \bibinfo {author} {\bibfnamefont
  {Z.}~\bibnamefont {Dong}}, \bibinfo {author} {\bibfnamefont {Q.}~\bibnamefont
  {Yu}}, \bibinfo {author} {\bibfnamefont {R.-K.}\ \bibnamefont {Zheng}},
  \bibinfo {author} {\bibfnamefont {T.-T.}\ \bibnamefont {Kang}},\ and\
  \bibinfo {author} {\bibfnamefont {K.}~\bibnamefont {Zhang}},\ }\bibfield
  {title} {\bibinfo {title} {Nontrivial giant linear magnetoresistance in
  nodal-line semimetal {ZrGeSe} {2D} layers},\ }\href
  {https://doi.org/10.1021/acs.nanolett.1c01647} {\bibfield  {journal}
  {\bibinfo  {journal} {Nano Lett.}\ }\textbf {\bibinfo {volume} {21}},\
  \bibinfo {pages} {10139} (\bibinfo {year} {2021})}\BibitemShut {NoStop}%
\bibitem [{\citenamefont {Wu}\ \emph {et~al.}(2019)\citenamefont {Wu},
  \citenamefont {Li}, \citenamefont {Pan}, \citenamefont {Jiang}, \citenamefont
  {Jin}, \citenamefont {Song}, \citenamefont {Wang},\ and\ \citenamefont
  {Wan}}]{WuJ_C_2019}%
  \BibitemOpen
  \bibfield  {author} {\bibinfo {author} {\bibfnamefont {J.}~\bibnamefont
  {Wu}}, \bibinfo {author} {\bibfnamefont {Y.}~\bibnamefont {Li}}, \bibinfo
  {author} {\bibfnamefont {D.}~\bibnamefont {Pan}}, \bibinfo {author}
  {\bibfnamefont {C.}~\bibnamefont {Jiang}}, \bibinfo {author} {\bibfnamefont
  {C.}~\bibnamefont {Jin}}, \bibinfo {author} {\bibfnamefont {F.}~\bibnamefont
  {Song}}, \bibinfo {author} {\bibfnamefont {G.}~\bibnamefont {Wang}},\ and\
  \bibinfo {author} {\bibfnamefont {J.}~\bibnamefont {Wan}},\ }\bibfield
  {title} {\bibinfo {title} {Effect of grain boundaries on charge transport in
  {CVD}-grown bilayer graphene},\ }\href
  {https://doi.org/https://doi.org/10.1016/j.carbon.2019.03.029} {\bibfield
  {journal} {\bibinfo  {journal} {Carbon}\ }\textbf {\bibinfo {volume} {147}},\
  \bibinfo {pages} {434} (\bibinfo {year} {2019})}\BibitemShut {NoStop}%
\bibitem [{\citenamefont {Niu}\ \emph {et~al.}(2017)\citenamefont {Niu},
  \citenamefont {Yu}, \citenamefont {Yip}, \citenamefont {Lim}, \citenamefont
  {Kotegawa}, \citenamefont {Matsuoka}, \citenamefont {Sugawara}, \citenamefont
  {Tou}, \citenamefont {Yanase},\ and\ \citenamefont {Goh}}]{NiuQ_NC_2017}%
  \BibitemOpen
  \bibfield  {author} {\bibinfo {author} {\bibfnamefont {Q.}~\bibnamefont
  {Niu}}, \bibinfo {author} {\bibfnamefont {W.~C.}\ \bibnamefont {Yu}},
  \bibinfo {author} {\bibfnamefont {K.~Y.}\ \bibnamefont {Yip}}, \bibinfo
  {author} {\bibfnamefont {Z.~L.}\ \bibnamefont {Lim}}, \bibinfo {author}
  {\bibfnamefont {H.}~\bibnamefont {Kotegawa}}, \bibinfo {author}
  {\bibfnamefont {E.}~\bibnamefont {Matsuoka}}, \bibinfo {author}
  {\bibfnamefont {H.}~\bibnamefont {Sugawara}}, \bibinfo {author}
  {\bibfnamefont {H.}~\bibnamefont {Tou}}, \bibinfo {author} {\bibfnamefont
  {Y.}~\bibnamefont {Yanase}},\ and\ \bibinfo {author} {\bibfnamefont {S.~K.}\
  \bibnamefont {Goh}},\ }\bibfield  {title} {\bibinfo {title} {Quasilinear
  quantum magnetoresistance in pressure-induced nonsymmorphic superconductor
  chromium arsenide},\ }\href {https://doi.org/10.1038/ncomms15358} {\bibfield
  {journal} {\bibinfo  {journal} {Nat. Commun.}\ }\textbf {\bibinfo {volume}
  {8}},\ \bibinfo {pages} {15358} (\bibinfo {year} {2017})}\BibitemShut
  {NoStop}%
\bibitem [{\citenamefont {Giraldo-Gallo}\ \emph {et~al.}(2018)\citenamefont
  {Giraldo-Gallo}, \citenamefont {Galvis}, \citenamefont {Stegen},
  \citenamefont {Modic}, \citenamefont {Balakirev}, \citenamefont {Betts},
  \citenamefont {Lian}, \citenamefont {Moir}, \citenamefont {Riggs},
  \citenamefont {Wu}, \citenamefont {Bollinger}, \citenamefont {He},
  \citenamefont {Bo{\v{z}}ovi{\'{c}}}, \citenamefont {Ramshaw}, \citenamefont
  {McDonald}, \citenamefont {Boebinger},\ and\ \citenamefont
  {Shekhter}}]{GiraldoGalloP_S_2018}%
  \BibitemOpen
  \bibfield  {author} {\bibinfo {author} {\bibfnamefont {P.}~\bibnamefont
  {Giraldo-Gallo}}, \bibinfo {author} {\bibfnamefont {J.~A.}\ \bibnamefont
  {Galvis}}, \bibinfo {author} {\bibfnamefont {Z.}~\bibnamefont {Stegen}},
  \bibinfo {author} {\bibfnamefont {K.~A.}\ \bibnamefont {Modic}}, \bibinfo
  {author} {\bibfnamefont {F.~F.}\ \bibnamefont {Balakirev}}, \bibinfo {author}
  {\bibfnamefont {J.~B.}\ \bibnamefont {Betts}}, \bibinfo {author}
  {\bibfnamefont {X.}~\bibnamefont {Lian}}, \bibinfo {author} {\bibfnamefont
  {C.}~\bibnamefont {Moir}}, \bibinfo {author} {\bibfnamefont {S.~C.}\
  \bibnamefont {Riggs}}, \bibinfo {author} {\bibfnamefont {J.}~\bibnamefont
  {Wu}}, \bibinfo {author} {\bibfnamefont {A.~T.}\ \bibnamefont {Bollinger}},
  \bibinfo {author} {\bibfnamefont {X.}~\bibnamefont {He}}, \bibinfo {author}
  {\bibfnamefont {I.}~\bibnamefont {Bo{\v{z}}ovi{\'{c}}}}, \bibinfo {author}
  {\bibfnamefont {B.~J.}\ \bibnamefont {Ramshaw}}, \bibinfo {author}
  {\bibfnamefont {R.~D.}\ \bibnamefont {McDonald}}, \bibinfo {author}
  {\bibfnamefont {G.~S.}\ \bibnamefont {Boebinger}},\ and\ \bibinfo {author}
  {\bibfnamefont {A.}~\bibnamefont {Shekhter}},\ }\bibfield  {title} {\bibinfo
  {title} {Scale-invariant magnetoresistance in a cuprate superconductor},\
  }\href {https://doi.org/10.1126/science.aan3178} {\bibfield  {journal}
  {\bibinfo  {journal} {Science}\ }\textbf {\bibinfo {volume} {361}},\ \bibinfo
  {pages} {479} (\bibinfo {year} {2018})}\BibitemShut {NoStop}%
\bibitem [{\citenamefont {Sarkar}\ \emph {et~al.}(2019)\citenamefont {Sarkar},
  \citenamefont {Mandal}, \citenamefont {Poniatowski}, \citenamefont {Chan},\
  and\ \citenamefont {Greene}}]{SarkarT_SA_2019}%
  \BibitemOpen
  \bibfield  {author} {\bibinfo {author} {\bibfnamefont {T.}~\bibnamefont
  {Sarkar}}, \bibinfo {author} {\bibfnamefont {P.~R.}\ \bibnamefont {Mandal}},
  \bibinfo {author} {\bibfnamefont {N.~R.}\ \bibnamefont {Poniatowski}},
  \bibinfo {author} {\bibfnamefont {M.~K.}\ \bibnamefont {Chan}},\ and\
  \bibinfo {author} {\bibfnamefont {R.~L.}\ \bibnamefont {Greene}},\ }\bibfield
   {title} {\bibinfo {title} {Correlation between scale-invariant normal-state
  resistivity and superconductivity in an electron-doped cuprate},\ }\href
  {https://doi.org/10.1126/sciadv.aav6753} {\bibfield  {journal} {\bibinfo
  {journal} {Sci. Adv.}\ }\textbf {\bibinfo {volume} {5}},\ \bibinfo {pages}
  {eaav6753} (\bibinfo {year} {2019})}\BibitemShut {NoStop}%
\bibitem [{\citenamefont {Zhang}\ \emph {et~al.}(2020)\citenamefont {Zhang},
  \citenamefont {Hu}, \citenamefont {Kuo}, \citenamefont {Kuo}, \citenamefont
  {Fang}, \citenamefont {Lai}, \citenamefont {Liu}, \citenamefont {Yip},
  \citenamefont {Sun}, \citenamefont {Balakirev}, \citenamefont {Lue},
  \citenamefont {Chen},\ and\ \citenamefont {Goh}}]{ZhangW_PRB_2020}%
  \BibitemOpen
  \bibfield  {author} {\bibinfo {author} {\bibfnamefont {W.}~\bibnamefont
  {Zhang}}, \bibinfo {author} {\bibfnamefont {Y.~J.}\ \bibnamefont {Hu}},
  \bibinfo {author} {\bibfnamefont {C.~N.}\ \bibnamefont {Kuo}}, \bibinfo
  {author} {\bibfnamefont {S.~T.}\ \bibnamefont {Kuo}}, \bibinfo {author}
  {\bibfnamefont {Y.-W.}\ \bibnamefont {Fang}}, \bibinfo {author}
  {\bibfnamefont {K.~T.}\ \bibnamefont {Lai}}, \bibinfo {author} {\bibfnamefont
  {X.~Y.}\ \bibnamefont {Liu}}, \bibinfo {author} {\bibfnamefont {K.~Y.}\
  \bibnamefont {Yip}}, \bibinfo {author} {\bibfnamefont {D.}~\bibnamefont
  {Sun}}, \bibinfo {author} {\bibfnamefont {F.~F.}\ \bibnamefont {Balakirev}},
  \bibinfo {author} {\bibfnamefont {C.~S.}\ \bibnamefont {Lue}}, \bibinfo
  {author} {\bibfnamefont {H.}~\bibnamefont {Chen}},\ and\ \bibinfo {author}
  {\bibfnamefont {S.~K.}\ \bibnamefont {Goh}},\ }\bibfield  {title} {\bibinfo
  {title} {Linear magnetoresistance with a universal energy scale in the
  strong-coupling superconductor {${\mathrm{Mo}}_{8}{\mathrm{Ga}}_{41}$}
  without quantum criticality},\ }\href
  {https://doi.org/10.1103/PhysRevB.102.241113} {\bibfield  {journal} {\bibinfo
   {journal} {Phys. Rev. B}\ }\textbf {\bibinfo {volume} {102}},\ \bibinfo
  {pages} {241113(R)} (\bibinfo {year} {2020})}\BibitemShut {NoStop}%
\bibitem [{\citenamefont {Maksimovic}\ \emph {et~al.}(2020)\citenamefont
  {Maksimovic}, \citenamefont {Hayes}, \citenamefont {Nagarajan}, \citenamefont
  {Analytis}, \citenamefont {Koshelev}, \citenamefont {Singleton},
  \citenamefont {Lee},\ and\ \citenamefont {Schenkel}}]{MaksimovicN_PRX_2020}%
  \BibitemOpen
  \bibfield  {author} {\bibinfo {author} {\bibfnamefont {N.}~\bibnamefont
  {Maksimovic}}, \bibinfo {author} {\bibfnamefont {I.~M.}\ \bibnamefont
  {Hayes}}, \bibinfo {author} {\bibfnamefont {V.}~\bibnamefont {Nagarajan}},
  \bibinfo {author} {\bibfnamefont {J.~G.}\ \bibnamefont {Analytis}}, \bibinfo
  {author} {\bibfnamefont {A.~E.}\ \bibnamefont {Koshelev}}, \bibinfo {author}
  {\bibfnamefont {J.}~\bibnamefont {Singleton}}, \bibinfo {author}
  {\bibfnamefont {Y.}~\bibnamefont {Lee}},\ and\ \bibinfo {author}
  {\bibfnamefont {T.}~\bibnamefont {Schenkel}},\ }\bibfield  {title} {\bibinfo
  {title} {Magnetoresistance scaling and the origin of {$H$}-linear resistivity
  in
  {${\mathrm{BaFe}}_{2}({\mathrm{As}}_{1\ensuremath{-}x}{\mathrm{P}}_{x}{)}_{2}$}},\
  }\href {https://doi.org/10.1103/PhysRevX.10.041062} {\bibfield  {journal}
  {\bibinfo  {journal} {Phys. Rev. X}\ }\textbf {\bibinfo {volume} {10}},\
  \bibinfo {pages} {041062} (\bibinfo {year} {2020})}\BibitemShut {NoStop}%
\bibitem [{\citenamefont {Kolincio}\ \emph {et~al.}(2020)\citenamefont
  {Kolincio}, \citenamefont {Roman},\ and\ \citenamefont
  {Klimczuk}}]{KolincioK_PRL_2020}%
  \BibitemOpen
  \bibfield  {author} {\bibinfo {author} {\bibfnamefont {K.~K.}\ \bibnamefont
  {Kolincio}}, \bibinfo {author} {\bibfnamefont {M.}~\bibnamefont {Roman}},\
  and\ \bibinfo {author} {\bibfnamefont {T.}~\bibnamefont {Klimczuk}},\
  }\bibfield  {title} {\bibinfo {title} {Enhanced mobility and large linear
  nonsaturating magnetoresistance in the magnetically ordered states of
  {${\mathrm{TmNiC}}_{2}$}},\ }\href
  {https://doi.org/10.1103/PhysRevLett.125.176601} {\bibfield  {journal}
  {\bibinfo  {journal} {Phys. Rev. Lett.}\ }\textbf {\bibinfo {volume} {125}},\
  \bibinfo {pages} {176601} (\bibinfo {year} {2020})}\BibitemShut {NoStop}%
\bibitem [{\citenamefont {Lei}\ \emph {et~al.}(2020)\citenamefont {Lei},
  \citenamefont {Zhou}, \citenamefont {Hao}, \citenamefont {Ma}, \citenamefont
  {Ma}, \citenamefont {Wang}, \citenamefont {Chen}, \citenamefont {Ye},
  \citenamefont {Wang}, \citenamefont {Ye}, \citenamefont {Wang}, \citenamefont
  {Mei},\ and\ \citenamefont {He}}]{LeiX_PRB_2020}%
  \BibitemOpen
  \bibfield  {author} {\bibinfo {author} {\bibfnamefont {X.}~\bibnamefont
  {Lei}}, \bibinfo {author} {\bibfnamefont {L.}~\bibnamefont {Zhou}}, \bibinfo
  {author} {\bibfnamefont {Z.~Y.}\ \bibnamefont {Hao}}, \bibinfo {author}
  {\bibfnamefont {X.~Z.}\ \bibnamefont {Ma}}, \bibinfo {author} {\bibfnamefont
  {C.}~\bibnamefont {Ma}}, \bibinfo {author} {\bibfnamefont {Y.~Q.}\
  \bibnamefont {Wang}}, \bibinfo {author} {\bibfnamefont {P.~B.}\ \bibnamefont
  {Chen}}, \bibinfo {author} {\bibfnamefont {B.~C.}\ \bibnamefont {Ye}},
  \bibinfo {author} {\bibfnamefont {L.}~\bibnamefont {Wang}}, \bibinfo {author}
  {\bibfnamefont {F.}~\bibnamefont {Ye}}, \bibinfo {author} {\bibfnamefont
  {J.~N.}\ \bibnamefont {Wang}}, \bibinfo {author} {\bibfnamefont {J.~W.}\
  \bibnamefont {Mei}},\ and\ \bibinfo {author} {\bibfnamefont {H.~T.}\
  \bibnamefont {He}},\ }\bibfield  {title} {\bibinfo {title} {Surface-induced
  linear magnetoresistance in the antiferromagnetic topological insulator
  {$\mathrm{Mn}{\mathrm{Bi}}_{2}{\mathrm{Te}}_{4}$}},\ }\href
  {https://doi.org/10.1103/PhysRevB.102.235431} {\bibfield  {journal} {\bibinfo
   {journal} {Phys. Rev. B}\ }\textbf {\bibinfo {volume} {102}},\ \bibinfo
  {pages} {235431} (\bibinfo {year} {2020})}\BibitemShut {NoStop}%
\bibitem [{\citenamefont {Campbell}\ \emph {et~al.}(2021)\citenamefont
  {Campbell}, \citenamefont {Collini}, \citenamefont {S{{\l}}awi{{\'{n}}}ska},
  \citenamefont {Autieri}, \citenamefont {Wang}, \citenamefont {Wang},
  \citenamefont {Wilfong}, \citenamefont {Eo}, \citenamefont {Neves},
  \citenamefont {Graf}, \citenamefont {Rodriguez}, \citenamefont {Butch},
  \citenamefont {Buongiorno~Nardelli},\ and\ \citenamefont
  {Paglione}}]{CampbellD_nQM_2021}%
  \BibitemOpen
  \bibfield  {author} {\bibinfo {author} {\bibfnamefont {D.~J.}\ \bibnamefont
  {Campbell}}, \bibinfo {author} {\bibfnamefont {J.}~\bibnamefont {Collini}},
  \bibinfo {author} {\bibfnamefont {J.}~\bibnamefont {S{{\l}}awi{{\'{n}}}ska}},
  \bibinfo {author} {\bibfnamefont {C.}~\bibnamefont {Autieri}}, \bibinfo
  {author} {\bibfnamefont {L.}~\bibnamefont {Wang}}, \bibinfo {author}
  {\bibfnamefont {K.}~\bibnamefont {Wang}}, \bibinfo {author} {\bibfnamefont
  {B.}~\bibnamefont {Wilfong}}, \bibinfo {author} {\bibfnamefont {Y.~S.}\
  \bibnamefont {Eo}}, \bibinfo {author} {\bibfnamefont {P.}~\bibnamefont
  {Neves}}, \bibinfo {author} {\bibfnamefont {D.}~\bibnamefont {Graf}},
  \bibinfo {author} {\bibfnamefont {E.~E.}\ \bibnamefont {Rodriguez}}, \bibinfo
  {author} {\bibfnamefont {N.~P.}\ \bibnamefont {Butch}}, \bibinfo {author}
  {\bibfnamefont {M.}~\bibnamefont {Buongiorno~Nardelli}},\ and\ \bibinfo
  {author} {\bibfnamefont {J.}~\bibnamefont {Paglione}},\ }\bibfield  {title}
  {\bibinfo {title} {Topologically driven linear magnetoresistance in
  helimagnetic {FeP}},\ }\href {https://doi.org/10.1038/s41535-021-00337-2}
  {\bibfield  {journal} {\bibinfo  {journal} {npj Quantum Mater.}\ }\textbf
  {\bibinfo {volume} {6}},\ \bibinfo {pages} {38} (\bibinfo {year}
  {2021})}\BibitemShut {NoStop}%
\bibitem [{\citenamefont {Klier}\ \emph {et~al.}(2015)\citenamefont {Klier},
  \citenamefont {Gornyi},\ and\ \citenamefont {Mirlin}}]{KlierJ_PRB_2015}%
  \BibitemOpen
  \bibfield  {author} {\bibinfo {author} {\bibfnamefont {J.}~\bibnamefont
  {Klier}}, \bibinfo {author} {\bibfnamefont {I.~V.}\ \bibnamefont {Gornyi}},\
  and\ \bibinfo {author} {\bibfnamefont {A.~D.}\ \bibnamefont {Mirlin}},\
  }\bibfield  {title} {\bibinfo {title} {Transversal magnetoresistance in
  {Weyl} semimetals},\ }\href {https://doi.org/10.1103/PhysRevB.92.205113}
  {\bibfield  {journal} {\bibinfo  {journal} {Phys. Rev. B}\ }\textbf {\bibinfo
  {volume} {92}},\ \bibinfo {pages} {205113} (\bibinfo {year}
  {2015})}\BibitemShut {NoStop}%
\bibitem [{\citenamefont {Xiao}\ \emph {et~al.}(2017)\citenamefont {Xiao},
  \citenamefont {Law},\ and\ \citenamefont {Lee}}]{XiaoX_PRB_2017}%
  \BibitemOpen
  \bibfield  {author} {\bibinfo {author} {\bibfnamefont {X.}~\bibnamefont
  {Xiao}}, \bibinfo {author} {\bibfnamefont {K.~T.}\ \bibnamefont {Law}},\ and\
  \bibinfo {author} {\bibfnamefont {P.~A.}\ \bibnamefont {Lee}},\ }\bibfield
  {title} {\bibinfo {title} {Magnetoconductivity in {Weyl} semimetals: Effect
  of chemical potential and temperature},\ }\href
  {https://doi.org/10.1103/PhysRevB.96.165101} {\bibfield  {journal} {\bibinfo
  {journal} {Phys. Rev. B}\ }\textbf {\bibinfo {volume} {96}},\ \bibinfo
  {pages} {165101} (\bibinfo {year} {2017})}\BibitemShut {NoStop}%
\bibitem [{\citenamefont {K\"onye}\ and\ \citenamefont
  {Ogata}(2018)}]{KoenyeV_PRB_2018}%
  \BibitemOpen
  \bibfield  {author} {\bibinfo {author} {\bibfnamefont {V.}~\bibnamefont
  {K\"onye}}\ and\ \bibinfo {author} {\bibfnamefont {M.}~\bibnamefont
  {Ogata}},\ }\bibfield  {title} {\bibinfo {title} {Magnetoresistance of a
  three-dimensional {Dirac} gas},\ }\href
  {https://doi.org/10.1103/PhysRevB.98.195420} {\bibfield  {journal} {\bibinfo
  {journal} {Phys. Rev. B}\ }\textbf {\bibinfo {volume} {98}},\ \bibinfo
  {pages} {195420} (\bibinfo {year} {2018})}\BibitemShut {NoStop}%
\bibitem [{\citenamefont {Rodionov}\ \emph {et~al.}(2020)\citenamefont
  {Rodionov}, \citenamefont {Kugel}, \citenamefont {Aronzon},\ and\
  \citenamefont {Nori}}]{RodionovY_PRB_2020}%
  \BibitemOpen
  \bibfield  {author} {\bibinfo {author} {\bibfnamefont {Y.~I.}\ \bibnamefont
  {Rodionov}}, \bibinfo {author} {\bibfnamefont {K.~I.}\ \bibnamefont {Kugel}},
  \bibinfo {author} {\bibfnamefont {B.~A.}\ \bibnamefont {Aronzon}},\ and\
  \bibinfo {author} {\bibfnamefont {F.}~\bibnamefont {Nori}},\ }\bibfield
  {title} {\bibinfo {title} {Effect of disorder on the transverse
  magnetoresistance of {Weyl} semimetals},\ }\href
  {https://doi.org/10.1103/PhysRevB.102.205105} {\bibfield  {journal} {\bibinfo
   {journal} {Phys. Rev. B}\ }\textbf {\bibinfo {volume} {102}},\ \bibinfo
  {pages} {205105} (\bibinfo {year} {2020})}\BibitemShut {NoStop}%
\bibitem [{\citenamefont {Parish}\ and\ \citenamefont
  {Littlewood}(2003)}]{ParishM_N_2003}%
  \BibitemOpen
  \bibfield  {author} {\bibinfo {author} {\bibfnamefont {M.~M.}\ \bibnamefont
  {Parish}}\ and\ \bibinfo {author} {\bibfnamefont {P.~B.}\ \bibnamefont
  {Littlewood}},\ }\bibfield  {title} {\bibinfo {title} {Non-saturating
  magnetoresistance in heavily disordered semiconductors},\ }\href
  {https://doi.org/10.1038/nature02073} {\bibfield  {journal} {\bibinfo
  {journal} {Nature}\ }\textbf {\bibinfo {volume} {426}},\ \bibinfo {pages}
  {162} (\bibinfo {year} {2003})}\BibitemShut {NoStop}%
\bibitem [{\citenamefont {Parish}\ and\ \citenamefont
  {Littlewood}(2005)}]{ParishM_PRB_2005}%
  \BibitemOpen
  \bibfield  {author} {\bibinfo {author} {\bibfnamefont {M.~M.}\ \bibnamefont
  {Parish}}\ and\ \bibinfo {author} {\bibfnamefont {P.~B.}\ \bibnamefont
  {Littlewood}},\ }\bibfield  {title} {\bibinfo {title} {Classical
  magnetotransport of inhomogeneous conductors},\ }\href
  {https://doi.org/10.1103/PhysRevB.72.094417} {\bibfield  {journal} {\bibinfo
  {journal} {Phys. Rev. B}\ }\textbf {\bibinfo {volume} {72}},\ \bibinfo
  {pages} {094417} (\bibinfo {year} {2005})}\BibitemShut {NoStop}%
\bibitem [{\citenamefont {Hu}\ \emph {et~al.}(2007)\citenamefont {Hu},
  \citenamefont {Parish},\ and\ \citenamefont {Rosenbaum}}]{HuJ_PRB_2007}%
  \BibitemOpen
  \bibfield  {author} {\bibinfo {author} {\bibfnamefont {J.}~\bibnamefont
  {Hu}}, \bibinfo {author} {\bibfnamefont {M.~M.}\ \bibnamefont {Parish}},\
  and\ \bibinfo {author} {\bibfnamefont {T.~F.}\ \bibnamefont {Rosenbaum}},\
  }\bibfield  {title} {\bibinfo {title} {Nonsaturating magnetoresistance of
  inhomogeneous conductors: Comparison of experiment and simulation},\ }\href
  {https://doi.org/10.1103/PhysRevB.75.214203} {\bibfield  {journal} {\bibinfo
  {journal} {Phys. Rev. B}\ }\textbf {\bibinfo {volume} {75}},\ \bibinfo
  {pages} {214203} (\bibinfo {year} {2007})}\BibitemShut {NoStop}%
\bibitem [{\citenamefont {Xu}\ \emph {et~al.}(2008)\citenamefont {Xu},
  \citenamefont {Zhang}, \citenamefont {Yang}, \citenamefont {Li},\ and\
  \citenamefont {Pan}}]{XuJ_JoAP_2008}%
  \BibitemOpen
  \bibfield  {author} {\bibinfo {author} {\bibfnamefont {J.}~\bibnamefont
  {Xu}}, \bibinfo {author} {\bibfnamefont {D.}~\bibnamefont {Zhang}}, \bibinfo
  {author} {\bibfnamefont {F.}~\bibnamefont {Yang}}, \bibinfo {author}
  {\bibfnamefont {Z.}~\bibnamefont {Li}},\ and\ \bibinfo {author}
  {\bibfnamefont {Y.}~\bibnamefont {Pan}},\ }\bibfield  {title} {\bibinfo
  {title} {A three-dimensional resistor network model for the linear
  magnetoresistance of {${\mathrm{Ag}}_{2+\delta}\mathrm{Se}$} and
  {${\mathrm{Ag}}_{2+\delta}\mathrm{Te}$} bulks},\ }\href
  {https://doi.org/10.1063/1.3035834} {\bibfield  {journal} {\bibinfo
  {journal} {J. Appl. Phys.}\ }\textbf {\bibinfo {volume} {104}},\ \bibinfo
  {pages} {113922} (\bibinfo {year} {2008})}\BibitemShut {NoStop}%
\bibitem [{\citenamefont {Ramakrishnan}\ \emph {et~al.}(2017)\citenamefont
  {Ramakrishnan}, \citenamefont {Lai}, \citenamefont {Lara}, \citenamefont
  {Parish},\ and\ \citenamefont {Adam}}]{RamakrishnanN_PRB_2017}%
  \BibitemOpen
  \bibfield  {author} {\bibinfo {author} {\bibfnamefont {N.}~\bibnamefont
  {Ramakrishnan}}, \bibinfo {author} {\bibfnamefont {Y.~T.}\ \bibnamefont
  {Lai}}, \bibinfo {author} {\bibfnamefont {S.}~\bibnamefont {Lara}}, \bibinfo
  {author} {\bibfnamefont {M.~M.}\ \bibnamefont {Parish}},\ and\ \bibinfo
  {author} {\bibfnamefont {S.}~\bibnamefont {Adam}},\ }\bibfield  {title}
  {\bibinfo {title} {Equivalence of effective medium and random resistor
  network models for disorder-induced unsaturating linear magnetoresistance},\
  }\href {https://doi.org/10.1103/PhysRevB.96.224203} {\bibfield  {journal}
  {\bibinfo  {journal} {Phys. Rev. B}\ }\textbf {\bibinfo {volume} {96}},\
  \bibinfo {pages} {224203} (\bibinfo {year} {2017})}\BibitemShut {NoStop}%
\bibitem [{\citenamefont {Kisslinger}\ \emph {et~al.}(2017)\citenamefont
  {Kisslinger}, \citenamefont {Ott},\ and\ \citenamefont
  {Weber}}]{KisslingerF_PRB_2017}%
  \BibitemOpen
  \bibfield  {author} {\bibinfo {author} {\bibfnamefont {F.}~\bibnamefont
  {Kisslinger}}, \bibinfo {author} {\bibfnamefont {C.}~\bibnamefont {Ott}},\
  and\ \bibinfo {author} {\bibfnamefont {H.~B.}\ \bibnamefont {Weber}},\
  }\bibfield  {title} {\bibinfo {title} {Origin of nonsaturating linear
  magnetoresistivity},\ }\href {https://doi.org/10.1103/PhysRevB.95.024204}
  {\bibfield  {journal} {\bibinfo  {journal} {Phys. Rev. B}\ }\textbf {\bibinfo
  {volume} {95}},\ \bibinfo {pages} {024204} (\bibinfo {year}
  {2017})}\BibitemShut {NoStop}%
\bibitem [{\citenamefont {Chen}\ \emph {et~al.}(2022)\citenamefont {Chen},
  \citenamefont {Yang},\ and\ \citenamefont {Yang}}]{ChenS_CPB_2022}%
  \BibitemOpen
  \bibfield  {author} {\bibinfo {author} {\bibfnamefont {S.-S.}\ \bibnamefont
  {Chen}}, \bibinfo {author} {\bibfnamefont {Y.}~\bibnamefont {Yang}},\ and\
  \bibinfo {author} {\bibfnamefont {F.}~\bibnamefont {Yang}},\ }\bibfield
  {title} {\bibinfo {title} {Analytical formula describing the non-saturating
  linear magnetoresistance in inhomogeneous conductors},\ }\href
  {https://doi.org/10.1088/1674-1056/ac6582} {\bibfield  {journal} {\bibinfo
  {journal} {Chin. Phys. B}\ }\textbf {\bibinfo {volume} {31}},\ \bibinfo
  {pages} {087303} (\bibinfo {year} {2022})}\BibitemShut {NoStop}%
\bibitem [{\citenamefont {Alekseev}\ \emph {et~al.}(2015)\citenamefont
  {Alekseev}, \citenamefont {Dmitriev}, \citenamefont {Gornyi}, \citenamefont
  {Kachorovskii}, \citenamefont {Narozhny}, \citenamefont {Sch\"utt},\ and\
  \citenamefont {Titov}}]{AlekseevP_PRL_2015}%
  \BibitemOpen
  \bibfield  {author} {\bibinfo {author} {\bibfnamefont {P.~S.}\ \bibnamefont
  {Alekseev}}, \bibinfo {author} {\bibfnamefont {A.~P.}\ \bibnamefont
  {Dmitriev}}, \bibinfo {author} {\bibfnamefont {I.~V.}\ \bibnamefont
  {Gornyi}}, \bibinfo {author} {\bibfnamefont {V.~Y.}\ \bibnamefont
  {Kachorovskii}}, \bibinfo {author} {\bibfnamefont {B.~N.}\ \bibnamefont
  {Narozhny}}, \bibinfo {author} {\bibfnamefont {M.}~\bibnamefont {Sch\"utt}},\
  and\ \bibinfo {author} {\bibfnamefont {M.}~\bibnamefont {Titov}},\ }\bibfield
   {title} {\bibinfo {title} {Magnetoresistance in two-component systems},\
  }\href {https://doi.org/10.1103/PhysRevLett.114.156601} {\bibfield  {journal}
  {\bibinfo  {journal} {Phys. Rev. Lett.}\ }\textbf {\bibinfo {volume} {114}},\
  \bibinfo {pages} {156601} (\bibinfo {year} {2015})}\BibitemShut {NoStop}%
\bibitem [{\citenamefont {Alekseev}\ \emph {et~al.}(2017)\citenamefont
  {Alekseev}, \citenamefont {Dmitriev}, \citenamefont {Gornyi}, \citenamefont
  {Kachorovskii}, \citenamefont {Narozhny}, \citenamefont {Sch\"utt},\ and\
  \citenamefont {Titov}}]{AlekseevP_PRB_2017}%
  \BibitemOpen
  \bibfield  {author} {\bibinfo {author} {\bibfnamefont {P.~S.}\ \bibnamefont
  {Alekseev}}, \bibinfo {author} {\bibfnamefont {A.~P.}\ \bibnamefont
  {Dmitriev}}, \bibinfo {author} {\bibfnamefont {I.~V.}\ \bibnamefont
  {Gornyi}}, \bibinfo {author} {\bibfnamefont {V.~Y.}\ \bibnamefont
  {Kachorovskii}}, \bibinfo {author} {\bibfnamefont {B.~N.}\ \bibnamefont
  {Narozhny}}, \bibinfo {author} {\bibfnamefont {M.}~\bibnamefont {Sch\"utt}},\
  and\ \bibinfo {author} {\bibfnamefont {M.}~\bibnamefont {Titov}},\ }\bibfield
   {title} {\bibinfo {title} {Magnetoresistance of compensated semimetals in
  confined geometries},\ }\href {https://doi.org/10.1103/PhysRevB.95.165410}
  {\bibfield  {journal} {\bibinfo  {journal} {Phys. Rev. B}\ }\textbf {\bibinfo
  {volume} {95}},\ \bibinfo {pages} {165410} (\bibinfo {year}
  {2017})}\BibitemShut {NoStop}%
\bibitem [{\citenamefont {Song}\ \emph {et~al.}(2015)\citenamefont {Song},
  \citenamefont {Refael},\ and\ \citenamefont {Lee}}]{SongJ_PRB_2015}%
  \BibitemOpen
  \bibfield  {author} {\bibinfo {author} {\bibfnamefont {J.~C.~W.}\
  \bibnamefont {Song}}, \bibinfo {author} {\bibfnamefont {G.}~\bibnamefont
  {Refael}},\ and\ \bibinfo {author} {\bibfnamefont {P.~A.}\ \bibnamefont
  {Lee}},\ }\bibfield  {title} {\bibinfo {title} {Linear magnetoresistance in
  metals: Guiding center diffusion in a smooth random potential},\ }\href
  {https://doi.org/10.1103/PhysRevB.92.180204} {\bibfield  {journal} {\bibinfo
  {journal} {Phys. Rev. B}\ }\textbf {\bibinfo {volume} {92}},\ \bibinfo
  {pages} {180204(R)} (\bibinfo {year} {2015})}\BibitemShut {NoStop}%
\bibitem [{\citenamefont {Xiao}\ \emph {et~al.}(2020)\citenamefont {Xiao},
  \citenamefont {Chen}, \citenamefont {Gao}, \citenamefont {Xiao},
  \citenamefont {MacDonald},\ and\ \citenamefont {Niu}}]{XiaoC_PRB_2020}%
  \BibitemOpen
  \bibfield  {author} {\bibinfo {author} {\bibfnamefont {C.}~\bibnamefont
  {Xiao}}, \bibinfo {author} {\bibfnamefont {H.}~\bibnamefont {Chen}}, \bibinfo
  {author} {\bibfnamefont {Y.}~\bibnamefont {Gao}}, \bibinfo {author}
  {\bibfnamefont {D.}~\bibnamefont {Xiao}}, \bibinfo {author} {\bibfnamefont
  {A.~H.}\ \bibnamefont {MacDonald}},\ and\ \bibinfo {author} {\bibfnamefont
  {Q.}~\bibnamefont {Niu}},\ }\bibfield  {title} {\bibinfo {title} {Linear
  magnetoresistance induced by intra-scattering semiclassics of {Bloch}
  electrons},\ }\href {https://doi.org/10.1103/PhysRevB.101.201410} {\bibfield
  {journal} {\bibinfo  {journal} {Phys. Rev. B}\ }\textbf {\bibinfo {volume}
  {101}},\ \bibinfo {pages} {201410(R)} (\bibinfo {year} {2020})}\BibitemShut
  {NoStop}%
\bibitem [{\citenamefont {Khouri}\ \emph {et~al.}(2016)\citenamefont {Khouri},
  \citenamefont {Zeitler}, \citenamefont {Reichl}, \citenamefont {Wegscheider},
  \citenamefont {Hussey}, \citenamefont {Wiedmann},\ and\ \citenamefont
  {Maan}}]{KhouriT_PRL_2016}%
  \BibitemOpen
  \bibfield  {author} {\bibinfo {author} {\bibfnamefont {T.}~\bibnamefont
  {Khouri}}, \bibinfo {author} {\bibfnamefont {U.}~\bibnamefont {Zeitler}},
  \bibinfo {author} {\bibfnamefont {C.}~\bibnamefont {Reichl}}, \bibinfo
  {author} {\bibfnamefont {W.}~\bibnamefont {Wegscheider}}, \bibinfo {author}
  {\bibfnamefont {N.~E.}\ \bibnamefont {Hussey}}, \bibinfo {author}
  {\bibfnamefont {S.}~\bibnamefont {Wiedmann}},\ and\ \bibinfo {author}
  {\bibfnamefont {J.~C.}\ \bibnamefont {Maan}},\ }\bibfield  {title} {\bibinfo
  {title} {Linear magnetoresistance in a quasifree two-dimensional electron gas
  in an ultrahigh mobility {GaAs} quantum well},\ }\href
  {https://doi.org/10.1103/PhysRevLett.117.256601} {\bibfield  {journal}
  {\bibinfo  {journal} {Phys. Rev. Lett.}\ }\textbf {\bibinfo {volume} {117}},\
  \bibinfo {pages} {256601} (\bibinfo {year} {2016})}\BibitemShut {NoStop}%
\bibitem [{\citenamefont {Mallik}\ \emph {et~al.}(2022)\citenamefont {Mallik},
  \citenamefont {Ménard}, \citenamefont {Saïz}, \citenamefont {Gilmutdinov},
  \citenamefont {Vignolles}, \citenamefont {Proust}, \citenamefont {Gloter},
  \citenamefont {Bergeal}, \citenamefont {Gabay},\ and\ \citenamefont
  {Bibes}}]{MallikS_NL_2022}%
  \BibitemOpen
  \bibfield  {author} {\bibinfo {author} {\bibfnamefont {S.}~\bibnamefont
  {Mallik}}, \bibinfo {author} {\bibfnamefont {G.~C.}\ \bibnamefont {Ménard}},
  \bibinfo {author} {\bibfnamefont {G.}~\bibnamefont {Saïz}}, \bibinfo
  {author} {\bibfnamefont {I.}~\bibnamefont {Gilmutdinov}}, \bibinfo {author}
  {\bibfnamefont {D.}~\bibnamefont {Vignolles}}, \bibinfo {author}
  {\bibfnamefont {C.}~\bibnamefont {Proust}}, \bibinfo {author} {\bibfnamefont
  {A.}~\bibnamefont {Gloter}}, \bibinfo {author} {\bibfnamefont
  {N.}~\bibnamefont {Bergeal}}, \bibinfo {author} {\bibfnamefont
  {M.}~\bibnamefont {Gabay}},\ and\ \bibinfo {author} {\bibfnamefont
  {M.}~\bibnamefont {Bibes}},\ }\bibfield  {title} {\bibinfo {title} {From
  low-field {Sondheimer} oscillations to high-field very large and linear
  magnetoresistance in a {${\mathrm{SrTiO}}_3$}-based two-dimensional electron
  gas},\ }\href {https://doi.org/10.1021/acs.nanolett.1c03198} {\bibfield
  {journal} {\bibinfo  {journal} {Nano Lett.}\ }\textbf {\bibinfo {volume}
  {22}},\ \bibinfo {pages} {65} (\bibinfo {year} {2022})}\BibitemShut {NoStop}%
\bibitem [{\citenamefont {Vasileva}\ \emph {et~al.}(2016)\citenamefont
  {Vasileva}, \citenamefont {Smirnov}, \citenamefont {Ivanov}, \citenamefont
  {Vasilyev}, \citenamefont {Alekseev}, \citenamefont {Dmitriev}, \citenamefont
  {Gornyi}, \citenamefont {Kachorovskii}, \citenamefont {Titov}, \citenamefont
  {Narozhny},\ and\ \citenamefont {Haug}}]{VasilevaG_PRB_2016}%
  \BibitemOpen
  \bibfield  {author} {\bibinfo {author} {\bibfnamefont {G.~Y.}\ \bibnamefont
  {Vasileva}}, \bibinfo {author} {\bibfnamefont {D.}~\bibnamefont {Smirnov}},
  \bibinfo {author} {\bibfnamefont {Y.~L.}\ \bibnamefont {Ivanov}}, \bibinfo
  {author} {\bibfnamefont {Y.~B.}\ \bibnamefont {Vasilyev}}, \bibinfo {author}
  {\bibfnamefont {P.~S.}\ \bibnamefont {Alekseev}}, \bibinfo {author}
  {\bibfnamefont {A.~P.}\ \bibnamefont {Dmitriev}}, \bibinfo {author}
  {\bibfnamefont {I.~V.}\ \bibnamefont {Gornyi}}, \bibinfo {author}
  {\bibfnamefont {V.~Y.}\ \bibnamefont {Kachorovskii}}, \bibinfo {author}
  {\bibfnamefont {M.}~\bibnamefont {Titov}}, \bibinfo {author} {\bibfnamefont
  {B.~N.}\ \bibnamefont {Narozhny}},\ and\ \bibinfo {author} {\bibfnamefont
  {R.~J.}\ \bibnamefont {Haug}},\ }\bibfield  {title} {\bibinfo {title} {Linear
  magnetoresistance in compensated graphene bilayer},\ }\href
  {https://doi.org/10.1103/PhysRevB.93.195430} {\bibfield  {journal} {\bibinfo
  {journal} {Phys. Rev. B}\ }\textbf {\bibinfo {volume} {93}},\ \bibinfo
  {pages} {195430} (\bibinfo {year} {2016})}\BibitemShut {NoStop}%
\bibitem [{\citenamefont {Chandan}\ \emph {et~al.}(2020)\citenamefont
  {Chandan}, \citenamefont {Islam}, \citenamefont {Venkataraman}, \citenamefont
  {Ghosh},\ and\ \citenamefont {Angadi}}]{Chandan_JoPDAP_2020}%
  \BibitemOpen
  \bibfield  {author} {\bibinfo {author} {\bibnamefont {Chandan}}, \bibinfo
  {author} {\bibfnamefont {S.}~\bibnamefont {Islam}}, \bibinfo {author}
  {\bibfnamefont {V.}~\bibnamefont {Venkataraman}}, \bibinfo {author}
  {\bibfnamefont {A.}~\bibnamefont {Ghosh}},\ and\ \bibinfo {author}
  {\bibfnamefont {B.}~\bibnamefont {Angadi}},\ }\bibfield  {title} {\bibinfo
  {title} {Observation of linear magneto-resistance with small cross-over field
  at room temperature in bismuth},\ }\href
  {https://doi.org/10.1088/1361-6463/ab985c} {\bibfield  {journal} {\bibinfo
  {journal} {J. Phys. D: Appl. Phys.}\ }\textbf {\bibinfo {volume} {53}},\
  \bibinfo {pages} {425102} (\bibinfo {year} {2020})}\BibitemShut {NoStop}%
\bibitem [{\citenamefont {Wang}\ \emph
  {et~al.}(2016{\natexlab{a}})\citenamefont {Wang}, \citenamefont {Liu},
  \citenamefont {Liu}, \citenamefont {Pan}, \citenamefont {Zhang},
  \citenamefont {Zeng}, \citenamefont {Fu}, \citenamefont {Wang}, \citenamefont
  {Xu}, \citenamefont {Huang}, \citenamefont {Wang}, \citenamefont {Lu},
  \citenamefont {Xing}, \citenamefont {Wang}, \citenamefont {Wan},\ and\
  \citenamefont {Miao}}]{WangY_NC_2016}%
  \BibitemOpen
  \bibfield  {author} {\bibinfo {author} {\bibfnamefont {Y.}~\bibnamefont
  {Wang}}, \bibinfo {author} {\bibfnamefont {E.}~\bibnamefont {Liu}}, \bibinfo
  {author} {\bibfnamefont {H.}~\bibnamefont {Liu}}, \bibinfo {author}
  {\bibfnamefont {Y.}~\bibnamefont {Pan}}, \bibinfo {author} {\bibfnamefont
  {L.}~\bibnamefont {Zhang}}, \bibinfo {author} {\bibfnamefont
  {J.}~\bibnamefont {Zeng}}, \bibinfo {author} {\bibfnamefont {Y.}~\bibnamefont
  {Fu}}, \bibinfo {author} {\bibfnamefont {M.}~\bibnamefont {Wang}}, \bibinfo
  {author} {\bibfnamefont {K.}~\bibnamefont {Xu}}, \bibinfo {author}
  {\bibfnamefont {Z.}~\bibnamefont {Huang}}, \bibinfo {author} {\bibfnamefont
  {Z.}~\bibnamefont {Wang}}, \bibinfo {author} {\bibfnamefont {H.-Z.}\
  \bibnamefont {Lu}}, \bibinfo {author} {\bibfnamefont {D.}~\bibnamefont
  {Xing}}, \bibinfo {author} {\bibfnamefont {B.}~\bibnamefont {Wang}}, \bibinfo
  {author} {\bibfnamefont {X.}~\bibnamefont {Wan}},\ and\ \bibinfo {author}
  {\bibfnamefont {F.}~\bibnamefont {Miao}},\ }\bibfield  {title} {\bibinfo
  {title} {Gate-tunable negative longitudinal magnetoresistance in the
  predicted type-{II} {Weyl} semimetal {${\mathrm{WTe}}_2$}},\ }\href
  {https://doi.org/10.1038/ncomms13142} {\bibfield  {journal} {\bibinfo
  {journal} {Nat. Commun.}\ }\textbf {\bibinfo {volume} {7}},\ \bibinfo {pages}
  {13142} (\bibinfo {year} {2016}{\natexlab{a}})}\BibitemShut {NoStop}%
\bibitem [{\citenamefont {Ashcroft}\ and\ \citenamefont
  {Mermin}(1976)}]{Ashcroft_1976}%
  \BibitemOpen
  \bibfield  {author} {\bibinfo {author} {\bibfnamefont {N.~W.}\ \bibnamefont
  {Ashcroft}}\ and\ \bibinfo {author} {\bibfnamefont {N.~D.}\ \bibnamefont
  {Mermin}},\ }\href@noop {} {\emph {\bibinfo {title} {Solid state physics}}}\
  (\bibinfo  {publisher} {Holt, Rinehart and Winston},\ \bibinfo {address} {New
  York},\ \bibinfo {year} {1976})\BibitemShut {NoStop}%
\bibitem [{\citenamefont {Xu}\ \emph {et~al.}(2017)\citenamefont {Xu},
  \citenamefont {Chiu}, \citenamefont {Miao}, \citenamefont {He}, \citenamefont
  {Alpichshev}, \citenamefont {Kapitulnik}, \citenamefont {Biswas},\ and\
  \citenamefont {Wray}}]{XuY_NC_2017}%
  \BibitemOpen
  \bibfield  {author} {\bibinfo {author} {\bibfnamefont {Y.}~\bibnamefont
  {Xu}}, \bibinfo {author} {\bibfnamefont {J.}~\bibnamefont {Chiu}}, \bibinfo
  {author} {\bibfnamefont {L.}~\bibnamefont {Miao}}, \bibinfo {author}
  {\bibfnamefont {H.}~\bibnamefont {He}}, \bibinfo {author} {\bibfnamefont
  {Z.}~\bibnamefont {Alpichshev}}, \bibinfo {author} {\bibfnamefont
  {A.}~\bibnamefont {Kapitulnik}}, \bibinfo {author} {\bibfnamefont {R.~R.}\
  \bibnamefont {Biswas}},\ and\ \bibinfo {author} {\bibfnamefont {L.~A.}\
  \bibnamefont {Wray}},\ }\bibfield  {title} {\bibinfo {title} {Disorder
  enabled band structure engineering of a topological insulator surface},\
  }\href {https://doi.org/10.1038/ncomms14081} {\bibfield  {journal} {\bibinfo
  {journal} {Nat. Commun.}\ }\textbf {\bibinfo {volume} {8}},\ \bibinfo {pages}
  {14081} (\bibinfo {year} {2017})}\BibitemShut {NoStop}%
\bibitem [{\citenamefont {Wang}\ \emph {et~al.}(2018)\citenamefont {Wang},
  \citenamefont {Fu},\ and\ \citenamefont {Shen}}]{WangH_PRB_2018}%
  \BibitemOpen
  \bibfield  {author} {\bibinfo {author} {\bibfnamefont {H.-W.}\ \bibnamefont
  {Wang}}, \bibinfo {author} {\bibfnamefont {B.}~\bibnamefont {Fu}},\ and\
  \bibinfo {author} {\bibfnamefont {S.-Q.}\ \bibnamefont {Shen}},\ }\bibfield
  {title} {\bibinfo {title} {Intrinsic magnetoresistance in three-dimensional
  {Dirac} materials with low carrier density},\ }\href
  {https://doi.org/10.1103/PhysRevB.98.081202} {\bibfield  {journal} {\bibinfo
  {journal} {Phys. Rev. B}\ }\textbf {\bibinfo {volume} {98}},\ \bibinfo
  {pages} {081202(R)} (\bibinfo {year} {2018})}\BibitemShut {NoStop}%
\bibitem [{\citenamefont {Shen}(2017)}]{Shen_2017}%
  \BibitemOpen
  \bibfield  {author} {\bibinfo {author} {\bibfnamefont {S.-Q.}\ \bibnamefont
  {Shen}},\ }\href {https://doi.org/10.1007/978-981-10-4606-3} {\emph {\bibinfo
  {title} {Topological {Insulators}: {Dirac} {Equation} in {Condensed}
  {Matter}}}},\ \bibinfo {edition} {2nd}\ ed.,\ \bibinfo {series} {Springer
  {Series} in {Solid}-{State} {Sciences}}\ No.\ \bibinfo {number} {187}\
  (\bibinfo  {publisher} {Springer Singapore : Imprint: Springer},\ \bibinfo
  {address} {Singapore},\ \bibinfo {year} {2017})\BibitemShut {NoStop}%
\bibitem [{\citenamefont {Chen}\ \emph {et~al.}(2015)\citenamefont {Chen},
  \citenamefont {Chen}, \citenamefont {Song}, \citenamefont {Schneeloch},
  \citenamefont {Gu}, \citenamefont {Wang},\ and\ \citenamefont
  {Wang}}]{ChenR_PRL_2015}%
  \BibitemOpen
  \bibfield  {author} {\bibinfo {author} {\bibfnamefont {R.~Y.}\ \bibnamefont
  {Chen}}, \bibinfo {author} {\bibfnamefont {Z.~G.}\ \bibnamefont {Chen}},
  \bibinfo {author} {\bibfnamefont {X.-Y.}\ \bibnamefont {Song}}, \bibinfo
  {author} {\bibfnamefont {J.~A.}\ \bibnamefont {Schneeloch}}, \bibinfo
  {author} {\bibfnamefont {G.~D.}\ \bibnamefont {Gu}}, \bibinfo {author}
  {\bibfnamefont {F.}~\bibnamefont {Wang}},\ and\ \bibinfo {author}
  {\bibfnamefont {N.~L.}\ \bibnamefont {Wang}},\ }\bibfield  {title} {\bibinfo
  {title} {Magnetoinfrared spectroscopy of {Landau} levels and {Zeeman}
  splitting of three-dimensional massless {Dirac} fermions in
  {${\mathrm{ZrTe}}_{5}$}},\ }\href
  {https://doi.org/10.1103/PhysRevLett.115.176404} {\bibfield  {journal}
  {\bibinfo  {journal} {Phys. Rev. Lett.}\ }\textbf {\bibinfo {volume} {115}},\
  \bibinfo {pages} {176404} (\bibinfo {year} {2015})}\BibitemShut {NoStop}%
\bibitem [{\citenamefont {Manzoni}\ \emph {et~al.}(2016)\citenamefont
  {Manzoni}, \citenamefont {Gragnaniello}, \citenamefont {Aut\`es},
  \citenamefont {Kuhn}, \citenamefont {Sterzi}, \citenamefont {Cilento},
  \citenamefont {Zacchigna}, \citenamefont {Enenkel}, \citenamefont {Vobornik},
  \citenamefont {Barba}, \citenamefont {Bisti}, \citenamefont {Bugnon},
  \citenamefont {Magrez}, \citenamefont {Strocov}, \citenamefont {Berger},
  \citenamefont {Yazyev}, \citenamefont {Fonin}, \citenamefont {Parmigiani},\
  and\ \citenamefont {Crepaldi}}]{ManzoniG_PRL_2016}%
  \BibitemOpen
  \bibfield  {author} {\bibinfo {author} {\bibfnamefont {G.}~\bibnamefont
  {Manzoni}}, \bibinfo {author} {\bibfnamefont {L.}~\bibnamefont
  {Gragnaniello}}, \bibinfo {author} {\bibfnamefont {G.}~\bibnamefont
  {Aut\`es}}, \bibinfo {author} {\bibfnamefont {T.}~\bibnamefont {Kuhn}},
  \bibinfo {author} {\bibfnamefont {A.}~\bibnamefont {Sterzi}}, \bibinfo
  {author} {\bibfnamefont {F.}~\bibnamefont {Cilento}}, \bibinfo {author}
  {\bibfnamefont {M.}~\bibnamefont {Zacchigna}}, \bibinfo {author}
  {\bibfnamefont {V.}~\bibnamefont {Enenkel}}, \bibinfo {author} {\bibfnamefont
  {I.}~\bibnamefont {Vobornik}}, \bibinfo {author} {\bibfnamefont
  {L.}~\bibnamefont {Barba}}, \bibinfo {author} {\bibfnamefont
  {F.}~\bibnamefont {Bisti}}, \bibinfo {author} {\bibfnamefont
  {P.}~\bibnamefont {Bugnon}}, \bibinfo {author} {\bibfnamefont
  {A.}~\bibnamefont {Magrez}}, \bibinfo {author} {\bibfnamefont {V.~N.}\
  \bibnamefont {Strocov}}, \bibinfo {author} {\bibfnamefont {H.}~\bibnamefont
  {Berger}}, \bibinfo {author} {\bibfnamefont {O.~V.}\ \bibnamefont {Yazyev}},
  \bibinfo {author} {\bibfnamefont {M.}~\bibnamefont {Fonin}}, \bibinfo
  {author} {\bibfnamefont {F.}~\bibnamefont {Parmigiani}},\ and\ \bibinfo
  {author} {\bibfnamefont {A.}~\bibnamefont {Crepaldi}},\ }\bibfield  {title}
  {\bibinfo {title} {Evidence for a strong topological insulator phase in
  {${\mathrm{ZrTe}}_{5}$}},\ }\href
  {https://doi.org/10.1103/PhysRevLett.117.237601} {\bibfield  {journal}
  {\bibinfo  {journal} {Phys. Rev. Lett.}\ }\textbf {\bibinfo {volume} {117}},\
  \bibinfo {pages} {237601} (\bibinfo {year} {2016})}\BibitemShut {NoStop}%
\bibitem [{\citenamefont {Lu}\ \emph {et~al.}(2015)\citenamefont {Lu},
  \citenamefont {Zhang},\ and\ \citenamefont {Shen}}]{LuH_PRB_2015}%
  \BibitemOpen
  \bibfield  {author} {\bibinfo {author} {\bibfnamefont {H.-Z.}\ \bibnamefont
  {Lu}}, \bibinfo {author} {\bibfnamefont {S.-B.}\ \bibnamefont {Zhang}},\ and\
  \bibinfo {author} {\bibfnamefont {S.-Q.}\ \bibnamefont {Shen}},\ }\bibfield
  {title} {\bibinfo {title} {High-field magnetoconductivity of topological
  semimetals with short-range potential},\ }\href
  {https://doi.org/10.1103/PhysRevB.92.045203} {\bibfield  {journal} {\bibinfo
  {journal} {Phys. Rev. B}\ }\textbf {\bibinfo {volume} {92}},\ \bibinfo
  {pages} {045203} (\bibinfo {year} {2015})}\BibitemShut {NoStop}%
\bibitem [{\citenamefont {Zhang}\ \emph {et~al.}(2016)\citenamefont {Zhang},
  \citenamefont {Lu},\ and\ \citenamefont {Shen}}]{ZhangS_NJoP_2016}%
  \BibitemOpen
  \bibfield  {author} {\bibinfo {author} {\bibfnamefont {S.-B.}\ \bibnamefont
  {Zhang}}, \bibinfo {author} {\bibfnamefont {H.-Z.}\ \bibnamefont {Lu}},\ and\
  \bibinfo {author} {\bibfnamefont {S.-Q.}\ \bibnamefont {Shen}},\ }\bibfield
  {title} {\bibinfo {title} {Linear magnetoconductivity in an intrinsic
  topological {Weyl} semimetal},\ }\href
  {https://doi.org/10.1088/1367-2630/18/5/053039} {\bibfield  {journal}
  {\bibinfo  {journal} {New J. Phys.}\ }\textbf {\bibinfo {volume} {18}},\
  \bibinfo {pages} {053039} (\bibinfo {year} {2016})}\BibitemShut {NoStop}%
\bibitem [{\citenamefont {Armitage}\ \emph {et~al.}(2018)\citenamefont
  {Armitage}, \citenamefont {Mele},\ and\ \citenamefont
  {Vishwanath}}]{ArmitageN_RMP_2018}%
  \BibitemOpen
  \bibfield  {author} {\bibinfo {author} {\bibfnamefont {N.~P.}\ \bibnamefont
  {Armitage}}, \bibinfo {author} {\bibfnamefont {E.~J.}\ \bibnamefont {Mele}},\
  and\ \bibinfo {author} {\bibfnamefont {A.}~\bibnamefont {Vishwanath}},\
  }\bibfield  {title} {\bibinfo {title} {{Weyl} and {Dirac} semimetals in
  three-dimensional solids},\ }\href
  {https://doi.org/10.1103/RevModPhys.90.015001} {\bibfield  {journal}
  {\bibinfo  {journal} {Rev. Mod. Phys.}\ }\textbf {\bibinfo {volume} {90}},\
  \bibinfo {pages} {015001} (\bibinfo {year} {2018})}\BibitemShut {NoStop}%
\bibitem [{\citenamefont {Halperin}\ and\ \citenamefont
  {Lax}(1966)}]{HalperinB_PR_1966}%
  \BibitemOpen
  \bibfield  {author} {\bibinfo {author} {\bibfnamefont {B.~I.}\ \bibnamefont
  {Halperin}}\ and\ \bibinfo {author} {\bibfnamefont {M.}~\bibnamefont {Lax}},\
  }\bibfield  {title} {\bibinfo {title} {Impurity-band tails in the
  high-density limit. {I}. minimum counting methods},\ }\href
  {https://doi.org/10.1103/PhysRev.148.722} {\bibfield  {journal} {\bibinfo
  {journal} {Phys. Rev.}\ }\textbf {\bibinfo {volume} {148}},\ \bibinfo {pages}
  {722} (\bibinfo {year} {1966})}\BibitemShut {NoStop}%
\bibitem [{\citenamefont
  {Samathiyakanit}(1974)}]{SamathiyakanitV_JoPCSSP_1974}%
  \BibitemOpen
  \bibfield  {author} {\bibinfo {author} {\bibfnamefont {V.}~\bibnamefont
  {Samathiyakanit}},\ }\bibfield  {title} {\bibinfo {title} {Path-integral
  theory of a model disordered system},\ }\href
  {https://doi.org/10.1088/0022-3719/7/16/015} {\bibfield  {journal} {\bibinfo
  {journal} {Journal of Physics C: Solid State Physics}\ }\textbf {\bibinfo
  {volume} {7}},\ \bibinfo {pages} {2849} (\bibinfo {year} {1974})}\BibitemShut
  {NoStop}%
\bibitem [{\citenamefont {Saitoh}\ and\ \citenamefont
  {Edwards}(1974)}]{SaitohM_JoPCSSP_1974}%
  \BibitemOpen
  \bibfield  {author} {\bibinfo {author} {\bibfnamefont {M.}~\bibnamefont
  {Saitoh}}\ and\ \bibinfo {author} {\bibfnamefont {S.~F.}\ \bibnamefont
  {Edwards}},\ }\bibfield  {title} {\bibinfo {title} {A note on the density of
  states of a disordered system with {Gaussian} random potentials},\ }\href
  {https://doi.org/10.1088/0022-3719/7/21/014} {\bibfield  {journal} {\bibinfo
  {journal} {Journal of Physics C: Solid State Physics}\ }\textbf {\bibinfo
  {volume} {7}},\ \bibinfo {pages} {3937} (\bibinfo {year} {1974})}\BibitemShut
  {NoStop}%
\bibitem [{\citenamefont {Sa-yakanit}(1979)}]{SayakanitV_PRB_1979}%
  \BibitemOpen
  \bibfield  {author} {\bibinfo {author} {\bibfnamefont {V.}~\bibnamefont
  {Sa-yakanit}},\ }\bibfield  {title} {\bibinfo {title} {Electron density of
  states in a {Gaussian} random potential: Path-integral approach},\ }\href
  {https://doi.org/10.1103/PhysRevB.19.2266} {\bibfield  {journal} {\bibinfo
  {journal} {Phys. Rev. B}\ }\textbf {\bibinfo {volume} {19}},\ \bibinfo
  {pages} {2266} (\bibinfo {year} {1979})}\BibitemShut {NoStop}%
\bibitem [{\citenamefont {Vargiamidis}\ and\ \citenamefont
  {Polatoglou}(2005)}]{VargiamidisV_PRB_2005}%
  \BibitemOpen
  \bibfield  {author} {\bibinfo {author} {\bibfnamefont {V.}~\bibnamefont
  {Vargiamidis}}\ and\ \bibinfo {author} {\bibfnamefont {H.~M.}\ \bibnamefont
  {Polatoglou}},\ }\bibfield  {title} {\bibinfo {title} {Conductance of a
  quantum wire with a {Gaussian} impurity potential and variable
  cross-sectional shape},\ }\href {https://doi.org/10.1103/PhysRevB.71.075301}
  {\bibfield  {journal} {\bibinfo  {journal} {Phys. Rev. B}\ }\textbf {\bibinfo
  {volume} {71}},\ \bibinfo {pages} {075301} (\bibinfo {year}
  {2005})}\BibitemShut {NoStop}%
\bibitem [{\citenamefont {Yuan}\ \emph {et~al.}(2010)\citenamefont {Yuan},
  \citenamefont {De~Raedt},\ and\ \citenamefont {Katsnelson}}]{YuanS_PRB_2010}%
  \BibitemOpen
  \bibfield  {author} {\bibinfo {author} {\bibfnamefont {S.}~\bibnamefont
  {Yuan}}, \bibinfo {author} {\bibfnamefont {H.}~\bibnamefont {De~Raedt}},\
  and\ \bibinfo {author} {\bibfnamefont {M.~I.}\ \bibnamefont {Katsnelson}},\
  }\bibfield  {title} {\bibinfo {title} {Electronic transport in disordered
  bilayer and trilayer graphene},\ }\href
  {https://doi.org/10.1103/PhysRevB.82.235409} {\bibfield  {journal} {\bibinfo
  {journal} {Phys. Rev. B}\ }\textbf {\bibinfo {volume} {82}},\ \bibinfo
  {pages} {235409} (\bibinfo {year} {2010})}\BibitemShut {NoStop}%
\bibitem [{\citenamefont {Suzuura}\ and\ \citenamefont
  {Ando}(2016)}]{SuzuuraH_PRB_2016}%
  \BibitemOpen
  \bibfield  {author} {\bibinfo {author} {\bibfnamefont {H.}~\bibnamefont
  {Suzuura}}\ and\ \bibinfo {author} {\bibfnamefont {T.}~\bibnamefont {Ando}},\
  }\bibfield  {title} {\bibinfo {title} {Theory of {Hall} effect in
  two-dimensional giant {Rashba} systems},\ }\href
  {https://doi.org/10.1103/PhysRevB.94.035302} {\bibfield  {journal} {\bibinfo
  {journal} {Phys. Rev. B}\ }\textbf {\bibinfo {volume} {94}},\ \bibinfo
  {pages} {035302} (\bibinfo {year} {2016})}\BibitemShut {NoStop}%
\bibitem [{\citenamefont {Datta}(2009)}]{Datta_2009}%
  \BibitemOpen
  \bibfield  {author} {\bibinfo {author} {\bibfnamefont {S.}~\bibnamefont
  {Datta}},\ }\href@noop {} {\emph {\bibinfo {title} {Electronic transport in
  mesoscopic systems}}},\ \bibinfo {edition} {1st}\ ed.,\ \bibinfo {series}
  {Cambridge studies in semiconductor physics and microelectronic engineering}\
  No.~\bibinfo {number} {3}\ (\bibinfo  {publisher} {Cambridge Univ. Press},\
  \bibinfo {address} {Cambridge},\ \bibinfo {year} {2009})\BibitemShut
  {NoStop}%
\bibitem [{\citenamefont {Mahan}(2000)}]{Mahan_2000}%
  \BibitemOpen
  \bibfield  {author} {\bibinfo {author} {\bibfnamefont {G.~D.}\ \bibnamefont
  {Mahan}},\ }\href@noop {} {\emph {\bibinfo {title} {Many-particle
  physics}}},\ \bibinfo {edition} {3rd}\ ed.,\ Physics of solids and liquids\
  (\bibinfo  {publisher} {Kluwer Academic/Plenum Publishers},\ \bibinfo
  {address} {New York},\ \bibinfo {year} {2000})\BibitemShut {NoStop}%
\bibitem [{\citenamefont {Murzin}(2000)}]{MurzinS_P_2000}%
  \BibitemOpen
  \bibfield  {author} {\bibinfo {author} {\bibfnamefont {S.~S.}\ \bibnamefont
  {Murzin}},\ }\bibfield  {title} {\bibinfo {title} {Electron transport in the
  extreme quantum limit in applied magnetic field},\ }\href
  {https://doi.org/10.1070/pu2000v043n04abeh000691} {\bibfield  {journal}
  {\bibinfo  {journal} {Phys. Usp.}\ }\textbf {\bibinfo {volume} {43}},\
  \bibinfo {pages} {349} (\bibinfo {year} {2000})}\BibitemShut {NoStop}%
\bibitem [{\citenamefont {Ma}\ \emph {et~al.}(2019)\citenamefont {Ma},
  \citenamefont {Xiao},\ and\ \citenamefont {Chan}}]{MaG_NRP_2019}%
  \BibitemOpen
  \bibfield  {author} {\bibinfo {author} {\bibfnamefont {G.}~\bibnamefont
  {Ma}}, \bibinfo {author} {\bibfnamefont {M.}~\bibnamefont {Xiao}},\ and\
  \bibinfo {author} {\bibfnamefont {C.~T.}\ \bibnamefont {Chan}},\ }\bibfield
  {title} {\bibinfo {title} {Topological phases in acoustic and mechanical
  systems},\ }\href {https://doi.org/10.1038/s42254-019-0030-x} {\bibfield
  {journal} {\bibinfo  {journal} {Nat. Rev. Phys.}\ }\textbf {\bibinfo {volume}
  {1}},\ \bibinfo {pages} {281} (\bibinfo {year} {2019})}\BibitemShut {NoStop}%
\bibitem [{\citenamefont {Ozawa}\ \emph {et~al.}(2019)\citenamefont {Ozawa},
  \citenamefont {Price}, \citenamefont {Amo}, \citenamefont {Goldman},
  \citenamefont {Hafezi}, \citenamefont {Lu}, \citenamefont {Rechtsman},
  \citenamefont {Schuster}, \citenamefont {Simon}, \citenamefont {Zilberberg},\
  and\ \citenamefont {Carusotto}}]{OzawaT_RMP_2019}%
  \BibitemOpen
  \bibfield  {author} {\bibinfo {author} {\bibfnamefont {T.}~\bibnamefont
  {Ozawa}}, \bibinfo {author} {\bibfnamefont {H.~M.}\ \bibnamefont {Price}},
  \bibinfo {author} {\bibfnamefont {A.}~\bibnamefont {Amo}}, \bibinfo {author}
  {\bibfnamefont {N.}~\bibnamefont {Goldman}}, \bibinfo {author} {\bibfnamefont
  {M.}~\bibnamefont {Hafezi}}, \bibinfo {author} {\bibfnamefont
  {L.}~\bibnamefont {Lu}}, \bibinfo {author} {\bibfnamefont {M.~C.}\
  \bibnamefont {Rechtsman}}, \bibinfo {author} {\bibfnamefont {D.}~\bibnamefont
  {Schuster}}, \bibinfo {author} {\bibfnamefont {J.}~\bibnamefont {Simon}},
  \bibinfo {author} {\bibfnamefont {O.}~\bibnamefont {Zilberberg}},\ and\
  \bibinfo {author} {\bibfnamefont {I.}~\bibnamefont {Carusotto}},\ }\bibfield
  {title} {\bibinfo {title} {Topological photonics},\ }\href
  {https://doi.org/10.1103/RevModPhys.91.015006} {\bibfield  {journal}
  {\bibinfo  {journal} {Rev. Mod. Phys.}\ }\textbf {\bibinfo {volume} {91}},\
  \bibinfo {pages} {015006} (\bibinfo {year} {2019})}\BibitemShut {NoStop}%
\bibitem [{\citenamefont {Nielsen}\ and\ \citenamefont
  {Ninomiya}(1983)}]{NielsenH_PLB_1983}%
  \BibitemOpen
  \bibfield  {author} {\bibinfo {author} {\bibfnamefont {H.}~\bibnamefont
  {Nielsen}}\ and\ \bibinfo {author} {\bibfnamefont {M.}~\bibnamefont
  {Ninomiya}},\ }\bibfield  {title} {\bibinfo {title} {The {Adler-Bell-Jackiw}
  anomaly and {Weyl} fermions in a crystal},\ }\href
  {https://doi.org/10.1016/0370-2693(83)91529-0} {\bibfield  {journal}
  {\bibinfo  {journal} {Phys. Lett. B}\ }\textbf {\bibinfo {volume} {130}},\
  \bibinfo {pages} {389} (\bibinfo {year} {1983})}\BibitemShut {NoStop}%
\bibitem [{\citenamefont {Son}\ and\ \citenamefont
  {Spivak}(2013)}]{SonD_PRB_2013}%
  \BibitemOpen
  \bibfield  {author} {\bibinfo {author} {\bibfnamefont {D.~T.}\ \bibnamefont
  {Son}}\ and\ \bibinfo {author} {\bibfnamefont {B.~Z.}\ \bibnamefont
  {Spivak}},\ }\bibfield  {title} {\bibinfo {title} {Chiral anomaly and
  classical negative magnetoresistance of {Weyl} metals},\ }\href
  {https://doi.org/10.1103/PhysRevB.88.104412} {\bibfield  {journal} {\bibinfo
  {journal} {Phys. Rev. B}\ }\textbf {\bibinfo {volume} {88}},\ \bibinfo
  {pages} {104412} (\bibinfo {year} {2013})}\BibitemShut {NoStop}%
\bibitem [{\citenamefont {Burkov}(2014)}]{BurkovA_PRL_2014}%
  \BibitemOpen
  \bibfield  {author} {\bibinfo {author} {\bibfnamefont {A.~A.}\ \bibnamefont
  {Burkov}},\ }\bibfield  {title} {\bibinfo {title} {Chiral anomaly and
  diffusive magnetotransport in {Weyl} metals},\ }\href
  {https://doi.org/10.1103/PhysRevLett.113.247203} {\bibfield  {journal}
  {\bibinfo  {journal} {Phys. Rev. Lett.}\ }\textbf {\bibinfo {volume} {113}},\
  \bibinfo {pages} {247203} (\bibinfo {year} {2014})}\BibitemShut {NoStop}%
\bibitem [{\citenamefont {Dai}\ \emph {et~al.}(2017)\citenamefont {Dai},
  \citenamefont {Du},\ and\ \citenamefont {Lu}}]{DaiX_PRL_2017}%
  \BibitemOpen
  \bibfield  {author} {\bibinfo {author} {\bibfnamefont {X.}~\bibnamefont
  {Dai}}, \bibinfo {author} {\bibfnamefont {Z.~Z.}\ \bibnamefont {Du}},\ and\
  \bibinfo {author} {\bibfnamefont {H.-Z.}\ \bibnamefont {Lu}},\ }\bibfield
  {title} {\bibinfo {title} {Negative magnetoresistance without chiral anomaly
  in topological insulators},\ }\href
  {https://doi.org/10.1103/PhysRevLett.119.166601} {\bibfield  {journal}
  {\bibinfo  {journal} {Phys. Rev. Lett.}\ }\textbf {\bibinfo {volume} {119}},\
  \bibinfo {pages} {166601} (\bibinfo {year} {2017})}\BibitemShut {NoStop}%
\bibitem [{\citenamefont {Andreev}\ and\ \citenamefont
  {Spivak}(2018)}]{AndreevA_PRL_2018}%
  \BibitemOpen
  \bibfield  {author} {\bibinfo {author} {\bibfnamefont {A.~V.}\ \bibnamefont
  {Andreev}}\ and\ \bibinfo {author} {\bibfnamefont {B.~Z.}\ \bibnamefont
  {Spivak}},\ }\bibfield  {title} {\bibinfo {title} {Longitudinal negative
  magnetoresistance and magnetotransport phenomena in conventional and
  topological conductors},\ }\href
  {https://doi.org/10.1103/PhysRevLett.120.026601} {\bibfield  {journal}
  {\bibinfo  {journal} {Phys. Rev. Lett.}\ }\textbf {\bibinfo {volume} {120}},\
  \bibinfo {pages} {026601} (\bibinfo {year} {2018})}\BibitemShut {NoStop}%
\bibitem [{\citenamefont {Black-Schaffer}\ \emph {et~al.}(2015)\citenamefont
  {Black-Schaffer}, \citenamefont {Balatsky},\ and\ \citenamefont
  {Fransson}}]{BlackSchafferA_PRB_2015}%
  \BibitemOpen
  \bibfield  {author} {\bibinfo {author} {\bibfnamefont {A.~M.}\ \bibnamefont
  {Black-Schaffer}}, \bibinfo {author} {\bibfnamefont {A.~V.}\ \bibnamefont
  {Balatsky}},\ and\ \bibinfo {author} {\bibfnamefont {J.}~\bibnamefont
  {Fransson}},\ }\bibfield  {title} {\bibinfo {title} {Filling of
  magnetic-impurity-induced gap in topological insulators by potential
  scattering},\ }\href {https://doi.org/10.1103/PhysRevB.91.201411} {\bibfield
  {journal} {\bibinfo  {journal} {Phys. Rev. B}\ }\textbf {\bibinfo {volume}
  {91}},\ \bibinfo {pages} {201411(R)} (\bibinfo {year} {2015})}\BibitemShut
  {NoStop}%
\bibitem [{\citenamefont {Li}\ \emph {et~al.}(2017)\citenamefont {Li},
  \citenamefont {Wang}, \citenamefont {Zheng}, \citenamefont {Wang},
  \citenamefont {Li},\ and\ \citenamefont {Yang}}]{LiS_FoP_2017}%
  \BibitemOpen
  \bibfield  {author} {\bibinfo {author} {\bibfnamefont {S.}~\bibnamefont
  {Li}}, \bibinfo {author} {\bibfnamefont {C.}~\bibnamefont {Wang}}, \bibinfo
  {author} {\bibfnamefont {S.-H.}\ \bibnamefont {Zheng}}, \bibinfo {author}
  {\bibfnamefont {R.-Q.}\ \bibnamefont {Wang}}, \bibinfo {author}
  {\bibfnamefont {J.}~\bibnamefont {Li}},\ and\ \bibinfo {author}
  {\bibfnamefont {M.}~\bibnamefont {Yang}},\ }\bibfield  {title} {\bibinfo
  {title} {Dynamic conductivity modified by impurity resonant states in doping
  three-dimensional {Dirac} semimetals},\ }\href
  {https://doi.org/10.1007/s11467-017-0742-2} {\bibfield  {journal} {\bibinfo
  {journal} {Front. Phys.}\ }\textbf {\bibinfo {volume} {13}},\ \bibinfo
  {pages} {137303} (\bibinfo {year} {2017})}\BibitemShut {NoStop}%
\bibitem [{\citenamefont {Pires}\ \emph {et~al.}(2022)\citenamefont {Pires},
  \citenamefont {Jo{\~{a}}o}, \citenamefont {Ferreira}, \citenamefont
  {Amorim},\ and\ \citenamefont {Lopes}}]{PiresJ__2022}%
  \BibitemOpen
  \bibfield  {author} {\bibinfo {author} {\bibfnamefont {J.~P.~S.}\
  \bibnamefont {Pires}}, \bibinfo {author} {\bibfnamefont {S.~M.}\ \bibnamefont
  {Jo{\~{a}}o}}, \bibinfo {author} {\bibfnamefont {A.}~\bibnamefont
  {Ferreira}}, \bibinfo {author} {\bibfnamefont {B.}~\bibnamefont {Amorim}},\
  and\ \bibinfo {author} {\bibfnamefont {J.~M. V.~P.}\ \bibnamefont {Lopes}},\
  }\bibfield  {title} {\bibinfo {title} {Anomalous transport signatures in
  {Weyl} semimetals with point defects},\ }\href@noop {} {\  (\bibinfo {year}
  {2022})},\ \Eprint {https://arxiv.org/abs/2205.15123} {arXiv:2205.15123}
  \BibitemShut {NoStop}%
\bibitem [{\citenamefont {Wang}\ \emph
  {et~al.}(2016{\natexlab{b}})\citenamefont {Wang}, \citenamefont {Lu},\ and\
  \citenamefont {Shen}}]{WangC_PRL_2016}%
  \BibitemOpen
  \bibfield  {author} {\bibinfo {author} {\bibfnamefont {C.~M.}\ \bibnamefont
  {Wang}}, \bibinfo {author} {\bibfnamefont {H.-Z.}\ \bibnamefont {Lu}},\ and\
  \bibinfo {author} {\bibfnamefont {S.-Q.}\ \bibnamefont {Shen}},\ }\bibfield
  {title} {\bibinfo {title} {Anomalous phase shift of quantum oscillations in
  3{D} topological semimetals},\ }\href
  {https://doi.org/10.1103/PhysRevLett.117.077201} {\bibfield  {journal}
  {\bibinfo  {journal} {Phys. Rev. Lett.}\ }\textbf {\bibinfo {volume} {117}},\
  \bibinfo {pages} {077201} (\bibinfo {year} {2016}{\natexlab{b}})}\BibitemShut
  {NoStop}%
\end{thebibliography}%

\end{document}